\documentclass[12pt,a4paper]{article}
\usepackage{jheppub}

\definecolor{orange}{rgb}{1.0, 0.65, 0.0}
\definecolor{purple}{rgb}{0.58, 0.0, 0.83}
\definecolor{pink}{rgb}{1.0, 0.0, 0.5}

\newcommand{\CLM}{$\mathbb{C}$LM$\,$}

\newcommand{\CLMs}{$\mathbb{C}$LMs$\,$}

\newcommand{\junk}[1]{}

\renewcommand{\Im}{\mathrm{Im}}

\title{\Large 
Pseudospectra of holographic diffusion: gauge fields breaking free from the master scalar
}
\author[1,2]{David Garcia-Fariña,}
\author[1]{Karl Landsteiner,}
\author[1,2]{Pau G. Romeu,}
\author[1,3]{Pablo Saura-Bastida}
\affiliation[1]{Instituto de F\'isica Te\'orica UAM/CSIC, Calle Nicol\'as Cabrera 13-15, 28049 Madrid, Spain}
\affiliation[2]{Departamento de F\'isica Te\'orica, Universidad Aut{\'o}noma de Madrid, Campus de Cantoblanco, 28049 Madrid, Spain}
\affiliation[3]{Departamento de Autom\'atica, Ingenier\'ia El\'ectrica y Tecnolog\'ia Electr\'onica, Universidad Polit\'ecnica de Cartagena, Calle Dr. Fleming, S/N, 30202 Cartagena, Murcia}

\preprint{\begin{flushright} IFT-UAM/CSIC-26-10\ \end{flushright}}

\emailAdd{david.garciafarinna@estudiante.uam.es}
\emailAdd{karl.landsteiner@csic.es}
\emailAdd{pau.garcia@uam.es}
\emailAdd{pablo.saura@upct.es}

\abstract{
We study pseudospectra of quasinormal frequencies and complex linear momenta of a U(1) gauge field in a Schwarzschild black branes in Anti-de Sitter.
We present a novel approach for computing the pseudospectra which uses directly the gauge field variables and contrast it to a conventional master scalar field approach. Upon clarifying a subtlety in the energy norm of the master scalar we show that the pseudospectra of both approaches conincide.
In the hydrodynamic regime we find that the hydrodynamic quasinormal frequency, the diffusive mode, is spectrally stable to a very good approximation. On the other hand hydrodynamic complex linear momenta show enhanced spectral instability as a consequence of a pole-collision at zero frequency.}

\usepackage{physics} 
\usepackage{tensor} 
\usepackage{subcaption}
\usepackage{graphicx}

\newtheorem{definition}{Def.}[section]

\begin{document}

\maketitle

\section{Introduction}

Eigenvalue problems for linear operators are ubiquitous in all of physics, playing a fundamental role in our understanding of both classical and quantum systems. For conservative systems, the corresponding linear operators are self-adjoint. This ensures that eigenvalue analysis is sufficient to understand the dynamics associated with these linear operators. Moreover, under small perturbations of the operators the effect on the eigenvalues will also be small. This allows us to disregard small contributions and use simplified models while keeping control of the effect of the discarded terms. On the other hand, non-conservative systems are characterized by non-self-adjoint and potentially non-normal operators. The existence of non-normality drastically changes the above picture as eigenvalues are no longer stable and very small perturbations to the operators can change them dramatically. 
Modeling non-conservative systems is therefore much more difficult since approximations might have large effects.
Moreover, eigenvalues do not completely characterize a non-normal operator. It is necessary to also address their stability properties.

The standard way to quantify spectral stability is by studying pseudospectra \cite{Trefethen:2005, Sjostrand:2019,davies:2007}, which give bounds on the maximal displacement eigenvalues can suffer under arbitrary perturbations of a given size. 
Unlike eigenvalues, pseudospectra crucially depend on a norm. This is quite natural as in order to assess stability it is necessary to be able to measure sizes. 
It is crucial to choose a norm that captures the correct physical notion of size for the system. It has been argued, that from a physical standpoint the norm should be a functional reproducing the energy of the eigenfunctions \cite{Trefethen:2005}.

In the context of gravitational physics, non-self-adjoint operators naturally appear in the study of quasinormal modes (QNMs) of black holes \cite{Nollert:1999ji, Kokkotas:1999bd, Berti:2009kk}. In these geometries the presence of an event horizon acting as a perfectly absorbing membrane renders the time evolution of fluctuations non-conservative. The generator of time-translations, whose eigenvalues are the quasinormal frequencies (QNFs), is a non-self-adjoint differential operator. In asymptotically flat spacetimes, fluctuations are also outgoing at null infinity, further contributing to the non-self-adjointness. Guided by the above reasoning, in \cite{Jaramillo:2020tuu} the study of pseudospectra of black holes QNMs was pioneered, finding that the QNFs of a Schwarzshild black hole are unstable. Since then, this same type of instability has also been observed in other asymptotically flat black holes in \cite{Destounis:2021lum, Cheung:2021bol, Berti:2022xfj, Konoplya:2022pbc, alsheikh:tel-04116011, Destounis:2023ruj,Boyanov:2024fgc,Cai:2025irl,Siqueira:2025lww} (see \cite{Boyanov:2022ark,Destounis:2025dck} also for horizonless compact objects and \cite{Cao:2024oud,Chen:2024mon,Cao:2025qws} for black holes beyond General Relativity), as well as in de Sitter (dS) \cite{Sarkar:2023rhp, Destounis:2023nmb,Warnick:2024usx} and anti-de Sitter (AdS) \cite{Arean:2023ejh, Cownden:2023dam, Boyanov:2023qqf,Carballo:2024kbk,Carballo:2025ajx} geometries. Further studies have also covered the implications of this spectral instability on the strong cosmic censorship \cite{Courty:2023rxk} and on the gravitational waves emitted in black hole mergers \cite{Jaramillo:2021tmt, Jaramillo:2022kuv}. The stability of full transmission modes was also recently discussed in \cite{Zhou:2025xdo}.
Alternatively to QNFs one can also study complex momentum modes in which the frequency is kept real and one solves for complex (angular) momentum. 
In asymptotically flat spacetimes these are Regge poles \cite{Andersson:1994rk} and in asymptotically AdS they describe absorption lengths \cite{Amado:2007pv}. Spectral stability of such modes has been discussed in \cite{Torres:2023nqg, Rosato:2024arw, Oshita:2024fzf, Garcia-Farina:2024pdd,Wu:2024ldo,Rosato:2025lxb,Xie:2025jbr}. 

Gauge/gravity duality \cite{Maldacena:1997re, Aharony:1999ti, Ammon:2015wua, zaanen2015holographic, Hartnoll:2018xxg} postulates that black hole geometries in asymptotically AdS spacetimes are dual to thermal states in a strongly coupled quantum field theory.  
In this context, black hole quasinormal frequencies are interpreted as poles of retarded propagators of the thermal field theory at real momentum $k$ \cite{Horowitz:1999jd, Birmingham:2001pj, Kovtun:2005ev}. Consequently, the study of quasinormal frequencies in AdS black holes has led to important insights into hydrodynamics and transport theory in the relativistic regime \cite{Policastro:2001yc, Baier:2007ix, Bhattacharyya:2007vjd}, with some remarkable results being the extremely low specific shear viscosity of holographic models of the quark-gluon plasma \cite{Kovtun:2004de}, phase transitions towards superconducting states \cite{Gubser:2008px, Hartnoll:2008vx, Amado:2009ts, Herzog:2009ci} and strongly coupled quantum critical phases \cite{Herzog:2007ij, Cubrovic:2009ye, Iqbal:2011ae}. 
Quasinormal modes also play an important role in quantum chaos \cite{Grozdanov:2017ajz,Blake:2018leo} via the pole skipping phenomenon \cite{Amado:2008ji}.
Understanding the stability of QNFs in AdS is therefore crucial for its potential implications for strongly coupled plasmas. 

Characterizing the stability of those QNFs that vanish as the momentum goes to zero (hydrodynamic QNFs) is of central importance. From the dual QFT perspective, these QNFs are related to conserved charges associated with global symmetries. They dominate the dynamics at large scales in the hydrodynamic regime and are believed to be universal features of the theory. The presence of spectral instability would then challenge this view and thus should be studied carefully.

In this paper we consider the simplest setup containing hydrodynamic modes, the longitudinal sector of a U(1) gauge field with standard Maxwell action 
\begin{equation}\label{eq:Maxwell Action}
   S= -\frac{1}{4}\int d^{d+1}x\sqrt{-g}\,F_{MN}F^{MN}\,,
\end{equation}
in a Schwarzschild AdS$_{d+1}$ black brane described by the following metric in regular coordinates 
\begin{align}\label{eq:Generic Metric Branes}
    ds^2 &=\frac{l^2}{(1-\rho)^2}\left[-f dt^2+2(1-f) dtd\rho+(2-f)d\rho^2+dy_1^2+d\bold{y}_\perp^2\right]\,,\\
    f & =1-(1-\rho)^d\,,\nonumber
\end{align}
where $l$ is the AdS length scale, which we set to unity without loss of generality, and $y_1$ and $y_\perp$ are coordinates parallel and perpendicular to the momentum, respectively. In these coordinates, the event horizon lies at $\rho=0$ and the AdS boundary at $\rho=1$. We have used the conformal symmetry of the dual QFT to set the temperature $T=d/(4\pi)$. The longitudinal gauge field sector contains a hydrodynamic QNF associated with diffusion of the U(1) charge in the dual QFT. The standard way to treat this sector is to introduce an effective scalar field in two dimensions, called master scalar, which on-shell is equivalent to the above gauge field. Pseudospectra for the master scalar are constructed by writing the scalar equations of motion as an eigenvalue problem and by introducing a norm given by the energy of the two-dimensional master scalar on a constant time slice. This is the approach followed in \cite{Cownden:2023dam}, where the authors studied the gravitoelectric sector of a Reissner-Nordström black brane. It is however not immediatly clear if the energy
norm of the master scalar and the associated field space captures all the relevant off-shell field configurations of the gauge field. 
For this reason we introduce a novel approach in which we work directly in terms of the gauge field. This allows us to employ as our energy norm the energy of the Maxwell field on a constant time slice, thus avoiding potential pitfalls. We claim that this approach is more fundamental as it does not require the use of the master scalar. Moreover, this ensures that our approach can be extended to more complicated setups where a master scalar is not easily available. This is indeed the case when we study the pseudospectra of complex momentum modes.

We find that both frameworks, the one based on the master scalar (MS) and the one where we work directly with the gauge field (GF) can be related using Hodge duality and are thus equivalent as long as one is careful with boundary terms in the energy norm for the MS. For any number of dimensions, the original problem of the longitudinal gauge field can be consistently reduced to an equivalent problem of a longitudinal gauge field in an effective 3-dimensional spacetime. Hodge duality in this 3-dimensional spacetime implies that the dynamics of the gauge field should be equivalent to those of a massless scalar living in the same 3-dimensional spacetime. The problem of studying the stability of the Hodge dual scalar field can be seen to be equivalent to the problem formulated in the MS framework, up a boundary term living in the AdS boundary that appears in the energy norm. Whenever this boundary term does not vanish on the space of functions, the MS energy norm fails to be positive definite and thus we claim that in those cases one should instead take the energy norm defined via the Hodge dual field as it is more fundamental.

We observe that although the non-hydrodynamic QNFs are generically unstable, the hydrodynamic one is increasingly stable at low momenta. This indicates that the hydrodynamic description of the system is robust under perturbations, in agreement with the physical expectation that the hydrodynamic description should be universal and thus not excessively sensitive to small perturbations. 

A main point of this paper is the stability of complex linear momenta (\CLMs) for the longitudinal U(1) gauge field. While QNFs are eigenvalues of the generator of time translations, \CLMs correspond to eigenvalues of the generator of spatial translations along some direction parallel to the brane. QNFs and \CLMs are dual to poles of the retarded Green's function at fixed momentum (relaxation times) and fixed frequency (absorption lengths), respectively. 
\CLMs also play an essential role in studying causality of the boundary field theory \cite{Landsteiner:2012gn, Gavassino:2023mad}. At zero frequency, the non-hydrodynamic \CLMs are dual to the glueball masses of a dimensionally compactified toy model for QCD \cite{Witten:1998zw, Csaki:1998qr, deMelloKoch:1998vqw, Brower:2000rp, Bak:2007fk}. \CLMs have also appeared in the holographic context in \cite{Sonner:2017jcf,Novak:2018pnv,Heller:2020uuy,Janik:2021jbq}. 

The study of the pseudospectra of \CLMs of a scalar field in an AdS$_{4+1}$ black brane was carried out in \cite{Garcia-Farina:2024pdd}. There, it was found that \CLMs are generically more stable than QNFs. Moreover \CLM pseudospectra avoid the non-convergence problem present for QNF pseudospectra \footnote{Pseudospectra for QNFs do not converge in the energy norm for points sufficiently deep in the lower complex half plane in the limit $N\rightarrow \infty$ where $N$ is the gridsize of a given discretization via pseudospectral methods. For details on this see \cite{Warnick:2013hba, Boyanov:2022ark}.}.
We find\CLM pseudospectra to be convergent. However, we find that while non-hydro \CLMs are more stable than non-hydro QNFs, the hydrodynamic \CLMs are increasingly unstable as we approach the hydrodynamic regime. This suggests that the hydrodynamic description in the \CLM picture is unstable while in the QNF picture it is stable. Although this seems puzzling at first we argue that it is what one should expect. In the \CLM picture the two hydrodynamic \CLMs behave at small frequency as $k_\pm\propto\pm\sqrt{i\omega}$ indicating that collide at zero frequency forming an exceptional point. Exceptional points are characterized by their increased spectral instability \cite{Cao:2025afs} and thus we find that the hydro \CLMs are unstable as we approach the hydrodynamic regime. Hence the instability of the hydro \CLMs is signaling that in the vicinity of pole collisions there is an enhancement in spectral instability. On the other hand, for the hydro QNF there is no pole collision in the hydrodynamic regime and thus no enhancement in instability. Then, both pictures are consistent with the hydrodynamic description. Hydrodynamics predicts a pole collision for the \CLMs which in turn results in them being unstable, while in the QNF picture as there is no pole collision the hydro QNF is stable. The different stability properties found for the hydrodynamic mode depending on whether we describe it via QNFs or \CLMs agree with the observation of \cite{Garcia-Farina:2024pdd} that stability properties of QNFs and \CLMs are fundamentally different. From a physical standpoint perturbing while keeping the frequency fixed (\CLM picture) is fundamentally different from perturbing with fixed momentum (QNF picture) and thus one generically expects to find nonequivalent stability properties as we do. This behavior is not specific to the black hole realization but an expected generic feature of the diffusion mode. 

The paper is organized as follows. We begin in section \ref{sec:Pseudospectra and stability} with a brief review of pseudospectrum drawing heavily from \cite{Trefethen:2005}. We pay special attention to the pseudospectra of rectangular matrices which plays a central role in our novel formulation of pseudospectra based on gauge field components, which we discuss in section \ref{sec:Longitudinal gauge field}. 
In section \ref{sec:MS approach} we review the master scalar approach to pseudospectra of QNFs and discuss the associated energy norm.

In section \ref{sec:Hodge duality} we comment on how 3-dimensional Hodge duality offers a natural way to introduce the master scalar and its positive definite energy norm. We also show how to compute \CLMs pseudospectra in the master scalar framework. 

We present our results and give details on the boundary conditions in section \ref{sec:Holographic setup}. We discuss pseudospectra of \CLMs and QNFs for the black branes of equation \eqref{eq:Generic Metric Branes} in 6 and 5 dimensions.  We also showcase the validity of our formulation for spherical black holes by computing pseudospectra of QNFs of an AdS$_{4+1}$ black hole in subsection \ref{subsec:QNFs of SAdS5 black hole}.
In section \ref{sec: Conclusions} we summarize our findings, present our conclusions and give an outlook on further studies. 

Some remarks on the subtleties arising when dealing with pseudospectra of generalized eigenvalue problems are given in appendix \ref{app:Stability of general eigenvalue problems}. Further details on the numerical methods are provided in appendix \ref{app:Numerical methods}. Lastly, in appendices \ref{app:Convergence of the relative difference}-\ref{app:Convergence tests pseudo} we present some relevant convergence tests.

\section{Pseudospectra and stability}\label{sec:Pseudospectra and stability}

In this section we summarize some important definitions and results concerning pseudospectra and condition numbers, 
and their relevance for the study of the spectral stability of matrices such as the ones arising from the discretization of differential operators. We refer the reader to \cite{Trefethen:2005} for further details and for the generalization to operators on infinite-dimensional Hilbert spaces.

Given a square matrix $M\in \mathbb{C}^{N\times N}$, its spectrum $\sigma(M)$ is given by the set of eigenvalues $\left\{\lambda_i\right\}$ solutions to the equation
\begin{equation}
    (M-\lambda_i\mathbb{I})u_i=0 \,,
\end{equation}
with $u_i$ the corresponding eigenvector. Alternatively, the spectrum of a generic operator can be defined as the set of points $z\in\mathbb{C}$ where the the resolvent $\mathcal{R}(M;z)=(M-z\mathbb{I})^{-1}$ is not defined. For matrices both definitions are equivalent, although this is not generally the case for arbitrary operators.

An important property of eigenvalues is that for self-adjoint matrices (or, in more physical terms, conservative systems), the spectral theorem ensures that if we perturb the system with a perturbation of size $\varepsilon$ the eigenvalues of the perturbed operator cannot suffer a displacement greater than $\varepsilon$ \cite{kato2013perturbation, Courant:1989}. This property holds in general for any normal matrix $M$ satisfying $\left[M,M^\dagger\right]=0$. Let us briefly explain our notation: we denote the complex conjugate with a bar, complex conjugate transpose with an asterisk and adjoint, defined as $\langle M^\dagger u, v\rangle = \langle u,M v\rangle$, with a dagger. 

However, in non-conservative systems normality is not guaranteed and small perturbations could potentially alter the spectrum in a significant manner. For that reason, one concludes~\cite{Trefethen:2005} that in non-conservative systems eigenvalue analysis alone is insufficient as the spectrum might be unstable.

To probe the stability of eigenvalues one introduces the notion of $\varepsilon$-pseudospectrum, which can be defined in three mathematically equivalent ways \cite{Trefethen:2005}:
\begin{definition}[Resolvent norm approach]
    \label{def:Pseudo Definition 1}
    Given a square matrix $M\in \mathbb{C}^{N\times N}$, and $\varepsilon>0$, the $\varepsilon$-pseudospectrum $\sigma_\varepsilon(M)$ is
    \begin{equation}\label{eq:Pseudo Definition 1}
        \sigma_\varepsilon(M)=\{z\in \mathbb{C} : \lVert \mathcal{R}(z;M) \rVert>1/\varepsilon \}\,,
    \end{equation}
    with the convention $\lVert \mathcal{R}(z;M) \rVert=\infty$ for $z\in\sigma(M)$.
\end{definition}

\begin{definition}[Perturbative approach]
    \label{def:Pseudo Definition 2}
    Given a square matrix $M\in \mathbb{C}^{N\times N}$, and $\varepsilon>0$, the $\varepsilon$-pseudospectrum $\sigma_\varepsilon(M)$ is
    \begin{equation}\label{eq:Pseudo Definition 2}
        \sigma_\varepsilon(M)=\{z\in \mathbb{C}, \exists V,\rVert V \lVert<\varepsilon : z\in\sigma(M+V) \}\,.
    \end{equation}
\end{definition}

\begin{definition}[Pseudoeigenvalue approach]
    \label{def:Pseudo Definition 3}
    Given square matrix $M\in \mathbb{C}^{N\times N}$, and $\varepsilon>0$, the $\varepsilon$-pseudospectrum $\sigma_\varepsilon(M)$ is
    \begin{equation}\label{eq:Pseudo Definition 3}
        \sigma_\varepsilon(M)=%
        \{z\in \mathbb{C},\exists u^\varepsilon \in \mathbb{C}^N :\; \rVert (M-z\mathbb{I}) u^\varepsilon\lVert<\varepsilon\lVert u^\varepsilon \rVert\}\,,
    \end{equation}
    where $u^\varepsilon$ is a $\varepsilon$-pseudoeigenvector with $\varepsilon$-pseudoeigenvalue $z$.
\end{definition}

Note that, contrary to the spectrum, the pseudospectrum depends on the operator norm
\begin{equation}\label{eq:OperatorNorm_Generic}
    \rVert V \lVert=\max_{u\in \mathbb{C}^N}\frac{\rVert Vu \lVert}{\rVert u \lVert}\,,
\end{equation}
as, in order to quantify stability, it needs a notion of what constitutes a small perturbation.

Remarkably, definition \ref{def:Pseudo Definition 2} corresponds to the physical intuition we were seeking: the $\varepsilon$-pseudospectrum constitutes the maximal region containing all possible displacements of the eigenvalues under perturbations of size $\varepsilon$. It is then quite natural to represent the pseudospectrum as a contour map indicating the boundaries of these regions for multiple values of $\varepsilon$. However, computationally definition \ref{def:Pseudo Definition 2} is very inefficient as one should compute the spectra for all possible bounded perturbations. For that reason, when studying general instability properties one instead uses definition \ref{def:Pseudo Definition 1} which, despite lacking such a clear physical interpretation, is much more manageable to compute in a finite-dimensional setting such as the ones arising when employing numerical approximations to the operators. 

Another useful tool for studying the stability of eigenvalues is the set of condition numbers $\{\kappa_i\}$ defined as
\begin{equation}\label{eq:ConditionNumbers_squareMatrix}
        \kappa_i=\frac{\norm{v_i}\norm{u_i}}{|\expval{v_i,u_i}|}\,,
\end{equation}
with $\expval{\cdot,\cdot}$ the inner product associated with the norm $\norm{\cdot}$, $u_i$ the right-eigenvector satisfying $M u_i=\lambda_i u_i$ and $v_i$ the left-eigenvector satisfying $M^\dagger v_i=\bar{\lambda}_i v_i$. 
Condition numbers manage to quantify the effect of perturbations of size $\varepsilon$ through only the knowledge of the orthogonality between the eigenvectors of the unperturbed operator and its adjoint. Explicitly, for a perturbation $V$ of size $\varepsilon$ we have:
\begin{equation}\label{eq:condnumbers_displacement}
    |\lambda_i(\varepsilon)-\lambda_i|\leq \varepsilon  \kappa_i \,,
\end{equation}
where $\{\lambda_i(\varepsilon)\}$ are the eigenvalues of the perturbed operator $M(\varepsilon)=M+V$ and $||V||=\varepsilon$. For normal matrices all eigenvalues are stable and have condition number 1.

\subsection{Pseudospectra of rectangular matrices}\label{subsec:Pseudospectra of rectangular matrices}

As we shall see explicitly in section \ref{sec:Longitudinal gauge field}, rectangular matrices typically appear when discretizing an eigenvalue problem subject to a constraint. 
In these cases, pseudoresonances can become important and spectra often fail to give a complete description of the physics. In fact, a rectangular matrix can have an empty spectrum while presenting highly non-trivial dynamics associated with pseudoresponances.

The above discussion of pseudospectra for $N\times N$ square matrices can be easily extended to rectangular matrices (see chapter 8 of \cite{Trefethen:2005} for a more detailed discussion).
For the sake of generality, let us further assume that these rectangular matrices solve a generalized eigenvalue problem of the form\footnote{As we will see in the following sections, we typically do get standard eigenvalue problems when dealing with rectangular matrices. We construct the rectangular problem to be equivalent to a square problem restricted to a subsection of the function space satisfying a given constraint. Hence, the identity matrix is replaced by a matrix that generically differs from the rectangular identity but is equivalent to the restriction of the square identity.}
\begin{equation}
    (M-\lambda_i Q)u_i=0\,.
\end{equation}
Then, definitions \ref{def:Pseudo Definition 1}-\ref{def:Pseudo Definition 3} above hold provided we change $\mathbb{I}\rightarrow Q$ and properly take into account that a matrices $M\,,Q\in \mathbb{C}^{M\times N}$ ($M\geq N$) are operators from a $\mathbb{C}^N$ vector space with inner product $\expval{\cdot,\cdot}_{G_1}$ to a $\mathbb{C}^M$ vector space with inner product $\expval{\cdot,\cdot}_{G_2}$.\footnote{When dealing with generalized eigenvalue problems there are multiple alternative definitions for pseudospectrum (see e.g., chapter 45 of \cite{Trefethen:2005} for a discussion). Our choice is consistent with the one presented in \cite{Cownden:2023dam} which was argued to be the physically-motivated one in the context of pseudospectra.} This requires us to generalize the definition of the operator norm \eqref{eq:OperatorNorm_Generic} to
\begin{align}
    \rVert V \lVert&=\rVert V \lVert_{G_1\rightarrow G_2}=\max_{u\in \mathbb{C}^N}\frac{\rVert Vu \lVert_{G_2}}{\rVert u \lVert_{G_1}}\qquad \text{for } V:\mathbb{C}^N\rightarrow \mathbb{C}^M\,,\\
    \rVert W \lVert&=\rVert W \lVert_{G_2\rightarrow G_1}=\max_{u\in \mathbb{C}^M}\frac{\rVert Wu \lVert_{G_1}}{\rVert u \lVert_{G_2}}\qquad \text{for } W:\mathbb{C}^M\rightarrow \mathbb{C}^N\,,
\end{align}
and of the resolvent $\mathcal{R}(z;M): \mathbb{C}^M\rightarrow \mathbb{C}^N$ to
\begin{equation}
    \mathcal{R}(z;M)=\left[(M-z Q)^*(M-z Q)\right]^{-1}(M-z Q)^*=(M-z Q)^+
\end{equation}
where we have swapped the usual inverse for the pseudoinverse.

Regarding condition numbers, in order to preserve the physical interpretation of equation \eqref{eq:condnumbers_displacement}, we have to modify equation \eqref{eq:ConditionNumbers_squareMatrix} to the following
\begin{equation}\label{eq:ConditionNumbers_rectangularMatrix}
        \kappa_i=\frac{\norm{v_i}_{G_2}\norm{u_i}_{G_1}}{|\expval{v_i,Qu_i}_{G_2}|}\,,
\end{equation}
with $u_i$ the right-eigenvector satisfying $(M-Q\lambda_i) u_i=0$ and $v_i$ the left-eigenvector satisfying $(M^\dagger-\bar{\lambda}_i Q^\dagger)v_i=0$. Here, for an operator $V:\mathbb{C}^N\rightarrow \mathbb{C}^M$, we define the adjoint $V^\dagger:\mathbb{C}^M\rightarrow \mathbb{C}^N$ as a solution to\footnote{Note that $V^\dagger$ is only uniquely defined on the image of $V$. Nonetheless, this ambiguity does not affect condition numbers.}
\begin{equation}
    \expval{V^\dagger v,u}_{G_1}=\expval{v,Vu}_{G_2}\,,\qquad \text{for } v\in \mathbb{C}^M \,\text{and } u\in \mathbb{C}^N\,.
\end{equation}
Note that $v_i$ in equation \eqref{eq:ConditionNumbers_rectangularMatrix} is not unique as $(M^\dagger-\bar{\lambda}_i Q^\dagger)v_i=0$ is an underdetermined system of equations. This in turn implies that $\kappa_i$ depends on the particular choice of $v_i$. Hence, to impose the strictest constraint \eqref{eq:condnumbers_displacement}, we have to minimize $\kappa_i$ over all possible solutions $v_i$. This requires minimization over $M-N-1$ complex variables and thus is numerically very intensive.

With these modifications in mind, let us collect an important result which will allow us to greatly simplify numerical computations \cite{Trefethen:2005},
    Given two operators $M,\,Q\in \mathbb{C}^{N\times M}$ from a $\mathbb{C}^N$ vector space with inner product $\expval{\cdot,\cdot}_{G_1}$ to a $\mathbb{C}^M$ ($M\geq N$) vector space with inner product $\expval{\cdot,\cdot}_{G_2}$ where    \begin{equation}
        \label{eq:G-norm in th:psuedospectrum in different norms }  
        \expval{v,u}_{G_1}=\sum_{i,j=1}^N\bar{v}^i (G_{1})_{ij} \ u^j \,,\qquad \expval{v,u}_{G_2}=\sum_{i,j=1}^M\bar{v}^i (G_2)_{ij} \ u^j\,, 
    \end{equation}
    with $G_{1}=T^*T$ and $G_{2}=F^*F$ symmetric positive definite $N\times N$ and $M\times M$ matrices.
        The $\varepsilon$-pseudospectrum associated with the eigenvalue problem $(M-\lambda Q)u=0$ ca be computed as
        \begin{equation}
        \label{eq:pseudo relation in th:psuedospectrum in different norms}      
        \sigma_\varepsilon(M)=\left\{ z\in\mathbb{C}:s_{\min}\left[F(M-z Q)T^{-1}\right]>\varepsilon\right\}\,.
        \end{equation} 
        where $s_{\min}$ is defined as 
        \begin{equation}
            s_{\min}(X)=\min\sqrt{\sigma(X^*X)}\,.
        \end{equation}
        The condition numbers $\{\kappa_{i}\}$ of the eigenvalue problem $(M-\lambda Q)u=0$ are 
        \begin{equation}
        \label{eq:condition numbers relation in th:psuedospectrum in different norms} 
            \kappa_i=\frac{\norm{\tilde{v}_i}_{2(M)}\norm{\tilde{u}_i}_{2(N)}}{|\expval{\tilde{v}_i,FQT^{-1}\tilde{u}_i}_{2(M)}|}\,,
        \end{equation}
        where $\expval{\cdot,\cdot}_{2(L)}$ denotes the usual euclidean inner product in $\mathbb{C}^L$ and $\tilde{u}_i$ and $\tilde{v}_i$ satisfy
        \begin{equation}
            \left[F(M-\lambda_iQ)T^{-1}\right] \tilde{u}_i=0\,,\qquad \left[F(M-\lambda_iQ)T^{-1}\right]^*\tilde{v}_i=0\,,
        \end{equation}
        with $\tilde{v}_i$ chosen to be the solution minimizing $\kappa_i$.

\subsubsection{An example}

In order to illustrate the previous discussion on pseudospectra of rectangular eigenvalue problems, here we consider a simple example. Let us first start by motivating how a rectangular eigenvalue problem might arise from a constrained system. Consider a standard eigenvalue problem
\begin{equation}\label{eq:example eigenvalue problem rectangular 1}
    Au-\lambda \mathbb{I}u=\left(
\begin{array}{cccc}
 1.50 & 5.40 & 1.78 & -8.42 \\
 0.146 & -1.85 & 0.794 & 3.52 \\
 -0.104 & -2.89 & -1.40 & 6.27 \\
 -1.10 & -3.69 & -2.03 & 6.04 \\
\end{array}
\right)   \left(
\begin{array}{c}
 u_1  \\
 u_2 \\
 u_3 \\
 u_4 \\
\end{array}
\right)-\lambda\left(\begin{array}{c}
 u_1  \\
 u_2 \\
 u_3 \\
 u_4 \\
\end{array}\right)=0
\end{equation}
where the fields are not independent and instead are forced to satisfy the constraint 
\begin{equation}\label{eq:example eigenvalue problem rectangular constraint}
    2u_1+\sqrt{2}u_2+u_3+u_4=0\,.
\end{equation}
Hence, formally we would like to find solutions to the eigenvalue problem \eqref{eq:example eigenvalue problem rectangular 1} living in the subspace $\mathcal{K}\subset\mathbb{C}^4$ isomorphic to $\mathbb{C}^3$, satisfying the constraint \eqref{eq:example eigenvalue problem rectangular constraint}. A straightforward way to do this is to reformulate the problem as a rectangular eigenvalue problem. First, we start by define a rectangular matrix $\mathcal{C}$ 
\begin{equation}
    \mathcal{C}=\left(
\begin{array}{ccc}
 1 & 0 & 0  \\
 0 & 1 & 0  \\
 0 & 0 & 1  \\
 -2 & -\sqrt{2} & -1  \\
\end{array}
\right)  
\end{equation}
which embeds vectors in $\mathbb{C}^3$ into the subspace $\mathcal{K}$ of $\mathbb{C}^4$. With this we then proceed to solve the following rectangular eigenvalue problem
\begin{equation}\label{eq:toymodel_eigenvaleq_rectangular}
    A\mathcal{C}v-\lambda \mathcal{C} v=\left(
\begin{array}{ccc}
 18.3 & 17.3 & 10.2 \\
 -6.9 & -6.8 & -2.7 \\
 -12.6 & -11.8 & -7.7 \\
 -13.2 & -12.2 & -8.1 \\
\end{array}
\right)    \left(
\begin{array}{c}
 v_1  \\
 v_2 \\
v_3 \\
\end{array}
\right)-\lambda\left(\begin{array}{ccc}
 1 & 0 & 0  \\
 0 & 1 & 0  \\
 0 & 0 & 1  \\
 -2 & -\sqrt{2} & -1   \\
\end{array}\right)\left(\begin{array}{c}
 v_1  \\
 v_2 \\
v_3 \\
\end{array}\right)=0
\end{equation}
which, by construction is equivalent to \eqref{eq:example eigenvalue problem rectangular 1} restricted to $\mathcal{K}$ as $\mathcal{C}$ is a one-to-one map of $\mathbb{C}^3$ into $\mathcal{K}$. It is worth mentioning, that for generic rectangular eigenvalue problems, we are not guaranteed to find solutions. Nonetheless in this case we do find two eigenvalues $\lambda=0$ and $\lambda=1$.

Now, having identified the rectangular eigenvalue problem, we seek to study the corresponding pseudospectrum. To do so we need to define inner products on $\mathbb{C}^3$ and $\mathbb{C}^4$. For $\mathbb{C}^4$ we take the standard euclidean inner product $\expval{\cdot,\cdot}_{4}$
\begin{equation}
    \expval{u_{(1)},u_{(2)}}_{4}=\sum_{i=1}^4 \bar{u}_{(1)}^i u_{(2)}^i \,,\qquad u_{(1)},u_{(2)}\in \mathbb{C}^4\,,
\end{equation}
and, for consistency, on $\mathbb{C}^3$ we take the inner product $\expval{\cdot,\cdot}_{3}$ to be
\begin{equation}
    \expval{v_{(1)},v_{(2)}}_3=\expval{\mathcal{C}v_{(1)},\mathcal{C}v_{(2)}}_4\,,\qquad v_{(1)},v_{(2)}\in \mathbb{C}^3\,,
\end{equation}
so that after the embedding we recover the $\mathbb{C}^4$ inner product of the corresponding states in $\mathcal{K}$. 

\begin{figure}
\centering
    \includegraphics[width=0.7\textwidth]{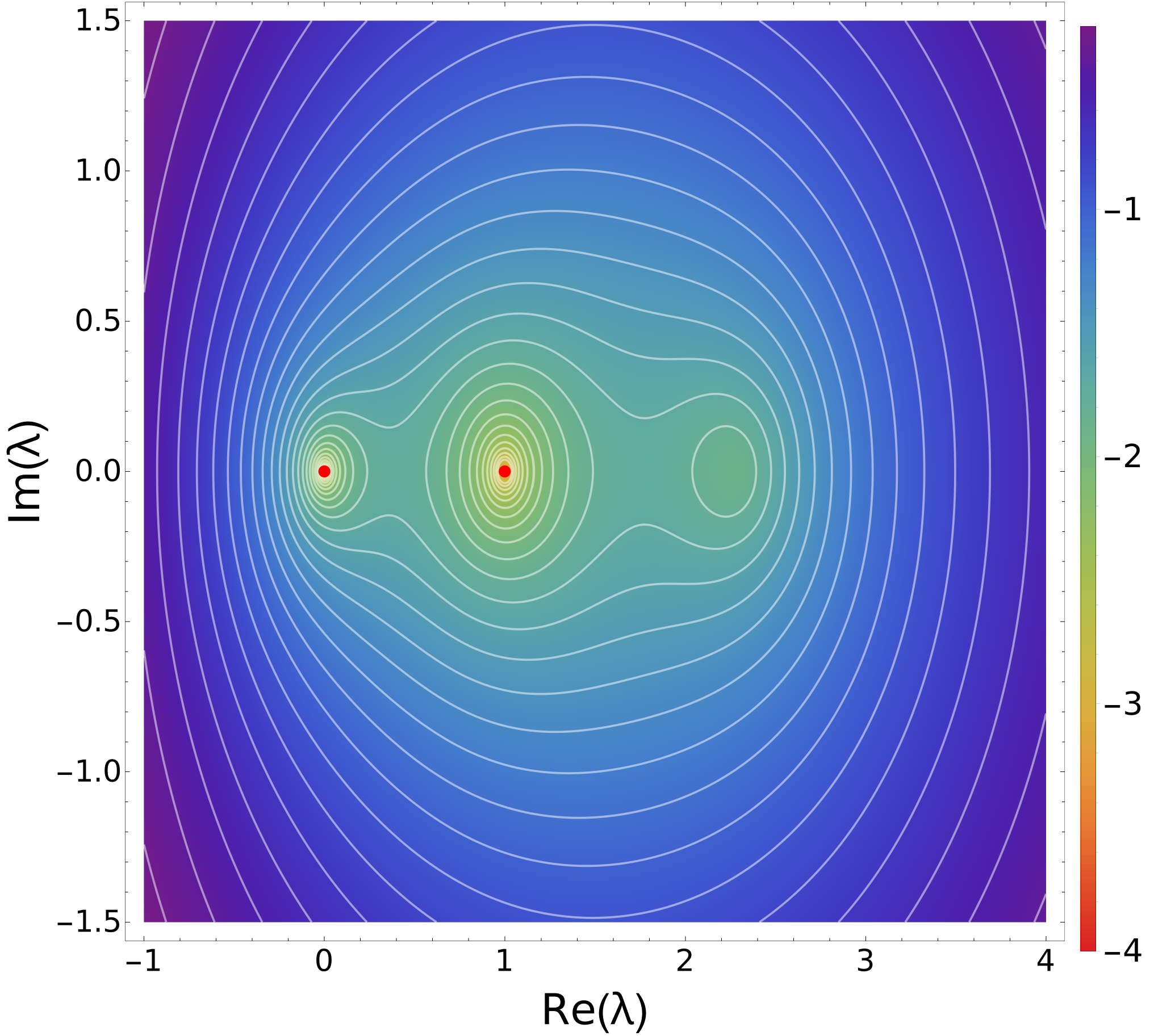}
    \caption{Pseudospectrum of $A\mathcal{C}-\lambda\mathcal{C}$ in \eqref{eq:toymodel_eigenvaleq_rectangular}. The red dots correspond to the eigenvalues, the white lines represent the boundaries of various $\varepsilon$-pseudospectra, and the heat map represent to the logarithm in base 10 of the inverse of the norm of the resolvent.}
    \label{fig:ToyModelRectangularPseudo}
\end{figure}

Hence, we can now identify the matrices $G_1=\mathcal{C}^*\mathbb{I}_3\mathcal{C}$ and $G_2=\mathbb{I}_3$ we use equations \eqref{eq:pseudo relation in th:psuedospectrum in different norms} and \eqref{eq:condition numbers relation in th:psuedospectrum in different norms} to compute the pseudospectrum and the condition numbers. In figure \ref{fig:ToyModelRectangularPseudo} we plot the pseudospectrum.
Remarkably, we observe that besides the two eigenvalues $\lambda=0$ and $\lambda=1$ there is also a pseudoresonance at $\lambda\sim 2.2$. Furthermore we can see that the restriction of $A$ to $\mathcal{K}$ is spectrally unstable, as small for $\varepsilon$ the boundary of the $\varepsilon$-pseudospectra contains both eigenvalues and the pseudoresponance. This instability is further confirmed by the condition numbers which are $9.3$ and $24.5$ for the eigenvalues $\lambda=0$ and $\lambda=1$, respectively.
The resolvent of a square matrix is an analytic function in $z$ and its norm obeys a maximum principle: it can take a maximum value only on the boundary of its domain of definition, i.e. the complex plane except for isolated singularities corresponding to the spectrum \cite{Trefethen:2005}. Therefore a pseudoresonance can not appear in this case. For rectangular matrices this is no longer true since the resolvent based on the pseudoinverse is not analytic in $z$ and therefore maxima (and minima) can appear in the pseudospectrum as in our simple example. In the following we will compute the pseudospectrum for longitudinal gauge fields in AdS black hole spacetimes. The presence of a constraint in the equations of motion will naturally lead us to a formulation in terms a rectangular eigenvalue problem.  This raises the question if pseudoresonances appear in this formulation. It would mean that the formulation in terms of a single masterfield equation leading to a square matrix problem was insufficient to address the pseudospectrum for gauge fields. As we shall see in the following this is not the case. No pseudoresonances appear in the rectangular formulation and indeed the pseudospectra of the approach based directly on the gauge fields and the approach based on a master field agree numerically to high accuracy.

\section{Longitudinal gauge field}\label{sec:Longitudinal gauge field}

Here we focus on the pseudospectra of QNFs and \CLMs of the longitudinal sector of U(1) gauge field with action \eqref{eq:Maxwell Action} in a generic black hole/brane background with metric of the form 
\begin{align}\label{eq:GenericBgMetric_regular}
    &ds^2=g_{\mu\nu}(\rho)dx^\mu dx^\nu+R(\rho)^2ds^2_{d-1} \,,\nonumber\\
    &ds^2_{d-1}=\gamma_{ij}(y)dy^idy^j\,,
\end{align}

This form of the metric is slightly more general than the black brane metrics in \eqref{eq:Generic Metric Branes}. The functions $g_{\mu\nu}$, $R$ and $\gamma_{ij}$ can be can be easily identified for the black brane. We prefer this slightly more general form since in section \ref{subsec:QNFs of SAdS5 black hole} we will briefly consider an AdS$_5$ black hole with spherical horizon topology. We take $x^\mu \in \{t,\rho\}$ and denote the rest of the coordinates by $y^i$, $i\in\{1,\dots,d-1\}$. 

As we shall see in section \ref{sec:MS approach}, one can study pseudospectra for this system introducing a master scalar field. However, in this approach the energy norm is postulated and apriori it is unclear whether it captures the notion of correct energy or if it instead misses features associated with the fundamental structure of the gauge field. For this reason, in this section we propose a new formulation where we work directly in terms of the gauge field variables. This allows us to avoid introducing new ingredients such as the master scalar and thus ensures we are capturing the right physics.
To make our presentation more transparent, we focus on the problem of studying the pseudospectrum of QNFs and we will discuss the necessary modifications for \CLMs in the next subsection. 

We start by decomposing a $(d+1)$-dimensional gauge field according to the isometries of $ds^2_{d-1}$ as
\begin{equation}\label{eq:General GF ansatz QNFs}
    A= A_\mu\,dx^\mu+\left(\partial_i A^{(L)}+A^{(T)}_i\right)\,dy^i\,,
\end{equation}
where $A^{(T)}_i$ satisfies $D^iA^{(T)}_i=0$ with $D$ the covariant derivative associated with $ds^2_{d-1}$. This decomposition makes explicit the existence of two distinct decoupled sectors. The first is the transverse sector, comprised of $A^{(T)}_i$, which is gauge invariant and transforms non-trivially under the isometries of $ds^2_{d-1}$. The second is the longitudinal sector, comprised of $\{A_\mu,A^{(L)}\}$ which behaves as a scalar under the isometries of $ds^2_{d-1}$ and transforms under gauge transformations in the following way
\begin{equation}\label{eq:ReducedgaugeTransf_preFourier}
    A_\mu\rightarrow A_\mu+\partial_\mu\chi\,,\qquad A^{(L)}\rightarrow A^{(L)}+\chi\,.
\end{equation}
From this point forward, we focus solely on the longitudinal sector and we completely remove the gauge freedom by fixing gauge to $A^{(L)}=0$.

We expand the gauge field components in eigenfunctions of the $ds^2_{d-1}$ Laplacian $\square_{d-1}$ 
\begin{equation} \label{eq:Basis}
     A_\mu = \sum_{\lambda^2,m} A^{(\lambda^2,m)}_\mu Y_{(\lambda^2,m)}  \quad \text{ with } \quad \square_{d-1} Y_{(\lambda^2,m)}
     = -\lambda^{2} \  Y_{(\lambda^2,m)} \, .
\end{equation}
where here $m$ is a multi-index resolving the degeneracy of $\lambda^2$ and the set $\{Y_{(\lambda^2,m)}\}$ is a complete basis of $\square_{d-1}$.\footnote{For a spherical black hole, $\lambda^2$ is the total angular momentum and $m$ labels the angular momenta of the smaller spheres forming the $S^{d-1}$. For a planar black hole $\lambda^2$ is the total linear momentum along the plane and $m$ labels the direction of the momentum. In this case the sum has to be exchanged by an integral.} Sectors with different $\lambda^2$ decouple and we can treat them independently. Henceforth, unless explicit spacetime dependence on $y^i$ is given, we always assume that the above decomposition has been performed and we drop the $(\lambda^2,m)$ superindex.

With this, in gauge $A_L=0$ and focusing on the sector with eigenvalue $-\lambda^2$ of the $ds^2_{d-1}$ Laplacian, we find that the Maxwell equations $\nabla_M F^{MN}=0$ can be written schematically an eigenvalue problem for the operator $i\partial_t$ plus a constraint independent of $i\partial_t$
\begin{align}\label{eq:GF_QNF_Eigenval_General}
    i\partial_t\Phi=\omega \Phi&=i\begin{pmatrix}
        \lambda^2 \frac{g}{R^2}\left(g^{t\rho}A_t+g^{\rho\rho}A_\rho\right)\\
        \left(\alpha_\rho+\partial_\rho A_t\right)\\
        -\frac{1}{g^{tt}}\left[ g^{t\rho}\left(\alpha_\rho+\partial_\rho A_t\right)+\frac{R^{3-d}}{\sqrt{-g}}\partial_\rho\left\{\sqrt{-g}R^{d-3}\left(g^{t\rho}A_t+g^{\rho\rho}A_\rho\right)\right\} \right]
    \end{pmatrix}\nonumber\\
    &=\begin{pmatrix}
        \mathcal{L}_{11}&\mathcal{L}_{12}&\mathcal{L}_{13}\\
        \mathcal{L}_{21}&\mathcal{L}_{22}&\mathcal{L}_{23}\\
        \mathcal{L}_{31}&\mathcal{L}_{32}&\mathcal{L}_{33}
    \end{pmatrix}\begin{pmatrix}
        \alpha_\rho\\A_\rho\\A_t
    \end{pmatrix}=\mathcal{L}\Phi\,,
\end{align}
\begin{equation}\label{eq:GF_QNF_Constraint_General}
    \mathcal{L}_0\Phi=\frac{1}{\sqrt{-g}R^{d-1}}\partial_\rho\left(\frac{R^{d-1}}{\sqrt{-g}}\alpha_\rho\right)-\frac{\lambda^2}{R^2}\left(g^{tt}A_t+g^{t\rho}A_\rho\right)=0\,,
\end{equation}
with $g=\det{g_{\mu\nu}}$, $\Psi=(\alpha_\rho,A_\rho,A_t)^T$, $\alpha_\rho=\partial_tA_\rho-\partial_\rho A_t$ an auxiliary variable introduced to make the problem linear in $\partial_t$ and $\omega$ the QNF. Although we have more equations than fields, the system itself is not overdetermined as the equations are related by the off-shell identity $\nabla_N\nabla_M F^{MN}=0$. Hence, naively one could try to forget the constraint and solve the eigenvalue problem \eqref{eq:GF_QNF_Eigenval_General}.
This leads however to unphysical modes appearing at $\omega=0$ associated with the fact that any solution $\mathcal{L}_C\Phi=c$ with $c$ a constant, fulfills the off-shell identity and the eigenvalue problem \eqref{eq:GF_QNF_Eigenval_General} despite failing to satisfy the constraint \eqref{eq:GF_QNF_Constraint_General}. Alternatively, we might try to work with a singular eigenvalue problem resulting from replacing a row in $(\mathcal{L}-\omega)\Phi$ by $\mathcal{L}_0\Phi$. However, this choice would complicate the physical interpretation as we would have to deal with a generalized eigenvalue problem (see appendix \ref{app:Stability of general eigenvalue problems} for a discussion) and can even lead to the appearance of spurious modes associated with the singular character of the problem which again would spoil the pseudospectrum making it difficult to discern physical from unphysical artifacts. Lastly, we could use \eqref{eq:GF_QNF_Constraint_General} to remove one of the gauge field variables (e.g. $A_t$) and build a standard eigenvalue problem for the remaining two by ignoring one of the equations in \eqref{eq:GF_QNF_Eigenval_General}. By construction, this new eigenvalue problem would be free from the previously discussed unphysical modes and thus could be used to study the pseudospectra of QNFs. However, this approach requires us to be able to solve analytically the constraint, which need not be possible in more complicated settings. In fact, for the \CLMs this is indeed not possible as in the constraint all the fields entering appear together with their derivatives. Moreover, from the perspective of being as fundamental as possible, this simplified approach seems potentially dangerous as it is using an on-shell equation and ignoring another, which suggests that we might be leaving out some bounded perturbations in our study of the pseudospectra.

In conclusion, to avoid making any assumptions that could restrict the perturbations allowed in pseudospectra, we would like to solve both the eigenvalue problem \eqref{eq:GF_QNF_Eigenval_General} and the constraint equation \eqref{eq:GF_QNF_Constraint_General} simultaneously without ignoring any of the equations. That is, we want to solve the eigenvalue problem \eqref{eq:GF_QNF_Eigenval_General} restricted to a subspace of the function space where $\Phi$ satisfies the constraint \eqref{eq:GF_QNF_Constraint_General}. Generically let us write 
\begin{equation}\label{eq:GF_QNF_Uplift_Generic}
    \Phi=\begin{pmatrix}
        \alpha_\rho\\A_\rho\\A_t
    \end{pmatrix}=\begin{pmatrix}
        1&0\\0&1\\\mathcal{C}_1&\mathcal{C}_2
    \end{pmatrix}\begin{pmatrix}
        \alpha_\rho\\A_\rho
    \end{pmatrix}=\mathcal{C}\Psi
\end{equation}
with $\mathcal{C}_1$ and $\mathcal{C}_2$ operators resulting from solving for $A_t$ in \eqref{eq:GF_QNF_Constraint_General}.\footnote{The constraint equation can be solved numerically in a grid to obtain a discretized version of $\mathcal{C}$ directly.} 
We can then study the following (generalized) rectangular eigenvalue problem
\begin{equation}
    \mathcal{L}_\mathcal{C}-\omega\,\mathcal{I}_\mathcal{C}=(\mathcal{L}-\omega\,\mathcal{I})\mathcal{C}\,,
\end{equation}
and its pseudospectrum (here $\mathcal{I}$ is the identity operator). 
This is equivalent to studying the spectrum and pseudospectrum of the restriction of $\mathcal{L}$ defined in \eqref{eq:GF_QNF_Eigenval_General} to the relevant subspace $\mathcal{K}$ of the function space $\mathcal{H}$ comprised of functions satisfying the constraint \eqref{eq:GF_QNF_Constraint_General}.\footnote{By construction $\mathcal{C}$ uplifts $\Psi=(\alpha_\rho,A_\rho)^T$ to a function $\Phi$ in $\mathcal{H}$ satisfying the constraint \eqref{eq:GF_QNF_Constraint_General}. Hence we abuse notation and use $\mathcal{K}$ to denote both the domain of $\mathcal{L_P}$ comprised of functions $\Psi$, as well as its uplift to functions $\Phi=\mathcal{C}\Psi$.} The eigenvalue problem for $\mathcal{L}_\mathcal{C}:\,\mathcal{K}\rightarrow\mathcal{H}$ is thus rectangular and, after discretization in a grid, the pseudospectrum can be studied according to our discussion in section \ref{subsec:Pseudospectra of rectangular matrices}.

Regarding the norms, we take the usual prescription and define the physically motivated norm from the energy. 
The energy momentum tensor of the gauge field is
\begin{equation}
    T_{MN}=F_{MS}\tensor{F}{_N^S}-\frac{1}{4}g_{MN}F_{SP}F^{SP}\,,
\end{equation}
and leads to the  following expression for the energy norm
\begin{equation}\label{eq:GF_QNF_Energy_General}
    E[A]=\int d\rho \sqrt{-g}R^{d-1}\left\{\frac{1}{|g|}\left|\partial_t A_\rho-\partial_\rho A_t\right|^2+\frac{\lambda^2}{R^2}\left(g^{\rho\rho}\left|A_\rho\right|^2+\left|g^{tt}\right| \left|A_t\right|^2\right)\right\}\,,
\end{equation}
where we have dropped the sum over the different eigenvalues $\{\lambda\}$ of the Laplacian of $ds^2_{d-1}$ as we in a sector with fixed $\lambda$. With this, introducing an auxiliary field $\alpha_\rho$ defined in equation \eqref{eq:GF_QNF_Constraint_General}, the final expression for the inner product is given by
\begin{equation}\label{eq:GF_QNF_InnerPdt_General}
    \expval{\Phi^{(1)},\Phi^{(2)}}_E=\int \sqrt{-g}R^{d-1}\left\{\frac{1}{|g|}\bar{\alpha}_\rho^{(1)}\alpha_\rho^{(2)}+\frac{\lambda^2}{R^2}\left(g^{\rho\rho}\bar{A}_\rho^{(1)}A_\rho^{(2)}+\left|g^{tt}\right| \bar{A}_t^{(1)}A_t^{(2)} \right)\right\}\,,
\end{equation}
In a bit more detail, we take this inner product to be the inner product defining $\mathcal{H}$, the space of functions $\Phi$ with finite energy. 
This is consistent with the usual approach for standard eigenvalue problems and allows for an adequate identification of the convergence bands of the QNF pseudospectrum \cite{Warnick:2013hba}. The inner product on $\mathcal{K}$ is given by the projection of the inner product on $\mathcal{H}$ to the subspace $\mathcal{K}$
\begin{equation}
    \expval{\Psi^{(1)},\Psi^{(2)}}_E=\expval{\mathcal{C}\Psi^{(1)},\mathcal{C}\Psi^{(2)}}_E \,, \qquad \text{for}\,\,\, \Psi^{(1)},\Psi^{(2)}\in \mathcal{K}\,.
\end{equation}

We finish this section by discussing the domain of convergence of the pseudospectra \cite{Warnick:2013hba,Boyanov:2023qqf,Garcia-Farina:2024pdd}.  
The lack of convergence of the pseudospectrum is related to the failure of the energy norm to properly impose infalling boundary conditions in the continuum limit. The analytic continuation of outgoing solutions deep into the lower half plane have finite energy and contribute to the pseudospectrum based on the energy norm. Let us assume that near the event horizon $\rho=0$ the metric takes the following form
\begin{equation}
    ds^2=(-4\pi T\rho\, dt^2+2dtd\rho+2d\rho^2)+s_{\text{BH}}\, ds^2_{d-1}
\end{equation}
with $s_{\text{BH}}$ and $T$, the entropy density and temperature of the black hole/brane, respectively. Then, from the near horizon behavior of $\expval{\mathcal{L}_\mathcal{C}\Psi^{(1)},\mathcal{L}_\mathcal{C}\Psi^{(1)}}$ it follows that modes with an outgoing component $\Psi^{(1)}\sim\rho^{-i\omega/(2\pi T)}$ yield a finite norm if $\Im(\omega)<-\pi T$. Hence, the domain of $\mathcal{L}_\mathcal{C}$ contains modes with $\Im(\omega)<-\pi T$ that have finite energy but are not infalling. Thus in the continuum limit the spectrum of $\mathcal{L}_\mathcal{C}$ is comprised of the QNFs with $\Im(\omega)>-\pi T$ together with the continuum $\Im(\omega)<-\pi T$. Hence, the pseudospectrum computed numerically in a grid only converges for $\Im(\omega)>-\pi T$. We comment on the physical implications of this lack of convergence in section \ref{subsubsec:QNFs of SAdS6 black brane - Numerical results} and we explicitly check the prediction for the band of convergence in appendix \ref{app:Convergence tests pseudo}.

\subsection{Modifications for \CLMs}\label{subsec: Modifications for CLMs}
To extend the approach described above to \CLMs of black branes we only have to make a few modifications. Firstly, we write the $ds^2_{d-1}$ portion of the metric in \eqref{eq:GenericBgMetric_regular} as
\begin{equation}
    ds^2_{d-1}=dy^2_1+d\bold{y}_\perp^2\,,
\end{equation}
and we use the rotational symmetry of $\mathbb{R}^{d-1}$ to fix the momentum $\boldsymbol{k}$ along $y_1$; i.e., we consider fluctuations depending only on $\{t,\rho,y_1\}$. Now, keeping in mind that for \CLMs we want to fix the frequency $\omega$ and study $k(\omega)$, we choose the following ansatz for the gauge field 
\begin{equation}
    A=A_t(t,\rho,y_1)\,dt+A_\rho(t,\rho,y_1)\,d\rho+ A_1 (t,\rho,y_1)\,dy_1\,,
\end{equation}
and we expand the time dependence in Fourier modes $e^{-i\omega t}$. As before, we have to deal with the gauge symmetry which for this ansatz reads
\begin{equation}
    A_t\rightarrow A_t+\partial_t\chi\,,\qquad A_\rho\rightarrow A_\rho+\partial_\rho\chi\,,\qquad A_1\rightarrow A_1+\partial_1\chi\,.
\end{equation}
In this case, we remove gauge redundancy by taking $A_t=0$ gauge.

With all this we conclude that, similar to the situation for QNFs, we can write the Maxwell equations as an eigenvalue problem for $-i\partial_1$ plus a constraint restricting the function space
\begin{align}\label{eq:GF_CLM_Eigenval_General}
    -i\partial_1\Phi=k \Phi&=-i\begin{pmatrix}
        R^2\left[i\omega g^{t\rho}\frac{\sqrt{-g}}{R^{d-1}}\partial_\rho\left(\frac{R^{d-1}}{\sqrt{-g}}A_\rho\right)+\omega^2 g^{tt} A_\rho\right]\\
        \alpha_\rho+\partial_\rho A_1\\
        R^2\left[-g^{\rho\rho}\frac{\sqrt{-g}}{R^{d-1}}\partial_\rho\left(\frac{R^{d-1}}{\sqrt{-g}}A_\rho\right)+i\omega g^{t\rho}A_\rho\right]
    \end{pmatrix}\nonumber\\
    &=\begin{pmatrix}
        \mathcal{L}_{11}&\mathcal{L}_{12}&\mathcal{L}_{13}\\
        \mathcal{L}_{21}&\mathcal{L}_{22}&\mathcal{L}_{23}\\
        \mathcal{L}_{31}&\mathcal{L}_{32}&\mathcal{L}_{33}
    \end{pmatrix}\begin{pmatrix}
        \alpha_\rho\\A_\rho\\A_1
    \end{pmatrix}=\mathcal{L}\Phi\,,
\end{align}
\begin{align}\label{eq:GF_CLM_Constraint_General}
    \mathcal{L}_0\Phi=&-i\omega \frac{g^{t\rho}}{R^2}\alpha_\rho + \frac{1}{\sqrt{-g}R^{d-1}}\partial_\rho\left(R^{d-3}\sqrt{-g}g^{\rho\rho} \alpha_\rho \right)\nonumber\\
    &+\omega^2 \frac{g^{tt}}{R^2}A_1+\frac{i\omega}{\sqrt{-g}R^{d-1}}\partial_\rho\left(R^{d-3}\sqrt{-g}g^{t\rho} A_1 \right)=0\,,
\end{align}
where here $\Phi=(\alpha_\rho,A_\rho,A_1)$ and $\alpha_\rho=\partial_1 A_\rho-\partial_\rho  A_1$. As before, we implement the constraint through an uplift $\mathcal{C}$ 
\begin{equation}
     \Phi=\begin{pmatrix}
        \alpha_\rho\\A_\rho\\A_1
    \end{pmatrix}=\begin{pmatrix}
        1&0\\0&1\\\mathcal{C}_1&\mathcal{C}_2
    \end{pmatrix}\begin{pmatrix}
        \alpha_\rho\\A_\rho
    \end{pmatrix}=\mathcal{C}\Psi\,,
\end{equation}
and instead study spectra and pseudospectra for the rectangular problem $(\mathcal{L}-k\mathcal{I})\mathcal{C}$.

With regards to the norm, we
follow \cite{Garcia-Farina:2024pdd} and take the energy norm as the time-average of the energy density along the $y^1$ direction.
Thus we take the following prescription for the inner product in $A_t=0$ gauge
\begin{equation}\label{eq:GF_CLM_InnerPdt_General}
    \expval{\Phi^{(1)},\Phi^{(2)}}_E=\int d\rho \sqrt{-g}R^{d-1}\left\{\frac{\omega^2}{|g|} \bar{A}_\rho^{(1)}{A}_\rho^{(2)} +\frac{1}{R^2}\left(g^{\rho\rho}\bar{\alpha}^{(1)}_\rho \alpha^{(2)}_\rho+\left|g^{tt}\right|\omega^2 \bar{A}^{(1)}_1 A^{(2)}_1\right)\right\}\,.
\end{equation}
As we shall see explicitly this inner product does capture the relevant physical notion of size as it reproduces an increasing stability for the gapped \CLMs as $\omega\rightarrow0$. 

Lastly we note that as for the master scalar, the fact that we are fixing $\Im(\omega)=0$ implies that the \CLM pseudospectrum will be convergent. Let us see this explicitly following the same approach as we did to identify the convergent region of the QNF pseudospectrum. We assume that near the event horizon $\rho=0$ the metric takes the form
\begin{equation}
    ds^2=(-4\pi T\rho\, dt^2+2dtd\rho+2d\rho^2)+s_{\text{BH}}\, ds^2_{d-1}
\end{equation}
where $s_{\text{BH}}$ and $T$ are the entropy density and the temperature of the black hole/brane, respectively. Then, from the near horizon behavior of $\expval{\mathcal{L}_\mathcal{C}\Psi^{(1)},\mathcal{L}_\mathcal{C}\Psi^{(1)}}$ it follows that modes with an outgoing component $\Psi^{(1)}\sim\rho^{-i\omega/(2\pi T)}$ yield a finite norm if $\Im(\omega)<-\pi T$. Hence, as we are working in the line $\Im(\omega)=0$, we are guaranteed that only infalling modes have finite energy. Thus the convergence problems present for the QNFs are absent when dealing with \CLMs \cite{Garcia-Farina:2024pdd}. The convergence properties of \CLM pseudospectra are explicitly checked in appendix \ref{app:Convergence tests pseudo}.

\section{Master scalar}\label{sec:MS approach}

The master field approach provides a systematic method to address the complexities associated with gauge redundancy. This framework expresses the $d+1$ components of the gauge vector in terms of $d-1$ independent, gauge-invariant degrees of freedom (see e.g. \cite{Santoni:2020xxx} for a review). In the case of a U(1) gauge field, this is achieved via the standard longitudinal and transverse decomposition described in section \ref{sec:Longitudinal gauge field}. Instead of imposing the gauge condition $A^{(L)}=0$, we work with gauge-invariant combinations
\begin{equation}
    a_\mu =  A_\mu  - \partial_\mu  A^{(L)} \,. 
\end{equation}
 
Beginning with the Maxwell action \eqref{eq:Maxwell Action}, the fields are expanded in a basis of eigenfunctions of the $(d - 1)$-dimensional Laplacian \eqref{eq:Basis}. Integrating over the transverse directions produces an effective two-dimensional action for each longitudinal mode
\begin{equation}
   S^{(L)} [a] = - \int d^2x \sqrt{-g} \left\{  R^{d-1}   \ g^{\mu \nu}\partial_\alpha \bar a_\mu  \partial^\alpha a_\nu+  \lambda^{2} R^{d-3}\  g^{\mu \nu} \bar a_\mu a_\nu \right\} \ .
\end{equation}
For notational simplicity, we have suppressed the mode labels $(\lambda^2,m)$. The two-dimensional field $a$ is massive and propagates one degree of freedom. To isolate this degree of freedom, we introduce an auxiliary scalar field $\psi$ through the interaction term
\begin{equation}
    S_c[\psi, a] = - \int d^2x \sqrt{-g} \left\{   R^{-2} \, \left| \lambda \ \psi+ R^{(d-1)/2} \ \epsilon^{\mu\nu}\nabla_\mu a_\nu \right|^2 \right\} \ .
\end{equation}
This term is chosen to cancel the kinetic term of $S^{(L)} [a]$. When the scalar $\psi$ is on shell, $S_c[\psi[a], a] = 0$ for any $a$. These choices allow us to find an exact solution for $a$ in the coupled system
\begin{equation} \label{eq:Map_MS}
    \delta_a \left( S^{(L)} [a] +S_c[\psi, a] \right) =0 \quad \to \quad a[\psi]_\mu= \frac{1}{ \lambda  \ R^{d-3}} \epsilon_{\mu \nu} \nabla^\nu\left(R^{(d-3)/2} \  \psi \right) \ .
\end{equation}
Substituting this relation back into the action yields a theory formulated solely in terms of the master scalar.
\begin{equation} \label{Action_MS}
\begin{aligned}
          S&^{(L)}  [a]  +S_c[\psi, a] =  S[\psi] = -\int d^2 x  \sqrt{-g} \left\{|\partial \psi |^2  + V(\phi, R) \left|\psi\right|^2 \right\} \ , \\
        V &= \frac{\lambda^2}{R^2}  + \frac{(d-3)}{2}  \left( \square \log\left( R\right) + \frac{(d-3)}{2} \nabla_\mu \log\left( R\right)  \nabla^\mu \log\left( R\right) \right) \ .
\end{aligned}
\end{equation}
So for any $\psi_{OS}$ extremizing \eqref{Action_MS}, using \eqref{eq:Map_MS}, we can obtain a solution for the longitudinal mode. Using the auxiliary field $\phi = \partial_t \psi $ we can write the EoMs as an eigenvalue problem 
\begin{equation}
\begin{aligned} \label{eq:MSevproblem}
     i \partial_t &\left(\begin{array}{c}
          \phi \\
          \psi 
    \end{array} \right) = \omega \left(\begin{array}{c}
          \phi \\
          \psi 
    \end{array} \right) = i \left(  \begin{array}{cc}
        \mathcal{L}^{(1)}  & \mathcal{L}^{(2)}  \\
         1 & 0
    \end{array}\right) \left(\begin{array}{c}
          \phi \\
          \psi 
    \end{array}\right) \,,\\ 
    & \mathcal{L}^{(1)}\phi =  -g^{tt} \left( \partial_\rho g_{t \rho}\phi + 2g_{t \rho} \partial_\rho \phi \right)\,, \\ 
    & \mathcal{L}^{(2)} \psi= -g^{tt} \left[ \partial_\rho \left( g_{tt} \partial_\rho \psi \right)+ V \psi \right] \,.
\end{aligned}
\end{equation}

\subsection{Boundary Conditions}
In the analysis of Quasinormal Modes, it is essential to consider not only the equations of motion but also the associated boundary conditions. Assuming we start from a gauge field on an AdS background, the potential can be expanded at large distances (large $R$), corresponding to the vicinity of the AdS boundary. Solving the equations of motion for the master scalar in this regime yields two solutions
\begin{equation} \label{eq:MS_NearBdy}
 \psi = \psi_{\pm} R^{\Delta_{\pm}} \, , \ \Delta_{\pm} = -\frac{1}{2} \pm \frac{d-4}{2}\ ,
\end{equation} 
Substituting this expansion into \eqref{eq:Map_MS} clarifies the correspondence: $\psi_+$ represents the source of the gauge field, whereas $\psi_-$ denotes the expectation value (vev) of the QFT operator dual to the gauge field.
On spaces with dimension five or greater, fixing the source of the gauge field corresponds to fixing the leading mode of the scalar (standard quantisation). For spaces with fewer than five dimensions, one should fix the subleading mode of the scalar (alternate quantisation \cite{Klebanov:1999tb}).
\subsection{Energy Norm}\label{subsec:Enorm MS}

We will use the energy norm to compute the pseudo-spectra for the eigenvalue problem of the master scalar.
There are two distinct methods for computing the energy. The standard approach involves calculating the energy directly from \eqref{Action_MS}, under the assumption that $V$ is independent of the metric, resulting in
\begin{equation}  \label{eq:EnergyNormMSnon}
     E[\psi]  =  \int d\rho \sqrt{-g} \, \left( -g^{tt}|\partial_t \psi|^2 + g^{\rho\rho}|\partial_\rho \psi|^2 +  V |\psi|^2  \right) \, .
\end{equation}
Alternatively, we can compute the energy from the original action \eqref{eq:Maxwell Action} and use the transformation \eqref{eq:Map_MS}, yielding
\begin{equation}  \label{eq:EnergyNormMS}
     E[a[\psi]]  =  \int d\rho \sqrt{-g} \, \left( -g^{tt}|\partial_t \psi|^2 + \frac{g^{\rho\rho}}{R^{d-3}}|\partial_\rho R^{(d-3)/2} \psi|^2 +  \frac{\lambda^2}{R^2} \left| \psi \right|^2  \right) \, .
\end{equation}
Generally, these two approaches differ by a boundary term. In asymptotically AdS spacetimes, the boundary term contributes only at the conformal boundary.

\begin{equation} \label{eq:boundaryterm}
    E[a[\psi]] = E[\psi] + \frac{d-3}{2L^2} \left. \left|\sqrt{R} \ \psi \right|^2 \right|_{\partial\text{AdS}} ,
\end{equation}
Substituting the asymptotic expansion from \eqref{eq:MS_NearBdy} and setting the gauge field source to zero reveals that the boundary term contributes only in three and five dimensions, thereby shifting the positive definiteness of $E[\psi]$ 
\begin{equation} \label{eq:BoundEpsi}
    E[\psi] \geq \int d\rho \sqrt{-g} \left( \frac{\lambda^2}{R^2} \left| \psi \right|^2 \right) -  \frac{d-3}{2L^2} \left. \left|\sqrt{R} \ \psi \right|^2 \right|_{\partial\text{AdS}} .
\end{equation}
One can check that for any 5-dimensional, asymptotically AdS Black Hole the field $\psi= \psi_0 R^{-1/2}$ obeys the right boundary conditions and exactly saturates the bound \eqref{eq:BoundEpsi}. Therefore, the norm \eqref{eq:EnergyNormMSnon} is not positive definite for $\lambda^2 < (R_h/L)^2$ where $R_h$ is the radius of the horizon. 
 Instead, using \eqref{eq:EnergyNormMS} as the norm for the inner product, we ensure there are no negative-norm states.\footnote{We note that in \cite{Cownden:2023dam} the authors use the non-positive norm \eqref{eq:EnergyNormMSnon}. The presence of negative eigenvalues in the norm can be addressed numerically by applying appropriate techniques. While this can in principle be done, our approach circumvents these subtleties altogether.} The following definition is establishes the inner product
\begin{equation}\label{eq:MSinnerproduct}
\begin{aligned}
\langle \Phi^{(1)},\Phi^{(2)}\rangle =
\int d\rho\sqrt{-g} & \left( 
\frac{g^{\rho\rho}}{R^{D-4}} (\partial_\rho R^{(d-3)/2)}\bar\psi^{(1)})(\partial_\rho R^{(d-3)/2)}\psi^{(2)} ) \right. \\ 
& \left. \qquad -g^{tt}\bar\phi^{(1)}\phi^{(2)} + \frac{\lambda^2}{R^2}\bar\psi^{(1)}\psi^{(2)}
 \right)\,. \\ 
\end{aligned}
\end{equation}
Following the same approach as in section \ref{sec:Longitudinal gauge field}, we find that the convergence properties match those found in that section.

\section{Hodge duality}\label{sec:Hodge duality}

The master scalar presents a way to compute the QNMs of a gauge field in terms of a 2-dimensional scalar field. However, as we are reducing to 2 dimensions, this approach does not allow us to pose the problem of computing $\mathbb{C}$LMs. In order to consider that problem, let us reduce the gauge field to a 3-dimensional space where we include one flat coordinate $y_1$. In this section we will focus on the black brane case, that is where $\gamma_{ij}=\delta_{ij}$. We are relabeling the coordinates $x^\mu = \{t, \rho, y_1\}$ and $y^i$ denote the rest of coordinates on the plane $\mathbb{R}^{d-1}$ different from $y_1$. With these modifications, we consider the full metric \eqref{eq:GenericBgMetric_regular} and gauge field of the form of section \ref{subsec: Modifications for CLMs}: 

\begin{align}
    \begin{split}
        ds^2 = &  \tilde{h}_{\mu \nu} dx^\mu dx^\nu + R^2 ds^2_{d-2} \, ,
        \\
        A = & A_\mu(x^\mu) dx^\mu \, .
    \end{split}
\end{align}
Using this ansatz on \eqref{eq:Maxwell Action}, after an integration of the transverse space, we find the following 3-dimensional problem. For one mode:
\begin{equation}
    S^{(L)}[A] = -\frac{1}{4} \int d^3x \sqrt{-\tilde{h}} \,   R^{d-2}F^2_{(3)} \, .
\end{equation}
To transform this expression into the canonically normalized Maxwell action we may rescale the 3-dimensional metric as $h_{\mu \nu} = R^{-2(d-2)}\tilde{h}_{\mu \nu}$. In that metric the action reads:
\begin{equation}
    S^{(L)}[A] = -\frac{1}{4} \int d^3x \sqrt{-h} \,  F^2_{(3)} \, .
\end{equation}
We may notice that, under the symmetry assumptions we have taken, the problem of a $(d+1)$-dimensional gauge field gets reduced to study a 3-dimensional gauge field. A 3D gauge field carries one single degree of freedom, which we can extract by performing the following change: 
\begin{equation} \label{eq:HodgeDualRescaled}
    F_{\mu \nu} = \epsilon_{\mu\nu\sigma}\partial^\sigma \psi_\star \, , 
\end{equation}
\noindent where $\psi_\star$ is a scalar field. This change can be understood as Hodge duality, written in form language $F = \star d\psi_\star$. This relation is manifesting explicitly the counting of degrees of freedom of a 3-dimensional gauge field, being then dual to a massless scalar field. Since in this way we introduce a scalar that is off-shell equivalent to the degrees of freedom of the gauge field in 3-dimensions, we should expect that the pseudospectra of the gauge field and the master field coincide. Plugging this in the action, we finally arrive to:
\begin{equation} \label{eq:HodgeDual_Action}
    S^{(L)}[A] = S[\psi_\star] = \frac{1}{2} \int d^3x \sqrt{-h} \, (\partial\psi_\star)^2 \, .    
\end{equation}
We are interested in comparing this with the master scalar introduced in Section \ref{sec:MS approach}. By taking $\mu=y_1$ in equation (\ref{eq:HodgeDualRescaled}) we arrive to:
\begin{equation}
    (k_1A_\nu - \partial_\nu A_1) = \epsilon_{1 \nu \sigma}\partial^\sigma \psi_\star \, .
\end{equation}
As we are reducing to a 2-dimensional scalar, we may set $A_1 = 0$. Then, by defining $\epsilon_{\mu\nu1} = R^{3-d} \,\epsilon_{\mu \nu}^{(2)}$:
\begin{equation}
    A_\nu = \frac{1}{k_1 R^{d-3}}\epsilon^{(2)}_{\nu\sigma}\partial^\sigma \psi_{\star}
\end{equation}
\noindent where now greek indices run over $\nu = \{t,\rho \}$. Comparing with (\ref{eq:Map_MS}), we define $\psi_\star = R^{(d-3)/2}\psi$, arriving to
\begin{equation}
    A_\mu = \frac{1}{k_1 R^{d-3}}\epsilon^{(2)}_{\mu\nu}\nabla^\nu \left(R^{(d-3)/2}\psi\right) \, .
\end{equation}
Finally, we may notice that this relation is invariant under 2-dimensional rescalings of the 2-dimensional metric. Thus, we might scale the 2d metric back to recover exactly (\ref{eq:Map_MS}). 

The energy-momentum tensor of this 3-dimensional scalar field is given by:
\begin{equation} \label{eq:EM_Tensor_HodgeDual}
    T_{\mu \nu} = \partial_\mu \psi_\star \partial_\nu \psi_\star + \partial_\nu\psi_\star \partial_\mu \psi_\star - h_{\mu \nu} (\partial\psi_\star)^2 \, ,
\end{equation}

\noindent that gives the following expression for the energy norm:
\begin{equation}
    E[\psi_\star] = \int d\rho \,\sqrt{-h}\left( -h^{tt}|\partial_t \psi_\star|^2 + h^{\rho\rho}|\partial_\rho \psi_\star|^2 + h^{11} k_1^2|\psi_\star|^2  \right) \, .
\end{equation}
\noindent where we have expanded the field in Fourier modes in $y_1$. To connect with the master scalar approach, we may express the energy norm using the relation between the two scalars and recovering the original metric. In terms of the 2-dimensional section of the metric, that norm reads:
\begin{equation}
     E[\psi_\star[\psi]] = \int d\rho \,\sqrt{-g}\left( -g^{tt}|\partial_t \psi|^2 + g^{\rho\rho} \frac{1}{R^{d-3}}|\partial_\rho (R^{(d-3)/2} \psi)|^2 + \frac{k_1^2}{R^2}|\psi|^2  \right) \, ,
\end{equation}

\noindent so we recover the energy norm \eqref{eq:boundaryterm}. By recovering the master scalar approach from the reduction to a 3-dimensional Hodge dual scalar, we may notice that we recover in a natural way the well-defined norm for the pseudospectrum problem. We infer from this result that Hodge duality presents an honest way of building a master scalar, thus connecting it to the rectangular matrix pseudospectrum of gauge fields. 

\subsection{\CLMs from Hodge duality}\label{sec:hodgeclm}

The Hodge dual scalar approach presents a natural way for computing $\mathbb{C}$LMs for a gauge field. The equations of motion of the scalar \eqref{eq:HodgeDual_Action} are
\begin{equation} \label{eq:EOMhodge}
    \square \psi_\star = \frac{1}{\sqrt{-h}} \partial_\mu \left(\sqrt{-h}\,  h^{\mu \nu} \partial_\nu \psi_\star \right) = 0 \, .
\end{equation}
By considering the time dependence $\psi_\star(t,\rho,y_1) = e^{-i\omega t} \psi_\star(\rho, y_1)$ and introducing an auxiliary field $\tilde{\phi}_\star = \partial_1\psi_\star$, as argued in \cite{Garcia-Farina:2024pdd}, we can define a well-motivated energy norm starting from the time-averaged energy density along a surface of constant $t$ and $y_1$. In our case:
\begin{equation}
    \bar{\varrho}[\psi_\star] = \int d\rho \sqrt{-h} \left( -h^{tt} \omega^2 |\psi_\star|^2 + h^{\rho\rho}|\partial_\rho \psi_\star|^2 + h^{11} |\partial_1\psi_\star|^2  \right) \, .
\end{equation}
Based on this expression we define then the inner product for $\tilde{\Phi}_\star = (\tilde{\phi}_\star,\psi_\star)$:
\begin{equation}
    \langle\tilde{\Phi}_\star^{(1)}, \tilde{\Phi}_\star^{(2)}\rangle_E = \int d\rho \sqrt{-h} \left( -h^{tt} \omega^2 \bar{\psi}_\star^{(1)}\psi_\star^{(2)} + h^{\rho\rho}\partial_\rho\bar{\psi}_\star^{(1)}\partial_\rho\psi_\star^{(2)} + h^{11} \bar{\tilde{\phi}}_\star^{(1)}\tilde{\phi}_\star^{(2)}  \right) \, .
\end{equation}

We can rewrite this inner product in terms of the 2-dimensional scalar, by using again the relation $\psi_\star = R^{(d-3)/2}\psi$ and the 2-dimensional metric
\begin{equation}\label{eq:MSinnerproductHodgeDual}
\begin{aligned}
\langle \tilde{\Phi}^{(1)},\tilde{\Phi}^{(2)}\rangle_E =
\int d\rho\sqrt{-g} & \left( 
\frac{g^{\rho\rho}}{R^{d-3}} (\partial_\rho R^{(d-3)/2)}\bar\psi^{(1)})(\partial_\rho R^{(d-3)/2)}\psi^{(2)} ) \right. \\ 
& \left. \qquad -g^{tt} \omega^2 \bar\psi^{(1)}\psi^{(2)} + \frac{1}{R^2}\bar{\tilde{\phi}}^{(1)}\tilde{\phi}^{(2)}
 \right)\,. \\ 
\end{aligned}
\end{equation}
Finally, in terms of the 2-dimensional scalar, the equation of motion \eqref{eq:EOMhodge} can be recasted as an eigenvalue problem:
\begin{equation} \label{eq:HDeigenvalue}
\begin{aligned}
     -i \partial_1 &\left(\begin{array}{c}
          \tilde{\phi} \\
          \psi
    \end{array} \right) = k_1 \left(\begin{array}{c}
          \tilde{\phi} \\
          \psi
    \end{array} \right) =  -i \left(  \begin{array}{cc}
         0 & \mathcal{L}  \\
         1 & 0
    \end{array}\right) \left(\begin{array}{c}
          \tilde{\phi} \\
          \psi 
    \end{array}\right) \,,\\ 
    & \mathcal{L}\psi =  R^2\left[\omega^2  g^{tt} \psi + i \omega \left(2 \frac{g^{t \rho}}{R^{(d-3)/2}}  \partial_\rho (R^{(d-3)/2}\psi) + \frac{R^{d-3}}{\sqrt{-g}}\partial_\rho\left(R^{-(d-3)}\sqrt{-g} g^{\rho t}\right) \psi\right) \right.
    \\
     & \left. -  \frac{R^{(d-3)/2}}{\sqrt{-g}}\partial_\rho\left(R^{-(d-3)}\sqrt{-g}g^{\rho\rho}\partial_\rho (R^{(d-3)/2}\psi)\right) \right] \, .
\end{aligned}
\end{equation}
Following the same approach as in section \ref{subsec: Modifications for CLMs}, we find that the convergence properties match those found in that section.

\section{Holographic setup}\label{sec:Holographic setup}

Having presented both our novel formulation in terms of the gauge field (GF) as well as the standard approach based on the master scalar (MS) we now present and compare pseudospectra in both frameworks for the following scenarios: 
\CLMs and QNFs of a Schwarzschild black brane in AdS$_{5+1}$, \CLMs and QNFs of a Schwarzschild black brane in AdS$_{4+1}$ and QNFs of a Schwarzschild black hole in AdS$_{4+1}$. 
We take this choice of spacetimes to illustrate importance of choosing the right energy norm for the MS approach. While in AdS$_{5+1}$ the boundary term in \eqref{eq:boundaryterm} plays no role, 
in AdS$_{4+1}$ the energy norm \eqref{eq:EnergyNormMSnon} is not positive definite and and instead we must use the energy norm \eqref{eq:EnergyNormMS}. 
Naively, one would expect that MS and GF formulations should yield dramatically different pseudospectra. However, we find the same pseudospectra in both frameworks. 
This agreement can be understood from Hodge duality. As we argued in section \ref{sec:Hodge duality}, the $d$-dimensional Maxwell action is equivalent to a 3-dimensional Maxwell action in an effective spacetime. The existence of Hodge duality in this reduced spacetime then implies that the gauge field dynamics should be correctly captured by a 3-dimensional scalar field; which can be related to the master scalar. Hence, it follows that MS and GF formulations should coincide.

\subsection{\CLMs of SAdS$_{5+1}$ black brane}\label{subsec:CLMs of SAdS6 black brane}
In this subsection we study the \CLMs of the longitudinal sector of a gauge field with action \eqref{eq:Maxwell Action} in a $(5+1)$-dimensional black brane with metric $\eqref{eq:Generic Metric Branes}$ with $d=5$. Recall that we are working in units where the temperature is fixed to $T=5/(4\pi)$. 

We start by stating the boundary conditions taken in GF and MS frameworks; thus defining the functional spaces under consideration. After that we present and discuss our results showcasing that indeed we find an equivalence between MS and GF approaches.

\subsubsection{Boundary conditions in the GF framework}\label{subsubsec:CLMs of SAdS6 black brane - Boundary conditions GF}

In the GF framework we study the pseudospectra of the eigenvalue problem \eqref{eq:GF_CLM_Eigenval_General} subject to the constraint \eqref{eq:GF_CLM_Constraint_General} with respect to the inner product \eqref{eq:GF_CLM_InnerPdt_General}.

In order to properly set up this problem we need to define the functional space by specifying boundary conditions for the gauge field components $\{A_\rho,\alpha_\rho,A_1\}$. To do so, we follow AdS/CFT as our guideline. To have \CLMs we impose regularity on the event horizon to select infalling modes and on the AdS boundary we demand that there is no source from the point of view of the dual QFT. To assess this latter point, we solve the equations of motion in the $\rho\rightarrow1$ region for modes with momentum $k$ and frequency $\omega$ and find the following asymptotic behavior of the fields depending only on two parameters $\{s,v\}$ 
\begin{subequations}\label{eq: AdS6brane CMMs bcs GF}
\begin{align}
    A_\rho&=-ik s(1-\rho)+ \frac{3ik v}{k^2-\omega^2} (1-\rho)^2+...\,,\\
    \alpha_\rho&=\omega^2 s(1-\rho)-\frac{3v \omega^2}{k^2-\omega^2}(1-\rho)^2+...\,,\\
    A_1&=s-\frac{s}{2}(k^2-\omega^2) (1-\rho )^2+v(1-\rho)^3 +... \,.
\end{align}
\end{subequations}
Now AdS/CFT tells us that the leading mode of $A_1$ should be identified with the source from the QFT perspective and with this we conclude that sourceless excitations have $s=0$. We can easily impose this by defining 
\begin{subequations}
\begin{align}
    A_\rho&=(1-\rho) \hat{A}_\rho\,,\\
    \alpha_\rho&=(1-\rho) \hat{\alpha}_\rho\,,\\
    A_1&=(1-\rho)^2 \hat{A}_1\,,
\end{align}
\end{subequations}
and demanding Dirichlet boundary conditions for the rescaled fields  $\{\hat{A}_\rho,\hat{\alpha}_\rho,\hat{A}_1\}$. With all this, we thus define the function space for the GF framework as the space of regular functions $\{\hat{A}_\rho,\hat{\alpha}_\rho,\hat{A}_1\}$ satisfying Dirichlet boundary conditions at $\rho=1$. This rescaling is consistent with demanding that any element in the function space has finite GF energy norm. Note that we choose to work directly in terms of the hatted variables, which in turn modifies the eigenvalue problem \eqref{eq:GF_CLM_Eigenval_General} and the energy norm \eqref{eq:GF_CLM_InnerPdt_General} in a trivial manner. We also choose to perform a rescaling of the original eigenvalue problem in order to maintain a standard eigenvalue problem for the hatted variables. Note that as discussed in section \ref{subsec: Modifications for CLMs}, the \CLM pseudospectrum is convergent as all outgoing modes with real frequency $\omega$ have infinite energy norm.

\subsubsection{Setup in the MS framework}\label{subsubsec:CLMs of SAdS6 black brane - Boundary conditions MS}

We study the pseudospectra of the eigenvalue problem \eqref{eq:HDeigenvalue} with respect to the inner product \eqref{eq:MSinnerproductHodgeDual} and to have a well-defined problem we need to specify boundary conditions for the MS field $\psi$.  
Near the AdS boundary we have the following asymptotic behavior
\begin{subequations}\label{eq: AdS6brane CMMs bcs MS}
\begin{align}
    \psi&=s-\frac{3 v}{k^2-\omega^2}(1-\rho)+...\,,\\
    \phi&=ik s-\frac{3ik v}{k^2-\omega^2}(1-\rho)+...\,,
\end{align}
\end{subequations}
where $s$ is the leading mode of $A_1$ in equation \eqref{eq: AdS6brane CMMs bcs GF}. Thus, for consistency we with the GF framework, we designate the function space as the space of regular functions $\{\psi,\phi\}$ satisfying Dirichlet boundary conditions at $\rho=1$. This is consistent with demanding that any element in the function space has finite MS energy norm.

\subsubsection{Numerical results}\label{subsubsec:CLMs of SAdS6 black brane - Numerical results}
Now we turn onto the discussion of our numerical results. A detailed discussion of the numerical procedure can be found in appendix \ref{app:Numerical methods}. In this section we use a grid of 50 points and work with $5\times$MachinePrecision in both frameworks.

\begin{figure}[h!]
\centering
\begin{subfigure}{.49\textwidth}
    \includegraphics[height=.9\textwidth]{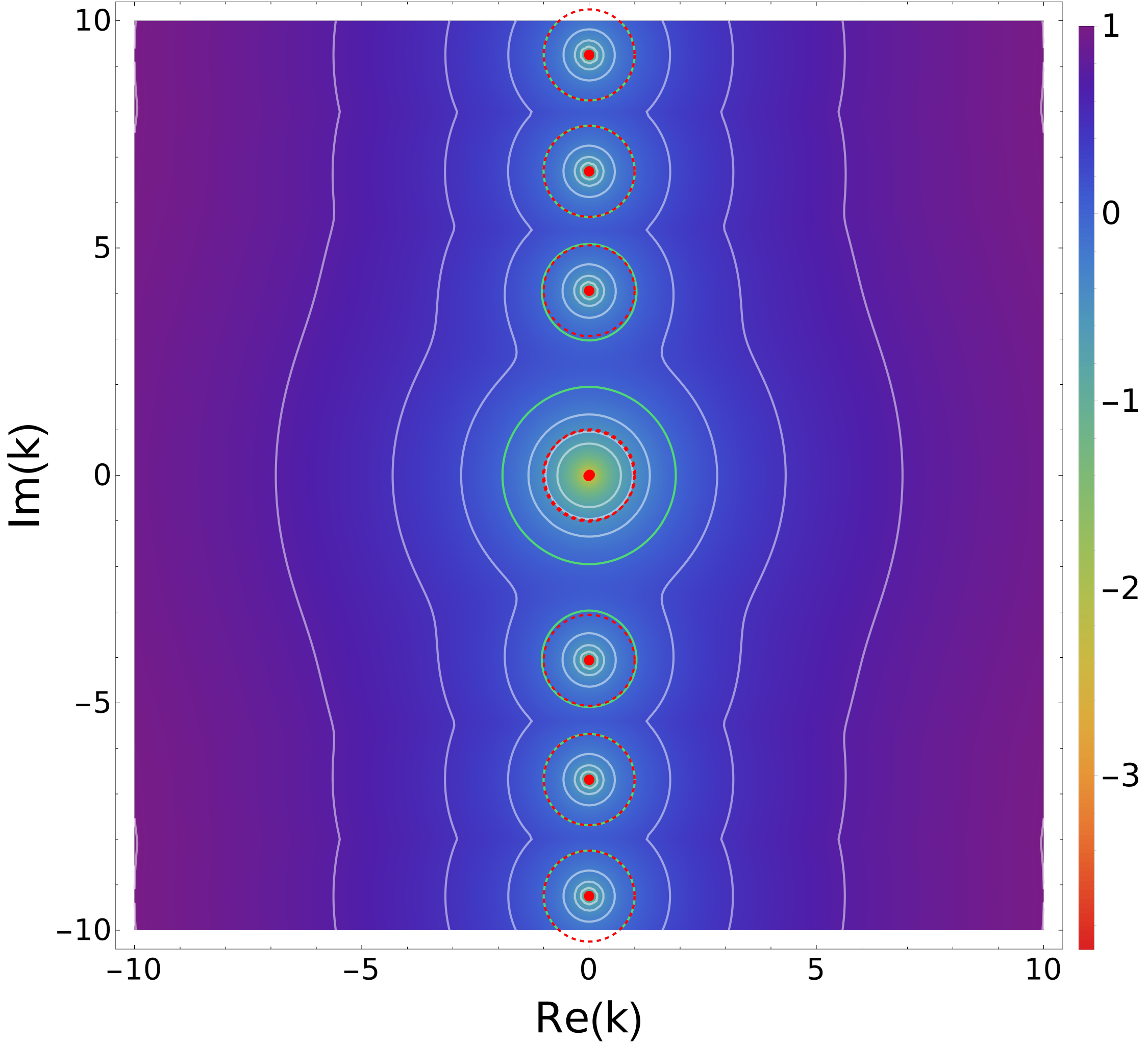}
    \label{fig:CLMPseudo_SAdS6Brane_GF_w1em4}
\end{subfigure}
\hfill
\begin{subfigure}{.49\textwidth}
    \includegraphics[height=.9\textwidth]{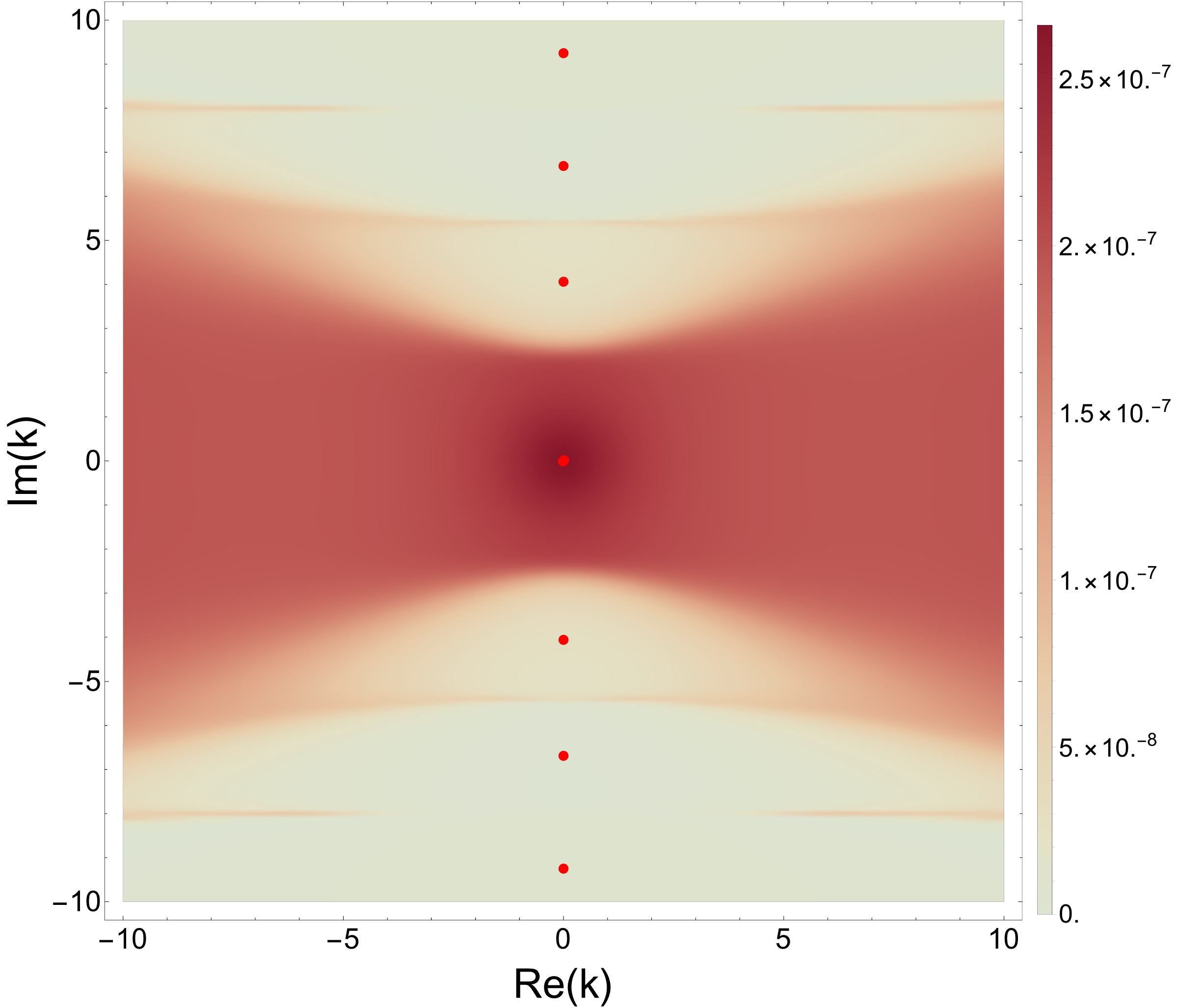}
    \label{fig:CLMPseudo_SAdS6Brane_Comparison_w1em4}
\end{subfigure}
\caption{(Left) \CLM pseudospectrum at $\omega=10^{-4}$ for the SAdS$_{5+1}$ black brane. The white lines denote the boundaries of different $\varepsilon$-pseudospectra and the heat map corresponds to the logarithm in base 10 of the inverse of the norm of the resolvent. The green circles correspond to the boundaries of the $\varepsilon=1$-pseudospectra and the dashed red circles are circles of radius $1$ centered on the \CLMs. For the higher \CLMs these coincide denoting spectral stabilty. (Right) Heatmap of the percentage difference between MS and GF frameworks. The red dots correspond to the \CLMs.}
\label{fig:CLMPseudo_SAdS6Brane_w1em4}
\end{figure}

\begin{figure}[h!]
\centering
\begin{subfigure}{.49\textwidth}
    \includegraphics[height=.9\textwidth]{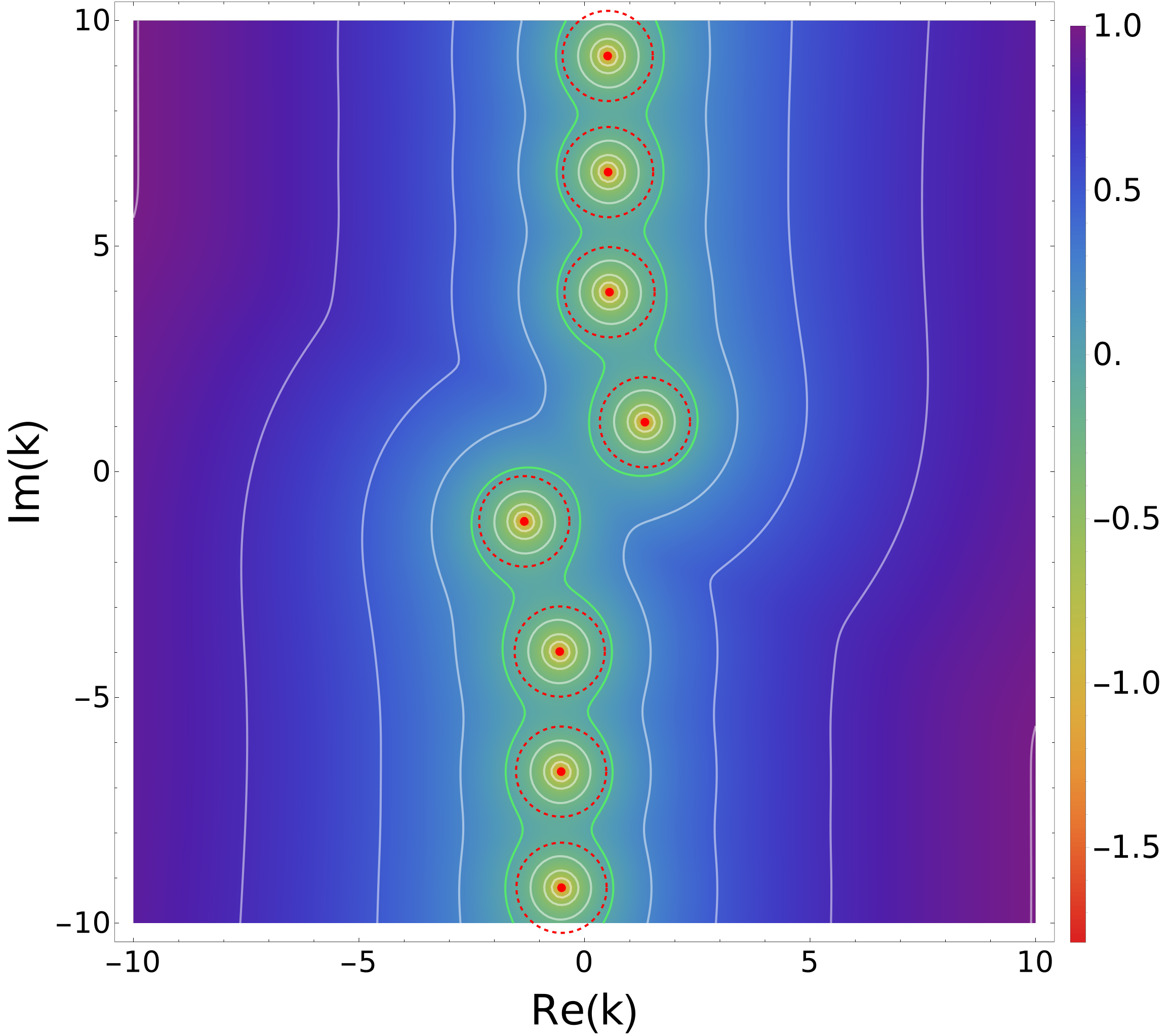}
    \label{fig:CLMPseudo_SAdS6Brane_GF_w1}
\end{subfigure}
\hfill
\begin{subfigure}{.49\textwidth}
    \includegraphics[height=.9\textwidth]{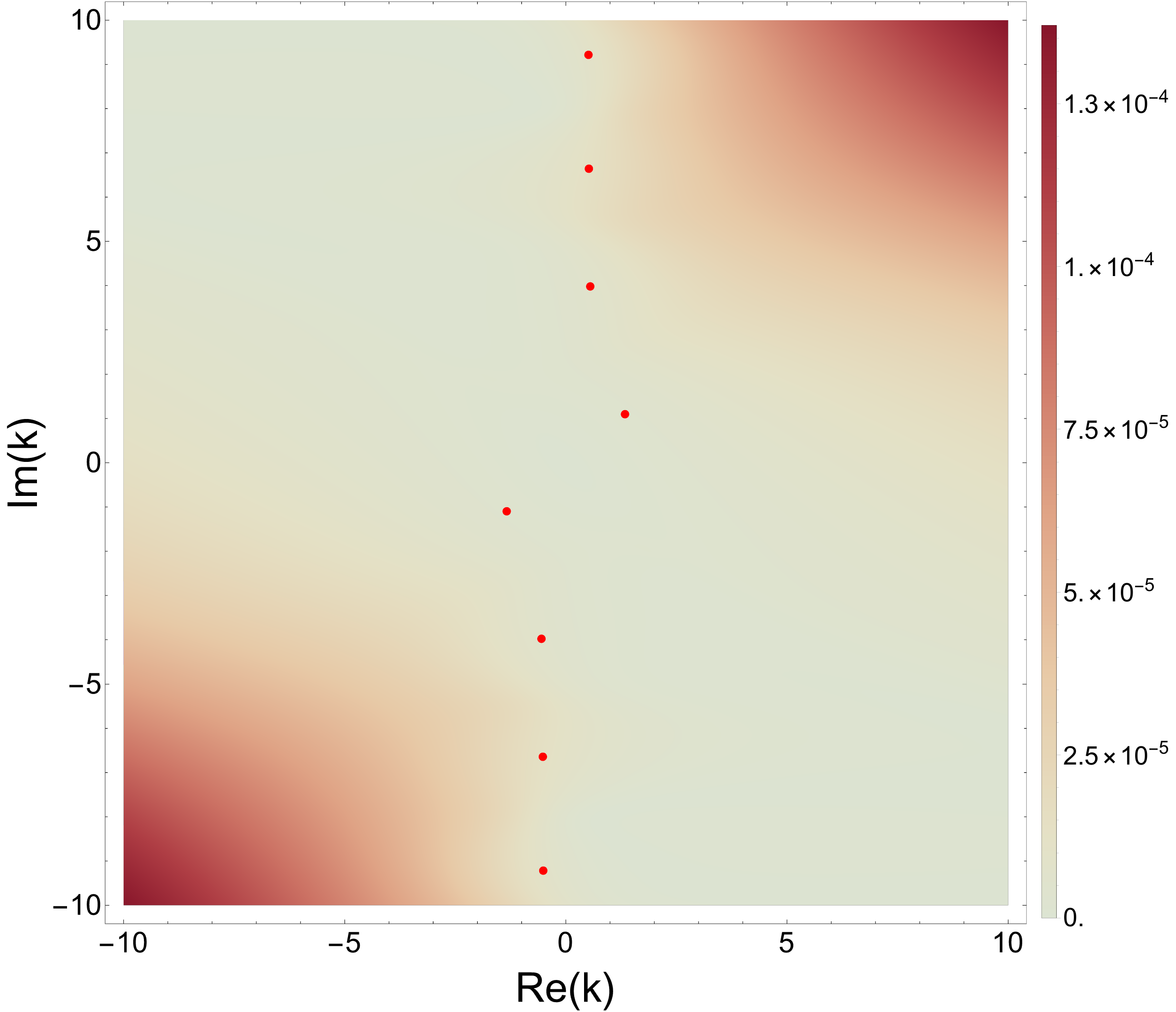}
    \label{fig:CLMPseudo_SAdS6Brane_Comparison_w1}
\end{subfigure}
\caption{(Left) \CLM pseudospectrum at $\omega=1$ for the SAdS$_{5+1}$ black brane. The white lines denote the boundaries of different $\varepsilon$-pseudospectra and the heat map corresponds to the logarithm in base 10 of the inverse of the norm of the resolvent. The green circles correspond to the boundaries of the $\varepsilon=1$-pseudospectra and the dashed red circles are circles of radius $1$ centered on the \CLMs. Notably, the \CLMs showcase mild spectral instability. (Right) Heatmap of the percentage difference between MS and GF frameworks. The red dots correspond to the \CLMs.}
\label{fig:CLMPseudo_SAdS6Brane_w1}
\end{figure}

\begin{figure}[h!]
\centering
\begin{subfigure}{.49\textwidth}
    \includegraphics[height=.9\textwidth]{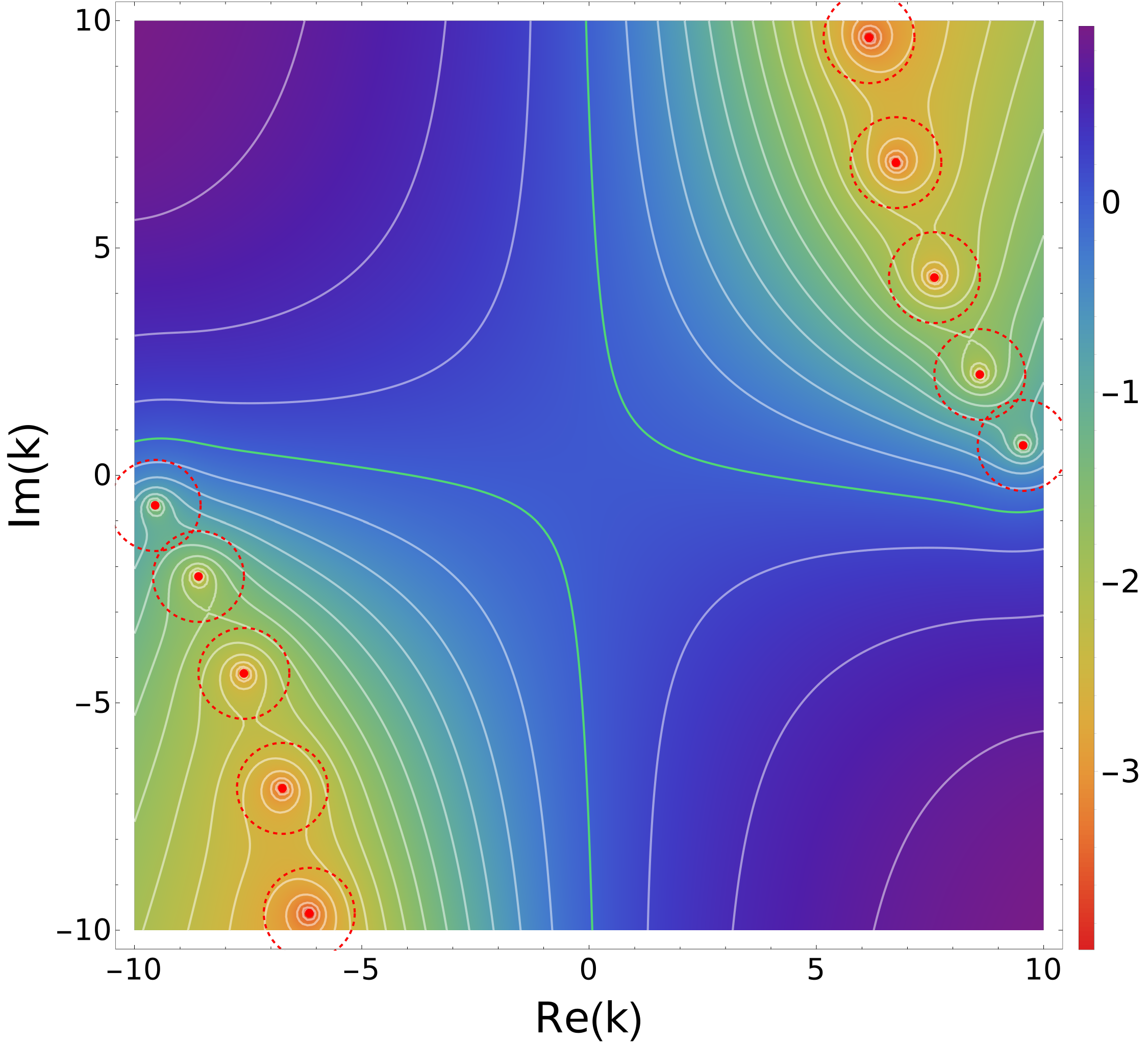}
    \label{fig:CLMPseudo_SAdS6Brane_GF_w10}
\end{subfigure}
\hfill
\begin{subfigure}{.49\textwidth}
    \includegraphics[height=.9\textwidth]{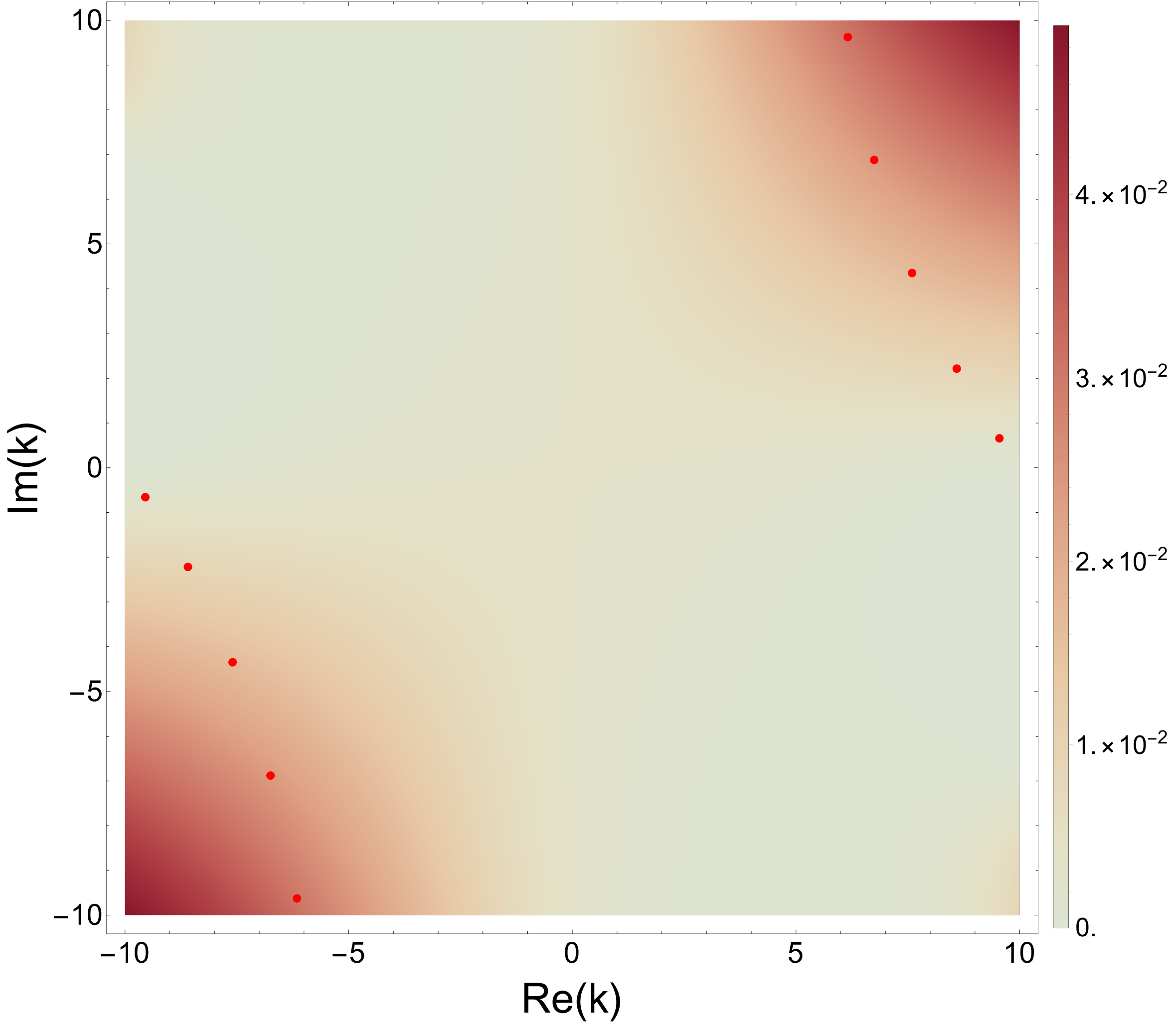}
    \label{fig:CLMPseudo_SAdS6Brane_Comparison_w10}
\end{subfigure}
\caption{(Left) \CLM pseudospectrum at $\omega=10$ for the SAdS$_{5+1}$ black brane. The white lines denote the boundaries of different $\varepsilon$-pseudospectra and the heat map corresponds to the logarithm in base 10 of the inverse of the norm of the resolvent. The green circles correspond to the boundaries of the $\varepsilon=1$-pseudospectra and the dashed red circles are circles of radius $1$ centered on the \CLMs. Notably, the \CLMs showcase spectral instability. (Right) Heatmap of the percentage difference between MS and GF frameworks. The red dots correspond to the \CLMs.}
\label{fig:CLMPseudo_SAdS6Brane_w10}
\end{figure}

In figures \ref{fig:CLMPseudo_SAdS6Brane_w1em4}-\ref{fig:CLMPseudo_SAdS6Brane_w10} we present the \CLMs pseudospectrum at frequencies $\omega=\{10^{-4},1,10\}$ and the percentage of relative difference between the two approaches. We observe that both frameworks provide equivalent results, with a maximal relative difference (in the plotted region of the complex $k$ plane) of $3\cdot10^{-7}\%$ for $\omega=10^{-4}$, of $10^{-4}\%$ for $\omega=1$ and of $5\cdot10^{-2}\%$ for $\omega=10$. 
We associate this small difference to numerical error and do not assign physical significance to it.

Furthermore, the increase in the relative difference at higher frequencies can be understood as consequence of the increased spectral instability which amplifies numerical errors.

\begin{figure}[h!]
\centering
\begin{subfigure}{.49\textwidth}
    \includegraphics[height=.65\textwidth]{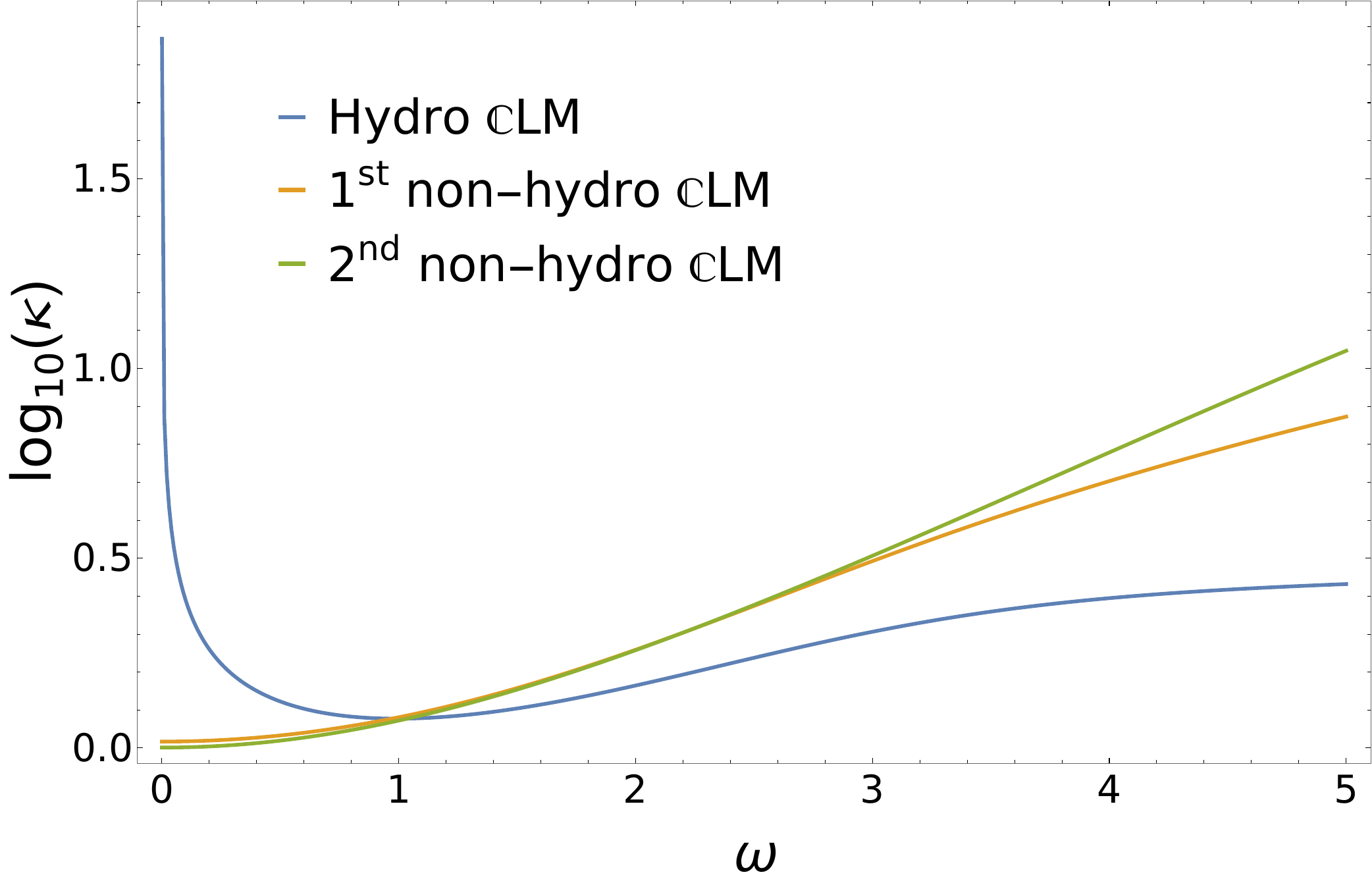}
    \label{fig:CLMKappa_SAdS6Brane}
\end{subfigure}
\hfill
\begin{subfigure}{.49\textwidth}
    \includegraphics[height=.65\textwidth]{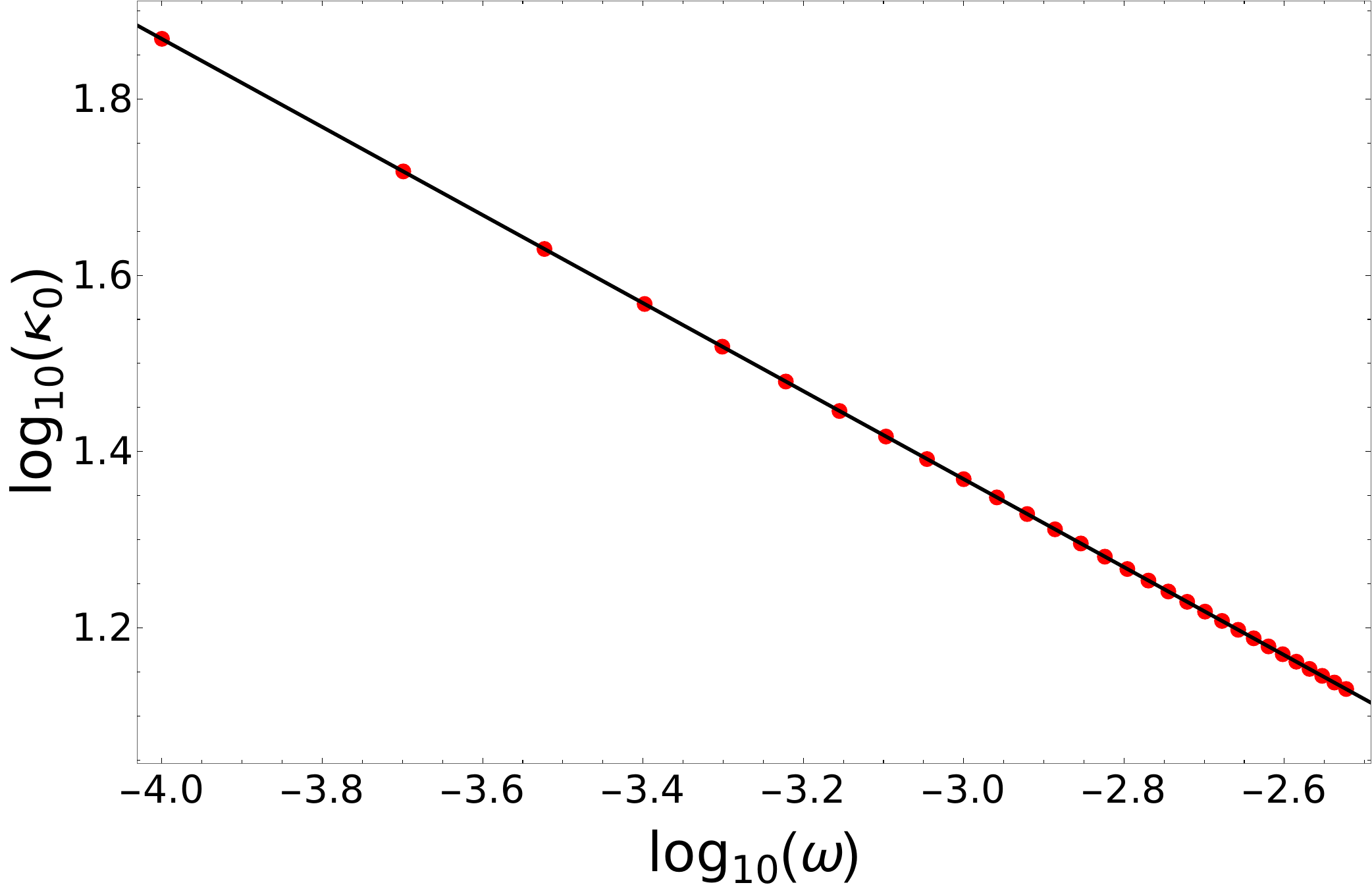}
    \label{fig:CLMKappaHydro_SAdS6Brane}
\end{subfigure}
\caption{(Left) Condition numbers for the hydro \CLM and the first two non-hydro \CLMs as a function of the frequency $\omega$  for the SAdS$_{5+1}$ black brane. (Right) Plot of the hydrodynamic \CLM condition number at small frequency $\omega$ for the SAdS$_{5+1}$ black brane. The red dots denote the numerical values of the condition numbers and the black line is the fit $\log_{10}(\kappa_0)=-0.130-0.500\log_{10}(\omega)$.}
\label{fig:CLMKappaPlots_SAdS6Brane}
\end{figure}

In figure \ref{fig:CLMPseudo_SAdS6Brane_w1em4} we observe that the non-hydro \CLMs are increasingly stable at small frequencies, with the boundaries of the $\varepsilon$-pseudospectra around them resembling a set of concentric circles of radius $\varepsilon$. This matches our intuition as in the limit $\omega=0$ all the \CLMs, except the hydrodynamical one, are dual to glueball spectrum of a Hermitian unitary theory arising from the dimensional compactification on the thermal circle \cite{Witten:1998zw,Csaki:1998qr, deMelloKoch:1998vqw}; which should have a stable spectrum. The hydrodynamic mode becomes a pole skipping mode at $\omega=0$ and thus leaves the physical spectrum, hence it need not be stable as $\omega$ is reduced. To quantify the spectral instability, we plot the condition numbers as a function of $\omega$ in the left panel of figure \ref{fig:CLMKappaPlots_SAdS6Brane} for the hydro \CLM and the first two non-hydro \CLMs.\footnote{The computation of the condition numbers in the GF approach is numerically too intensive and thus we have chosen to calculate the condition numbers in the MS approach for simplicity. The equivalence between both approaches guarantees that results are equivalent.} Remarkably we observe that while non-hydro \CLMs have monotonically increasing condition numbers, the condition number of the hydro \CLM is non-monotonic, it decreases to a minimum value before increasing. This indicates that, while for the non-hydro \CLMs the instability always increases with $\omega$, for the hydro \CLM the situation is more subtle: there is first a sharp decrease in instability up some minimum value and only after that do we observe that the instability increases with $\omega$. 

The increasing instability of the hydrodynamic \CLMs at small $\omega$ suggests that in the \CLM picture, the hydrodynamic description is strongly dependent on small perturbations at small frequency. Indeed this is what one would expect as a consequence of the pole collision taking place between the \CLMs at $\omega=0$. The existence of a pole collision at $\omega=0$ can be seen from the fact that the hydro \CLMs follow a Puiseux series around $\omega=0$ of the form 
\begin{equation}
    k_\pm(\omega)=\pm \sqrt{i\omega/D}+ O(\omega)
\end{equation}
with $D$ the diffusion constant. This implies that $\omega=0$ is an exceptional point of second order where $k_\pm$ coalesce and the eigenvalue problem becomes non-diagonalizable.  Exceptional points are characterized by their increased spectral instability: for sufficiently small $\epsilon$ the boundary of $\epsilon$-pseudospectrum is given by a circle of radius $\epsilon^{1/p}$ with $p$ the order of the exceptional point \cite{Cao:2025afs}. Then, as the pseudospectrum changes continuously with $\omega$, the increased spectral instability of the exceptional point must leave a non-trivial imprint at small finite $\omega$. This can be seen explicitly by writing the hydrodynamic condition number $\kappa_0(\omega)$ as the supremum over perturbations $V$
\begin{equation}
    \kappa_0(\omega)=\lim_{\epsilon\rightarrow0} \sup_{\norm{V}<\epsilon} \left|\frac{k_\pm(\omega;\epsilon)-k_\pm(\omega)}{\epsilon} \right|\,,
\end{equation}
with $k_\pm(\omega;\epsilon)$ the perturbed \CLMs.\footnote{Note that due to the symmetries of the pseudospectrum the condition number of $k_+$ coincides with that of $k_-$, hence we use $\kappa_0$ to denote both} Then it follows that if we construct a perturbation $V$ whose effect is to shift $\omega\rightarrow\epsilon$, the hydrodynamic condition number must satisfy the following bound
\begin{equation}
    \kappa_0(\omega)\geq\lim_{\epsilon\rightarrow0} \left|\frac{k_\pm(\omega+\epsilon)-k_\pm(\omega)}{\epsilon} \right|\approx \lim_{\epsilon\rightarrow0} \left|\frac{\sqrt{i(\omega+\epsilon)/D}-\sqrt{i\omega/D}}{\epsilon} \right|\approx \frac{1}{\sqrt{2D\omega}}\,.
\end{equation}
Hence, the existence of the exceptional point of order $2$ implies that at small $\omega$, $\kappa_0(\omega)$ must diverge at least as fast as $1/\sqrt{\omega}$. In the right panel of figure \ref{fig:CLMKappaPlots_SAdS6Brane}, we explicitly check this by plotting $\log_{10}(\kappa_0)$ against $\log_{10}(\omega)$. We find that the condition number diverges as $\kappa_0\propto\omega^{-1/2}$.

We claim that the instability of the hydro \CLMs at small $\omega$ is simply signaling the non-analytic behavior in $\omega$ resulting from a pole collision. This suggests that pseudospectra could predict the existence of pole collisions as in their vicinity there is enhanced spectral instability. 
We stress that the instability of the hydro \CLMs is in agreement with the expectation from hydrodynamics. The presence of the non-analytic behavior in the \CLM description of diffusion alone is what makes the system very susceptible to small perturbations. From a physical standpoint, the increased instability is indicating the presence of an exceptional point and signaling that perturbations are enhanced in its vicinity. Thus, the instability is purely of hydrodynamic origin and is not directly related to the realization of this diffusive mode in the black hole background.

\subsection{QNFs of SAdS$_{5+1}$ black brane}\label{subsec:QNFs of SAdS6 black brane}

Now we turn to the study of the spectral stability of the QNFs of the longitudinal sector of a gauge field with action \eqref{eq:Maxwell Action} in the (5+1)-dimensional black brane introduced in the previous subsection, whose metric is given by equation \eqref{eq:Generic Metric Branes} with $d=5$.

\subsubsection{Boundary conditions in the GF framework}\label{subsubsec:QNFs of SAdS6 black brane - Boundary conditions GF}

In the GF framework we study the pseudospectra of the eigenvalue problem \eqref{eq:GF_QNF_Eigenval_General} subject to the constraint \eqref{eq:GF_QNF_Constraint_General} with respect to the inner product \eqref{eq:GF_QNF_Energy_General}.
As before, to define the boundary conditions we follow AdS/CFT as our guideline. We impose regularity on the event horizon to select infalling modes and on the AdS boundary we demand that there is no source from the point of view of the dual QFT. Solving the equations of motion in the $\rho\rightarrow1$ region for modes with momentum $k$ and frequency $\omega$ we find the following asymptotic behavior of the fields depending only on two parameters $\{s,v\}$ 
\begin{subequations}\label{eq: AdS6brane QNMs bcs GF}
\begin{align}
    A_\rho&=-i\omega s(1-\rho)+ \frac{3i\omega v}{k^2-\omega^2} (1-\rho)^2+...\,,\\
    \alpha_\rho&=-k^2 s(1-\rho)+\frac{3 k^2 v}{k^2-\omega^2}(1-\rho)^2+...\,,\\
    A_t&=s-\frac{s}{2}(k^2-\omega^2) (1-\rho )^2+v(1-\rho)^3 +... \,.
\end{align}
\end{subequations}
AdS/CFT identifies the leading mode of $A_t$ with the source and thus sourceless excitations have $s=0$. This can be imposed by defining 
\begin{subequations}
\begin{align}
    A_\rho&=(1-\rho) \hat{A}_\rho\,,\\
    \alpha_\rho&=(1-\rho) \hat{\alpha}_\rho\,,\\
    A_t&=(1-\rho)^2 \hat{A}_t\,, 
\end{align}
\end{subequations}
and demanding Dirichlet boundary conditions for the rescaled fields  $\{\hat{A}_\rho,\hat{\alpha}_\rho,\hat{A}_1\}$. We thus define the function space as the space of regular functions $\{\hat{A}_\rho,\hat{\alpha}_\rho,\hat{A}_1\}$ satisfying Dirichlet boundary conditions at $\rho=1$. This rescaling is consistent with demanding that any element in the function space has finite GF energy norm. As for the \CLMs, we choose to work directly in terms of the hatted variables and we rescale the original eigenvalue problem to maintain a standard eigenvalue problem for the hatted variables. 

\subsubsection{Boundary conditions in the MS framework}\label{subsubsec:QNFs of SAdS6 black brane - Boundary conditions MS}

Now, we study the pseudospectra of the eigenvalue problem \eqref{eq:MSevproblem} with respect to the inner product \eqref{eq:MSinnerproduct}. Near the AdS boundary we have the following asymptotic behavior for the fields
\begin{subequations}\label{eq: AdS6brane QNMs bcs MS}
\begin{align}
    \psi&= s-\frac{3 v}{k^2-\omega^2}(1-\rho)+...\,,\\
    \phi&=-i\omega s+\frac{3 i\omega v}{k^2-\omega^2}(1-\rho)+...\,,
\end{align}
\end{subequations}
where $s$ is the leading mode of $A_t$ in equation \eqref{eq: AdS6brane QNMs bcs GF}. Thus, for consistency with the GF framework, we designate the function space as the space of regular functions $\{\psi,\phi\}$ satisfying Dirichlet boundary conditions at $\rho=1$. This is consistent with demanding that any element in the function space has finite MS energy norm.

\subsubsection{Numerical results}\label{subsubsec:QNFs of SAdS6 black brane - Numerical results}
Now we present our numerical results for the QNF pseudospectra in the GF and MS approaches. We use a grid of 50 points and work with $5\times$MachinePrecision in both frameworks.

\begin{figure}[h!]
\centering
\begin{subfigure}{.49\textwidth}
    \includegraphics[height=.89\textwidth]{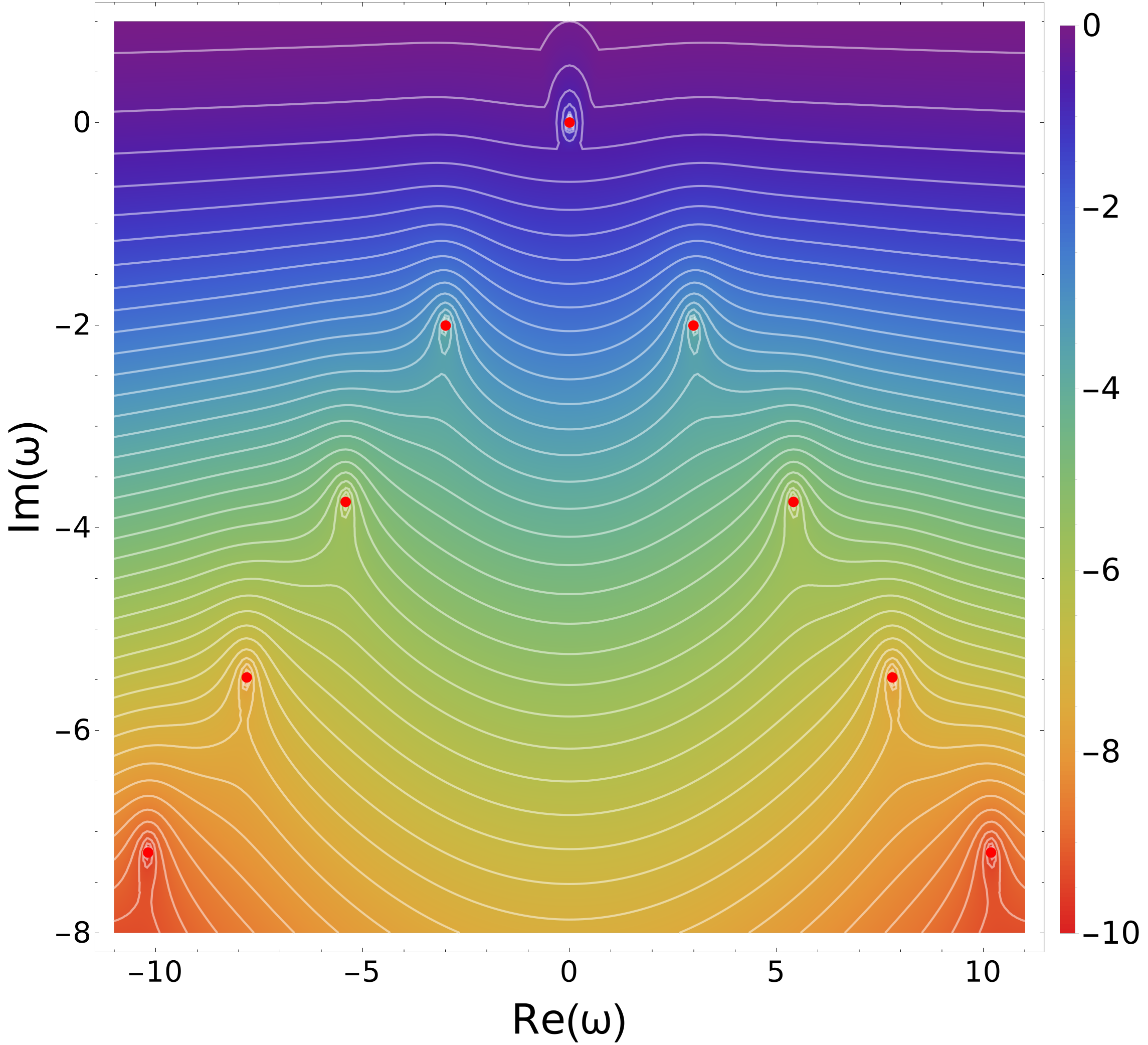}
    \label{fig:QNFPseudo_SAdS6Brane_GF_w1em4}
\end{subfigure}
\hfill
\begin{subfigure}{.49\textwidth}
    \includegraphics[height=.89\textwidth]{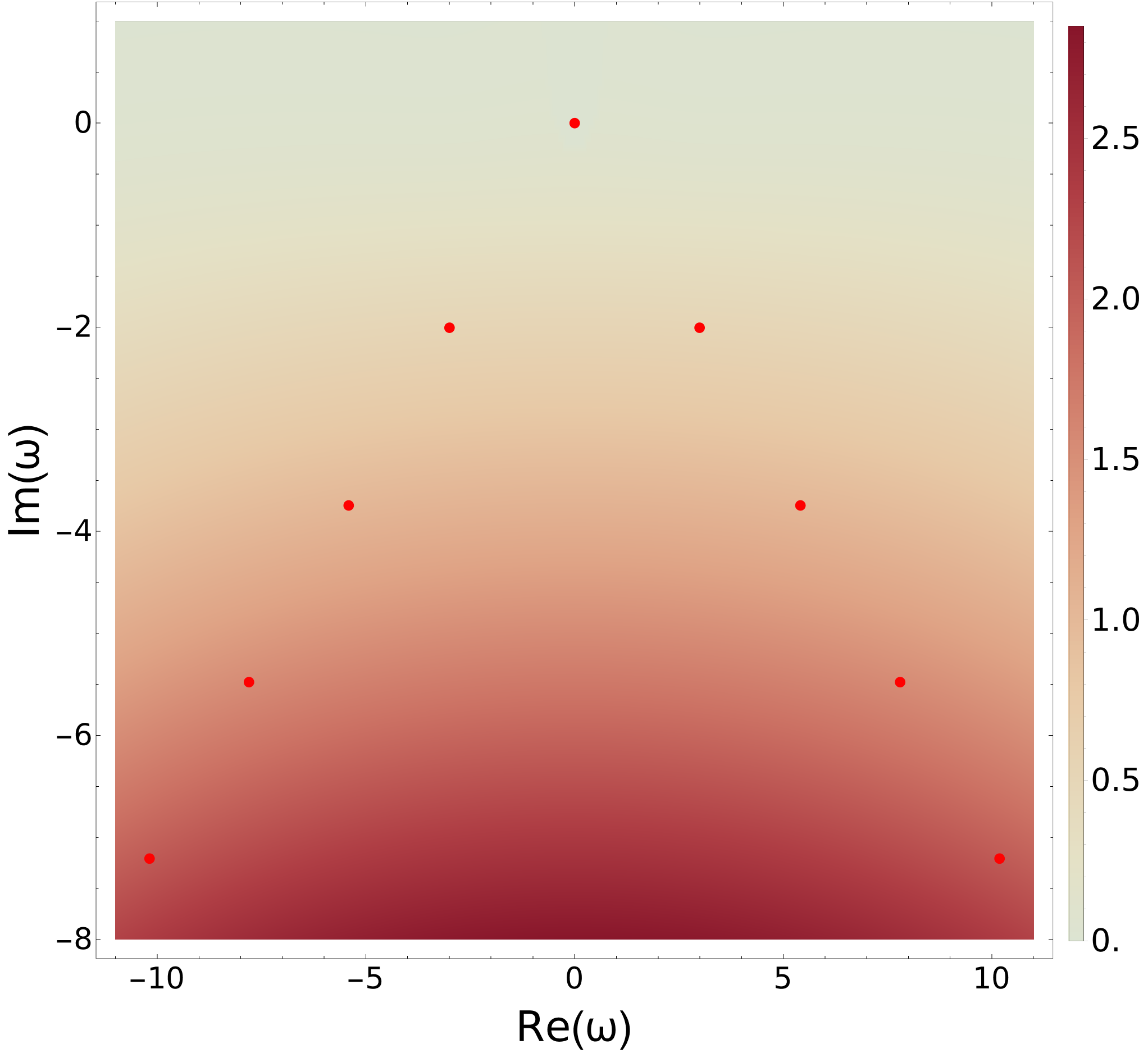}
    \label{fig:QNFPseudo_SAdS6Brane_Comparison_w1em4}
\end{subfigure}
\caption{(Left) QNF pseudospectrum at $k=10^{-4}$ for the SAdS$_{5+1}$ black brane. The white lines denote the boundaries of different $\varepsilon$-pseudospectra and the heat map corresponds to the logarithm in base 10 of the inverse of the norm of the resolvent. (Right) Heatmap of the percentage difference between MS and GF frameworks. The red dots correspod to the QNFs.}
\label{fig:QNFPseudo_SAdS6Brane_w1em4}
\end{figure}

\begin{figure}[h!]
\centering
\begin{subfigure}{.49\textwidth}
    \includegraphics[height=.89\textwidth]{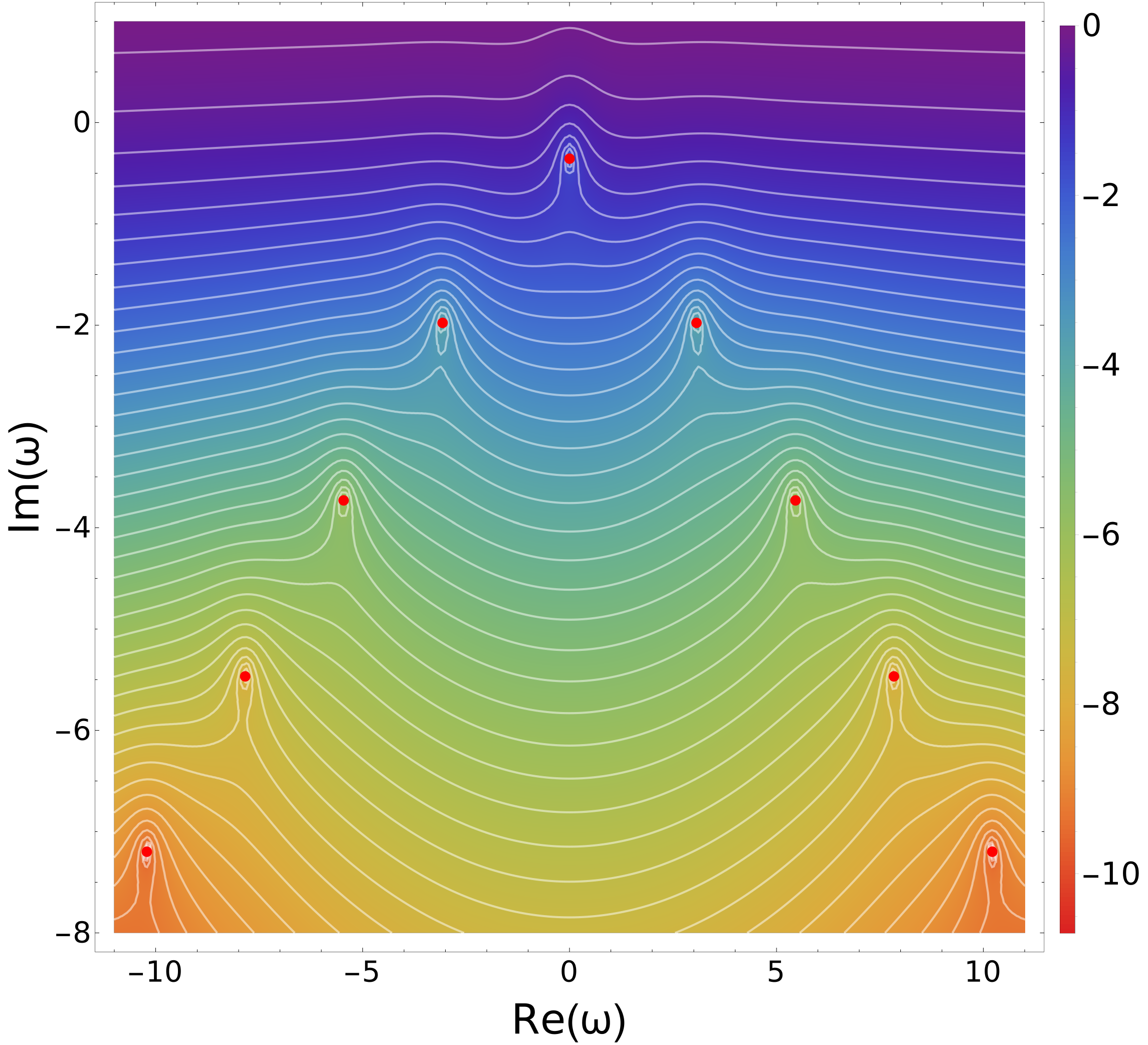}
    \label{fig:QNFPseudo_SAdS6Brane_GF_w1}
\end{subfigure}
\hfill
\begin{subfigure}{.49\textwidth}
    \includegraphics[height=.89\textwidth]{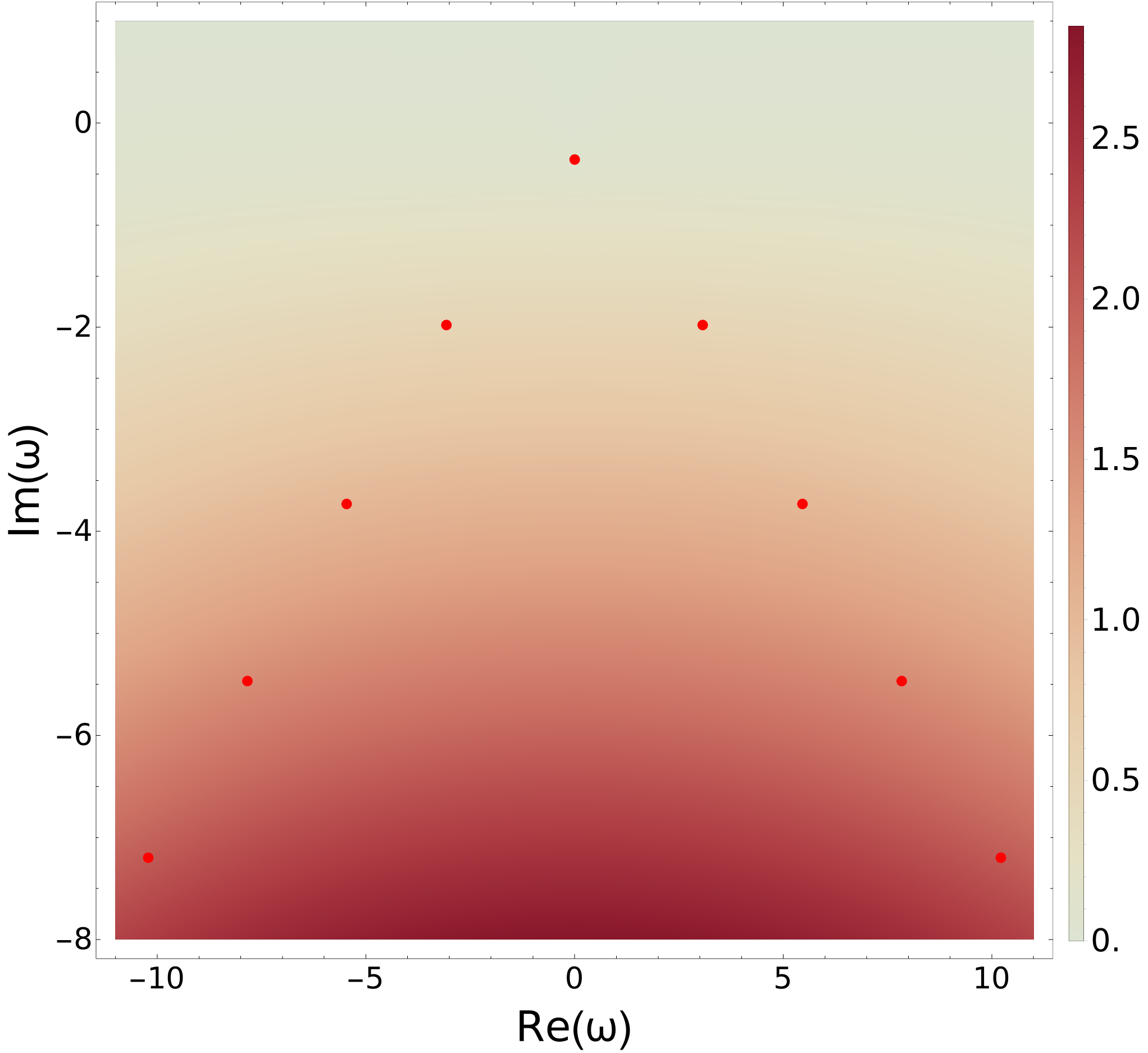}
    \label{fig:QNFPseudo_SAdS6Brane_Comparison_w1}
\end{subfigure}
\caption{(Left) QNF pseudospectrum at $k=1$ for the SAdS$_{5+1}$ black brane. The white lines denote the boundaries of different $\varepsilon$-pseudospectra and the heat map corresponds to the logarithm in base 10 of the inverse of the norm of the resolvent. (Right) Heatmap of the percentage difference between MS and GF frameworks. The red dots correspod to the QNFs.}
\label{fig:QNFPseudo_SAdS6Brane_w1}
\end{figure}

\begin{figure}[h!]
\centering
\begin{subfigure}{.49\textwidth}
    \includegraphics[height=.89\textwidth]{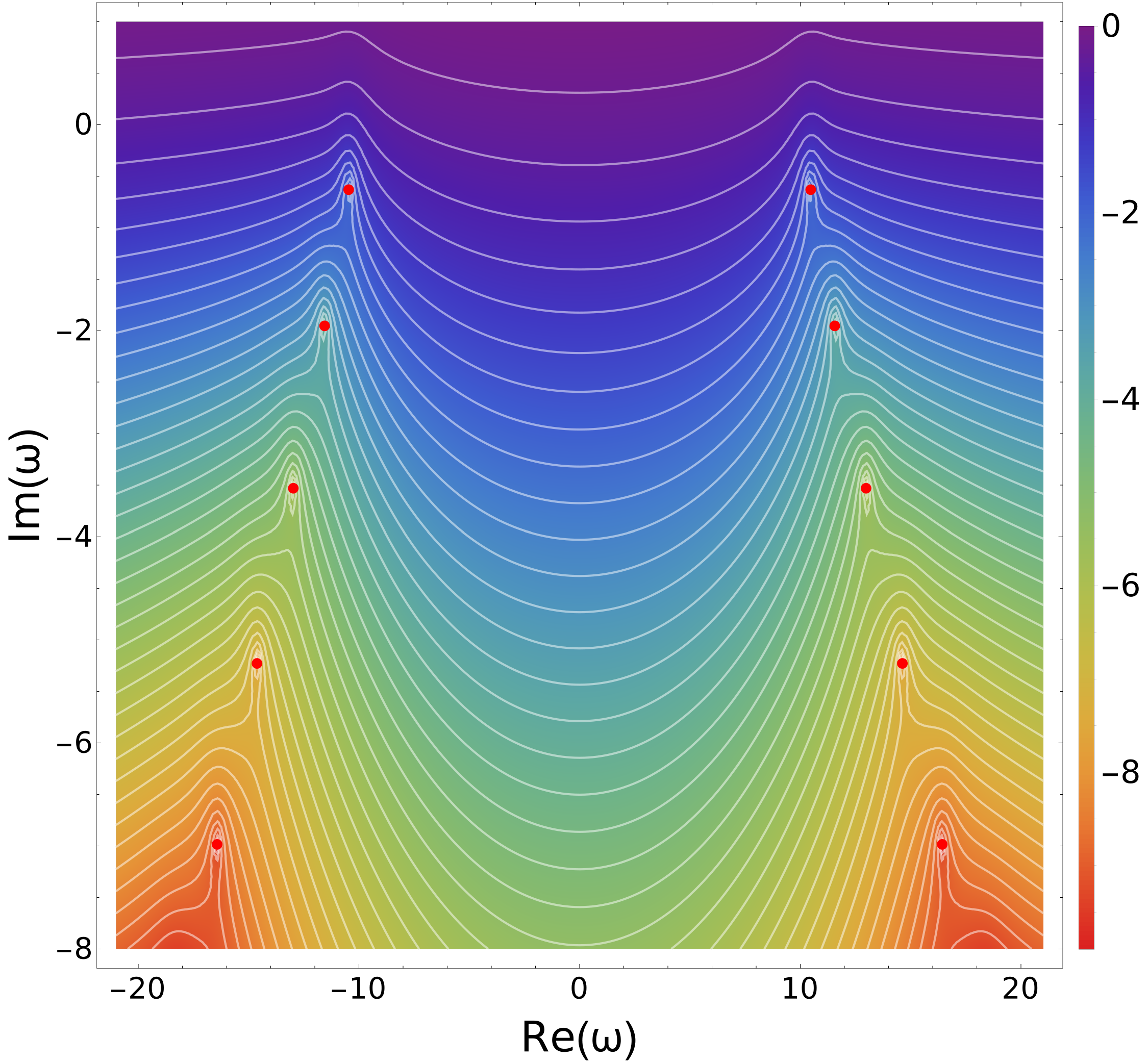}
    \label{fig:QNFPseudo_SAdS6Brane_GF_w10}
\end{subfigure}
\hfill
\begin{subfigure}{.49\textwidth}
    \includegraphics[height=.89\textwidth]{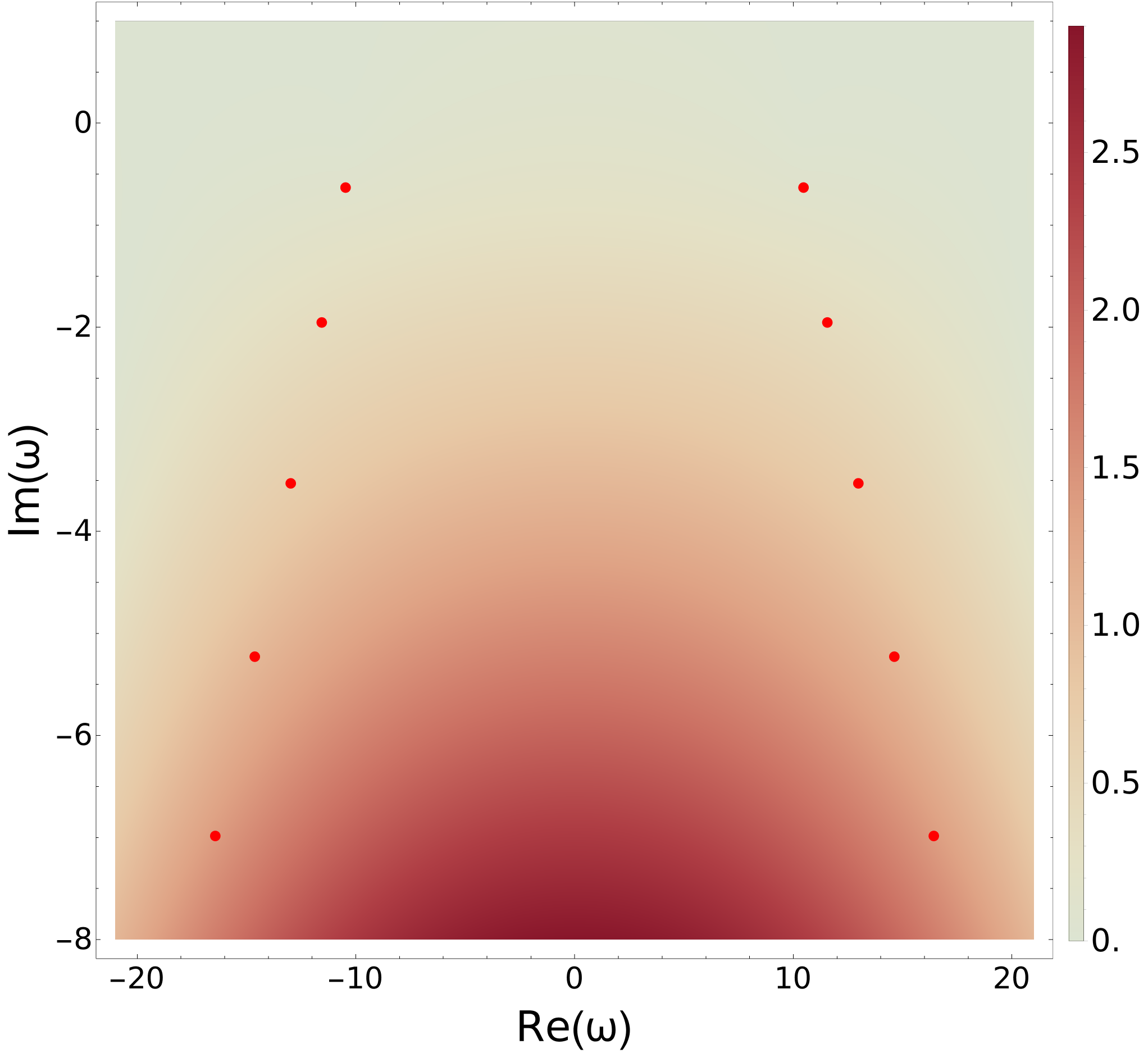}
    \label{fig:QNFPseudo_SAdS6Brane_Comparison_w10}
\end{subfigure}
\caption{(Left) QNF pseudospectrum at $k=10$ for the SAdS$_{5+1}$ black brane. The white lines denote the boundaries of different $\varepsilon$-pseudospectra and the heat map corresponds to the logarithm in base 10 of the inverse of the norm of the resolvent. (Right) Heatmap of the percentage difference between MS and GF frameworks. The red dots correspod to the QNFs.}
\label{fig:QNFPseudo_SAdS6Brane_w10}
\end{figure}

\begin{figure}[h!]
\centering
\begin{subfigure}{.49\textwidth}
    \includegraphics[width=\textwidth]{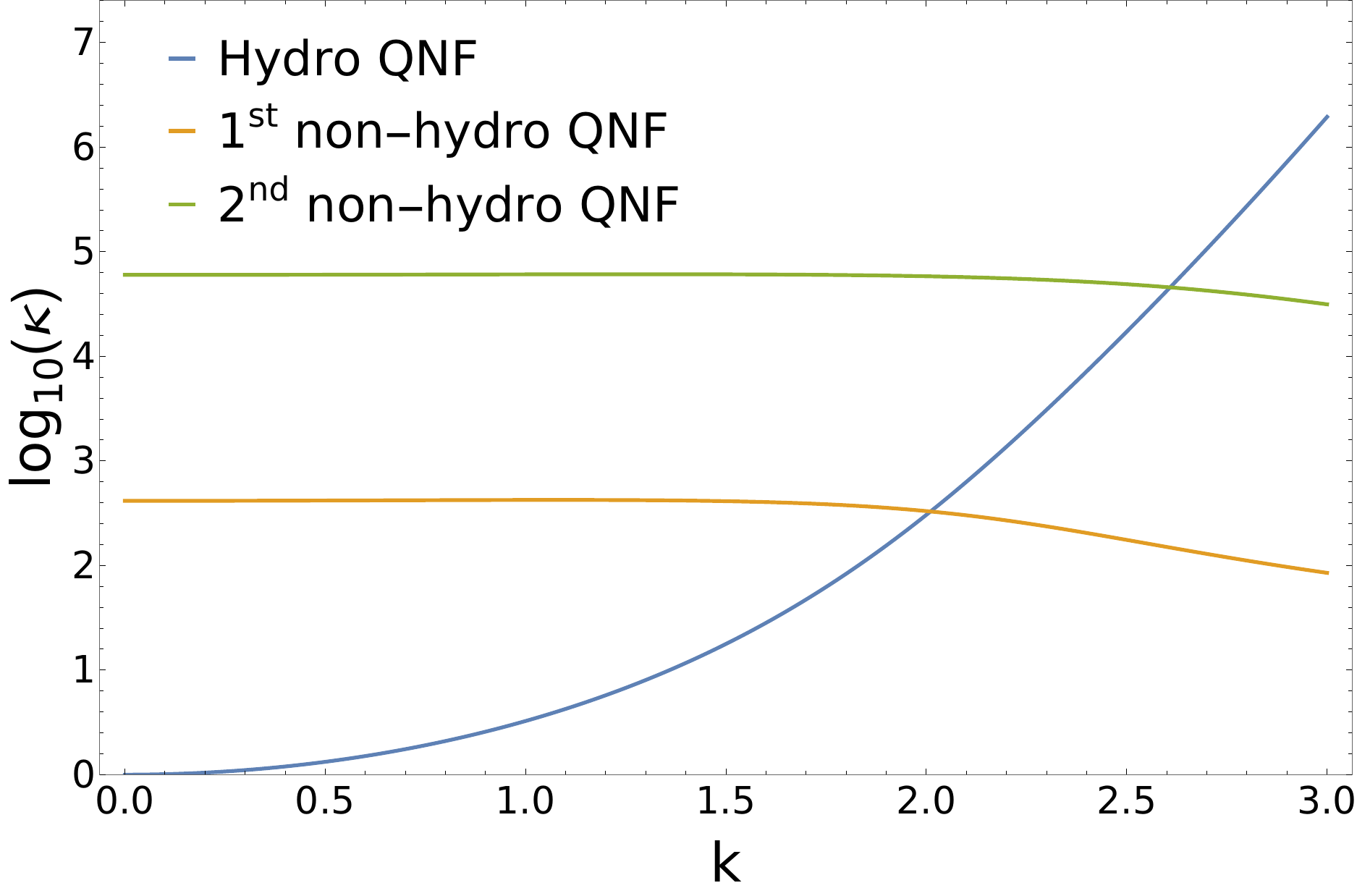}
    \caption{Condition numbers.}
    \label{fig:QNFKappa_SAdS6Brane}
\end{subfigure}
\hfill
\begin{subfigure}{.49\textwidth}
    \includegraphics[width=\textwidth]{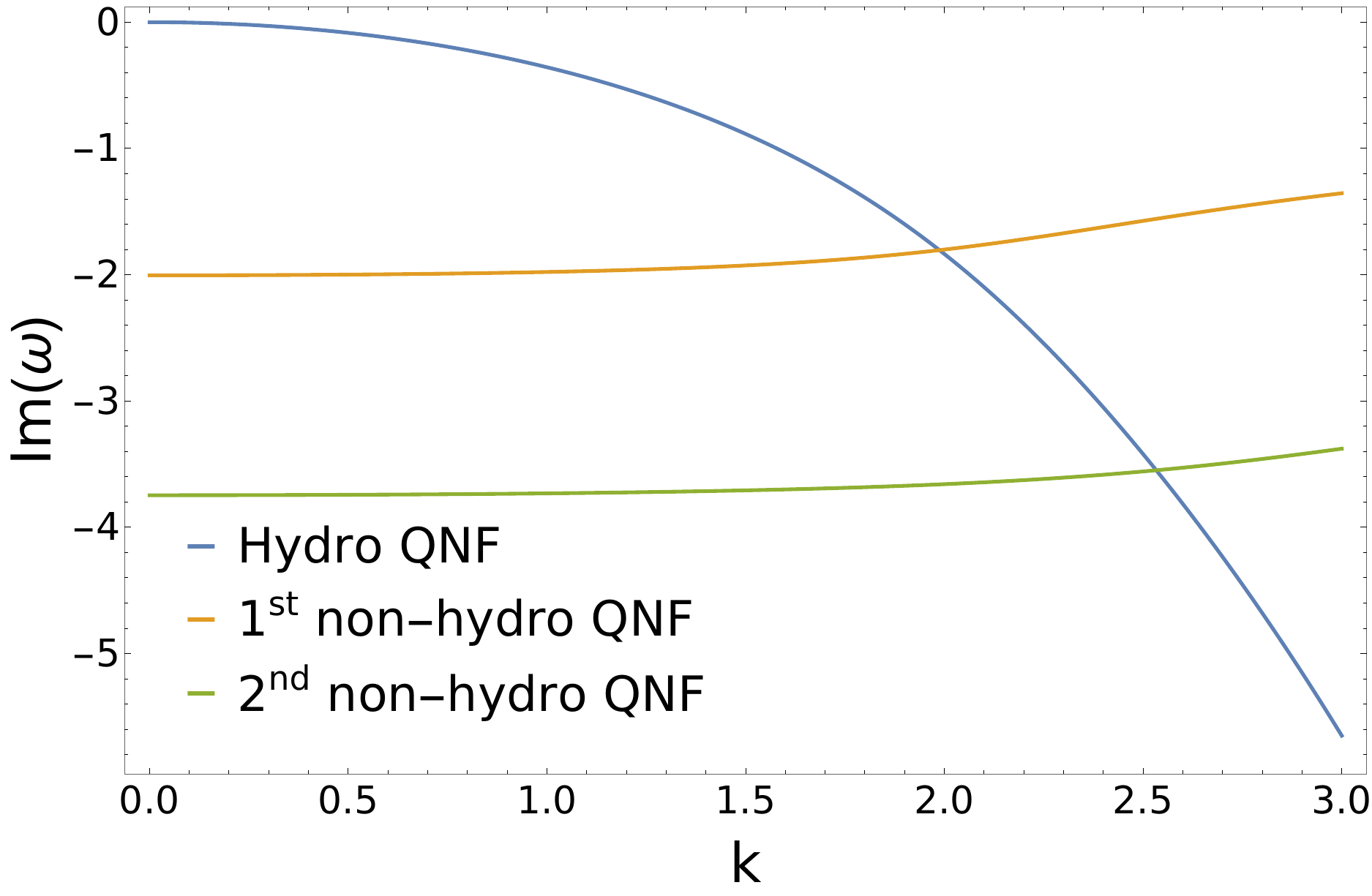}
    \caption{Imaginary parts.}
    \label{fig:ImQNF_SAdS6Brane}
\end{subfigure}
\caption{Condition numbers and imaginary part of the hydro QNF and the first two non-hydro QNFs as a function of the momentum $k$ for the SAdS$_{5+1}$ black brane. We see that the hydro QNF becomes more unstable than the non-hydro QNFs at values of the momentum similar to those for which the imaginary part of the former crosses that of the latter.}
\label{fig:QNF_SAdS6Brane}
\end{figure}

In figures \ref{fig:QNFPseudo_SAdS6Brane_w1em4}-\ref{fig:QNFPseudo_SAdS6Brane_w10} we present the QNF pseudospectrum at $k=10^{-4}$, $k=1$ and $k=10$ and the percentage of relative difference between the MS and GF approaches. We observe that both frameworks provide equivalent results, with a maximal relative difference (in the plotted region of the complex $\omega$ plane) of approximately $3\%$ in all cases. As for the \CLMs, this relative difference is associated with the numerical procedure. However, in this case it is significantly larger as the QNFs are much more unstable. Notably, we also find that the relative difference increases towards the lower half of the complex $\omega$ plane where the QNFs are seen to be more spectrally unstable, further confirming that the observed discrepancy is indeed a numerical artifact.

We observe that QNFs are more spectrally stable as they get closer to the real axis. Moreover the spectral instability of the QNFs far away from the real axis is much greater than the spectral instability observed for \CLMs. This has extensively been observed in other setups and seems to be closely related to the lack of convergence of the QNF pseudospectrum \cite{Boyanov:2022ark,Warnick:2013hba}. In detail, as we go deeper in the complex $\omega$ plane, outgoing modes are increasingly close in the continuum function space to a sum of Chebyshev polynomials living in our truncated space.\footnote{Here by close we mean that the energy norm of the difference becomes smaller} Hence those modes behave as $\varepsilon$-pseudoeigenmodes with $\varepsilon=\varepsilon(N)$. Thus the pseudospectrum does not converge on the entire complex plane, only above the band $\Im(\omega)>-\pi T=-5/4$, and it showcases large instabilities deep in the lower half plane. We want to stress however, that this lack of convergence is a fundamental feature of the problem and consequently not unphysical. In the continuum limit we cannot eliminate outgoing modes with an energy norm derived from a theory of two derivatives, such as GR. Indeed, as argued in \cite{Warnick:2013hba} to properly define the set of all QNMs one needs a Sobolev norm with an infinite number of derivatives. Hence we can interpret the lack of convergence as a hint of the breakdown of GR as a two-derivative EFT signaling the need for higher derivative corrections. Within this picture, we then view the grid size $N$ as an UV cutoff telling us the smallest scale one can possibly discern within the two-derivative EFT. Thus we claim that the results are physical but computed at a given value of the UV cutoff. Also, it is worth mentioning that for sufficiently small $k$ the hydrodynamic mode is in the convergent region $\Im(\omega)>-\pi T=-5/4$ and its stability properties do not depend on the cutoff. This is checked numerically in appendix \ref{app:Convergence tests pseudo}.

Regarding the hydrodynamic mode, we observe that it is stable in the limit of $k\rightarrow0$. In fact, for $k=10^{-4}$ we find that its condition number is $\kappa_{\text{hydro}}=1+10^{-8}$. This implies that the hydrodynamic description is stable at sufficiently small momenta, indicating that models differing by small perturbations predict the same hydrodynamics up to errors of the same order. This matches the physical expectations that the hydrodynamic description should be universal. On the other hand, non-hydro QNFs are always unstable. In figure \ref{fig:QNF_SAdS6Brane} we plot the condition number as a function of $k$ for the hydro QNF and the first two non-hydro QNFs. Remarkably, we observe that while the non-hydro QNFs become more stable with $k$ (they get closer to the real axis) the hydro QNF becomes more unstable (it gets further away to the real axis). Furthermore, we can see that the value of $k$ at which the hydro QNF becomes more unstable than the non-hydro QNFs, is close to the $k$ for which the imaginary part of the former becomes larger in absolute value than that of the latter. Lastly, note that the rapid increase of $\kappa_{\text{hydro}}(k)$ implies that the hydrodynamic description rapidly becomes unstable as it feels shorter scales, confirming that the QNF instability starts playing a role only when start feeling the UV.

\subsection{\CLMs of  SAdS$_{4+1}$ black brane}\label{subsec:CLMs of SAdS5 black brane}

In this subsection we study the \CLMs of the longitudinal sector of a gauge field with action \eqref{eq:Maxwell Action} in a $(4+1)$-dimensional black brane with metric \eqref{eq:Generic Metric Branes} with $d=4$. Recall that we work in units where the temperature is fixed to $T=1/\pi$.

\subsubsection{Boundary conditions in the GF framework}\label{subsubsec:CLMs of SAdS5 black brane - Boundary conditions GF}

In the GF framework we study the pseudospectra of the eigenvalue problem \eqref{eq:GF_CLM_Eigenval_General} subject to the constraint \eqref{eq:GF_CLM_Constraint_General} with respect to the inner product \eqref{eq:GF_CLM_InnerPdt_General}.
For the boundary conditions, we impose regularity on the event horizon to select infalling modes and on the AdS boundary we demand that there is no source from the point of view of the dual QFT. Solving the equations of motion in the $\rho\rightarrow1$ region for modes with momentum $k$ and frequency $\omega$ and we find the following asymptotic behavior of the fields depending only on two parameters $\{s,v\}$ 

\begin{subequations}\label{eq: AdS5brane CMMs bcs GF}
\begin{align}
    A_\rho&=\frac{iks}{2}(1-\rho)+\frac{2ikv}{k^2-\omega^2}(1-\rho)+iks(1-\rho)\log(1-\rho)+...\,,\\
    \alpha_\rho&=-\frac{s\omega^2}{2}(1-\rho)-\frac{2v\omega^2}{k^2-\omega^2}(1-\rho)-s\omega^2(1-\rho)\log(1-\rho)+...\,,\\
    A_1&=s+v(1-\rho)^2+\frac{s}{2}(k^2-\omega^2)(1-\rho)^2\log(1-\rho)+... \,,
\end{align}
\end{subequations}
where the AdS/CFT dictionary establishes that $s$ is the source of the QFT perspective. With this we conclude that sourceless excitations have $s=0$ and thus we impose this by defining 
\begin{subequations}
\begin{align}
    A_\rho&=\hat{A}_\rho\,,\\
    \alpha_\rho&=\hat{\alpha}_\rho\,,\\
    A_1&=(1-\rho)\hat{A}_1\,,
\end{align}    
\end{subequations}
and demanding Dirichlet boundary conditions for the hatted fields  $\{\hat{A}_\rho,\hat{\alpha}_\rho,\hat{A}_1\}$. Thus we define the function space for the as the space of regular functions $\{\hat{A}_\rho,\hat{\alpha}_\rho,\hat{A}_1\}$ satisfying Dirichlet boundary conditions at $\rho=1$. This rescaling is consistent with demanding that any element in the function space has finite GF energy norm. As before, we work directly in terms of the hatted variables and rescale the original eigenvalue problem to have a standard eigenvalue problem for the hatted variables.

\subsubsection{Setup in the MS framework}\label{subsubsec:CLMs of SAdS5 black brane - Boundary conditions MS}

We study the pseudospectra of the eigenvalue problem \eqref{eq:HDeigenvalue} with respect to the inner product \eqref{eq:MSinnerproductHodgeDual} and to have a well-defined problem we need to specify boundary conditions for the MS field $\psi$. 

Near the AdS boundary we have the following asymptotic behavior 
\begin{subequations}\label{eq: AdS5brane CMMs bcs MS}
\begin{align}
    \psi&=\frac{ks}{2}\sqrt{1-\rho}+\frac{2kv}{k^2-\omega^2}\sqrt{1-\rho}+ks\sqrt{1-\rho}\log(1-\rho)+...\,,\\
    \phi&=\frac{isk^2}{2}\sqrt{1-\rho}+\frac{2ivk^2}{k^2-\omega^2}\sqrt{1-\rho}+isk^2\sqrt{1-\rho}\log(1-\rho)+...\,,
\end{align}
\end{subequations}
where $s$ is the leading mode of $A_1$ in equation \eqref{eq: AdS5brane CMMs bcs GF}. Thus, for consistency we with the GF framework we designate the function space as the space of regular functions $\{\hat\psi,\hat\phi\}$, where the hatted fields are defined by
\begin{subequations}
\begin{align}
    \psi=\sqrt{1-\rho}\,\hat{\psi}\,,\\
    \phi=\sqrt{1-\rho}\,\hat{\phi}\,,
\end{align}
\end{subequations}
so that the non-analytic factors of $\sqrt{1-\rho}$ present in $\{\psi,\phi\}$ disappear.

This is precisely the case in which the boundary term in the energy norm of the master scalar prevents positive definiteness. We base our analysis therefore on the norm \eqref{eq:EnergyNormMS} and inner product \eqref{eq:MSinnerproduct}.

\subsubsection{Numerical results}\label{subsubsec:CLMs of SAdS5 black brane - Numerical results}
Now we turn onto the discussion of our numerical results. In this section we use a grid of 50 points and work with $5\times$MachinePrecision in both frameworks. 

\begin{figure}[h!]
\centering
\begin{subfigure}{.49\textwidth}
    \includegraphics[height=.9\textwidth]{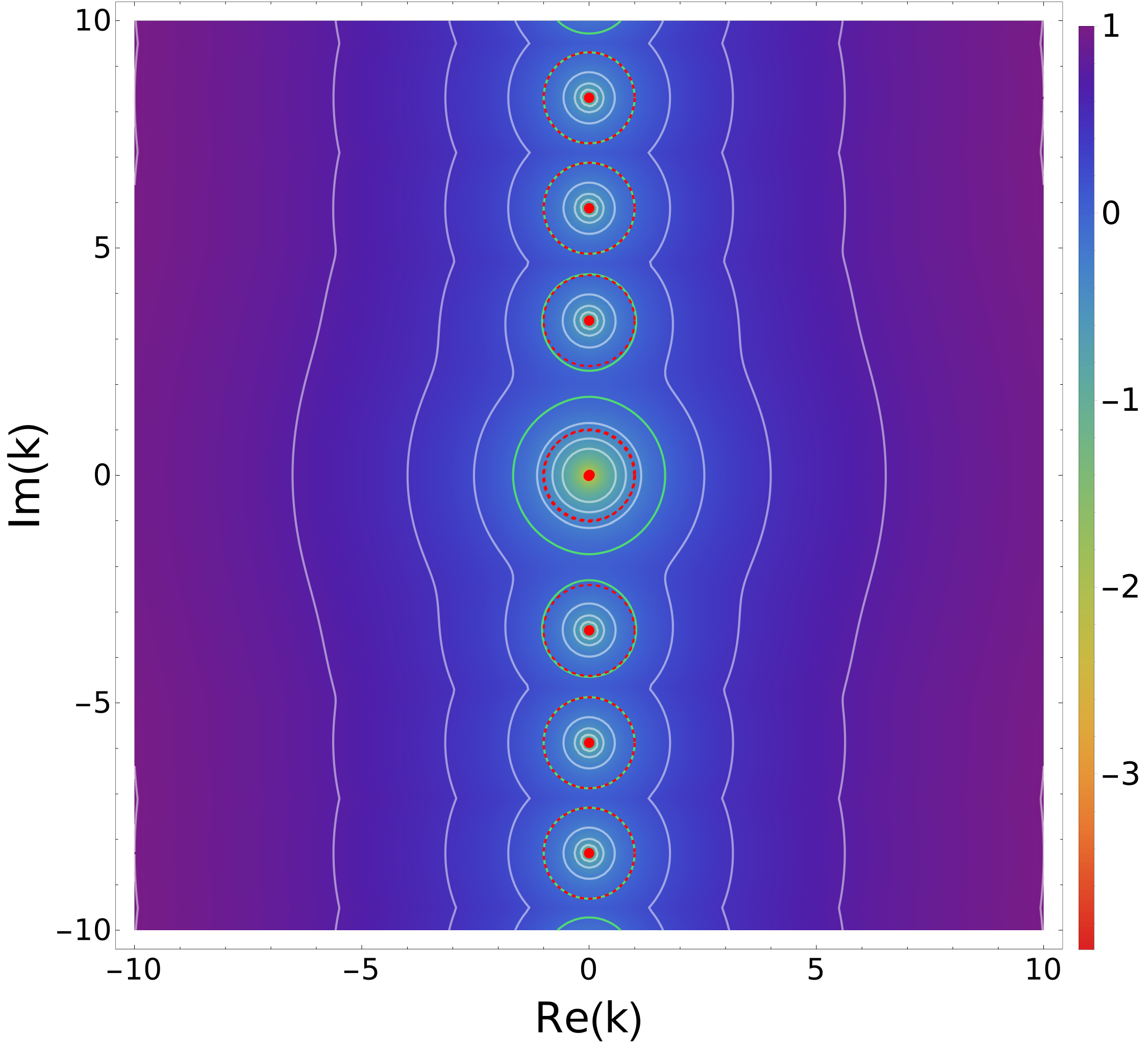}
    \label{fig:CLMPseudo_SAdS5Brane_GF_w1em4}
\end{subfigure}
\hfill
\begin{subfigure}{.49\textwidth}
    \includegraphics[height=.9\textwidth]{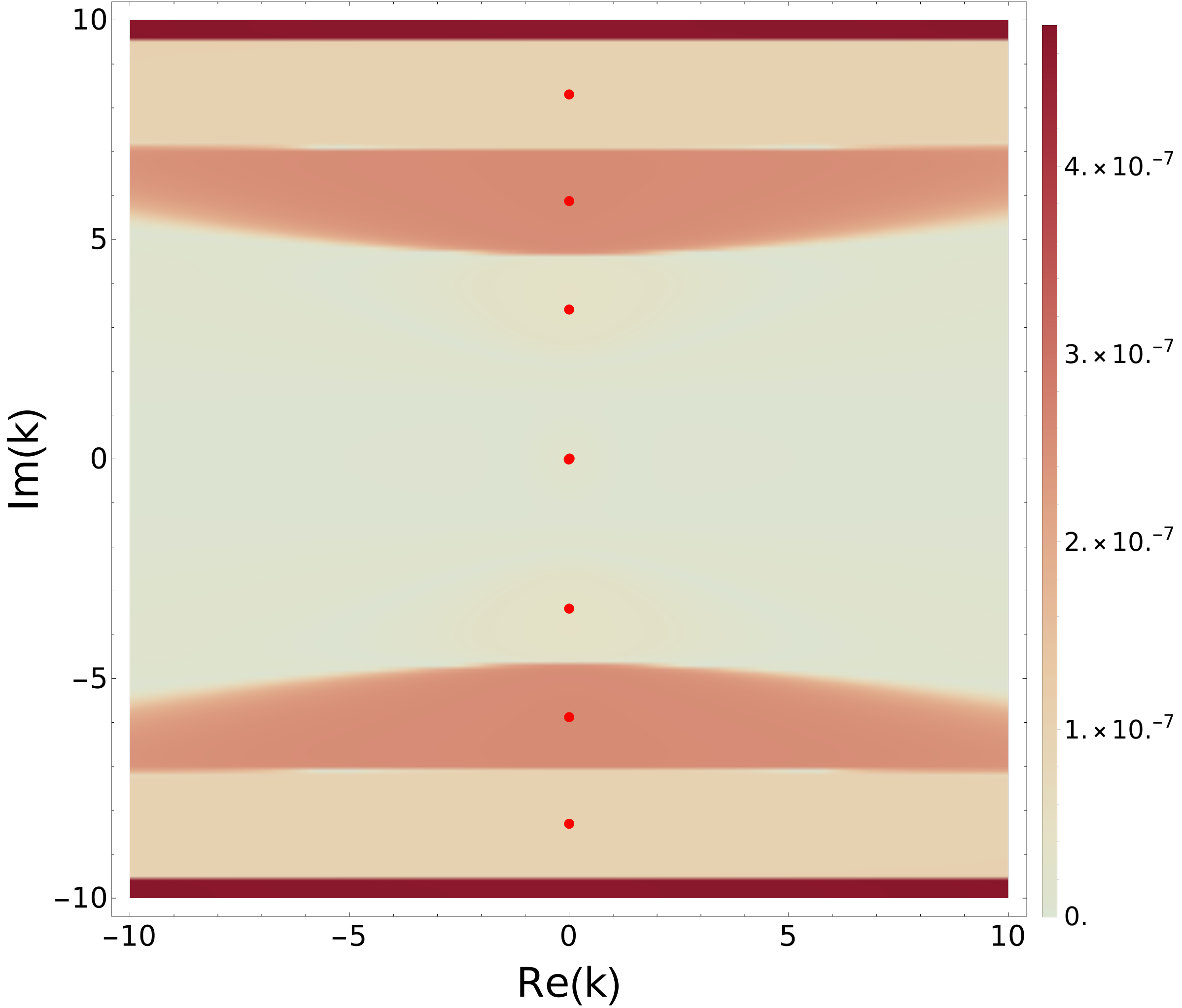}
    \label{fig:CLMPseudo_SAdS5Brane_Comparison_w1em4}
\end{subfigure}
\caption{(Left) \CLM pseudospectrum at $\omega=10^{-4}$ for the SAdS$_{4+1}$ black brane. The white lines denote the boundaries of different $\varepsilon$-pseudospectra and the heat map corresponds to the logarithm in base 10 of the inverse of the norm of the resolvent. The green circles correspond to the boundaries of the $1$-pseudospectra and the dashed red circles are circles of radius $1$ centered on the \CLMs. For the higher \CLMs these coincide denoting spectral stabilty. (Right) Heatmap of the percentage difference between MS and GF frameworks. The red dots correspond to the \CLMs.}
\label{fig:CLMPseudo_SAdS5Brane_w1em4}
\end{figure}

\begin{figure}[h!]
\centering
\begin{subfigure}{.49\textwidth}
    \includegraphics[height=.9\textwidth]{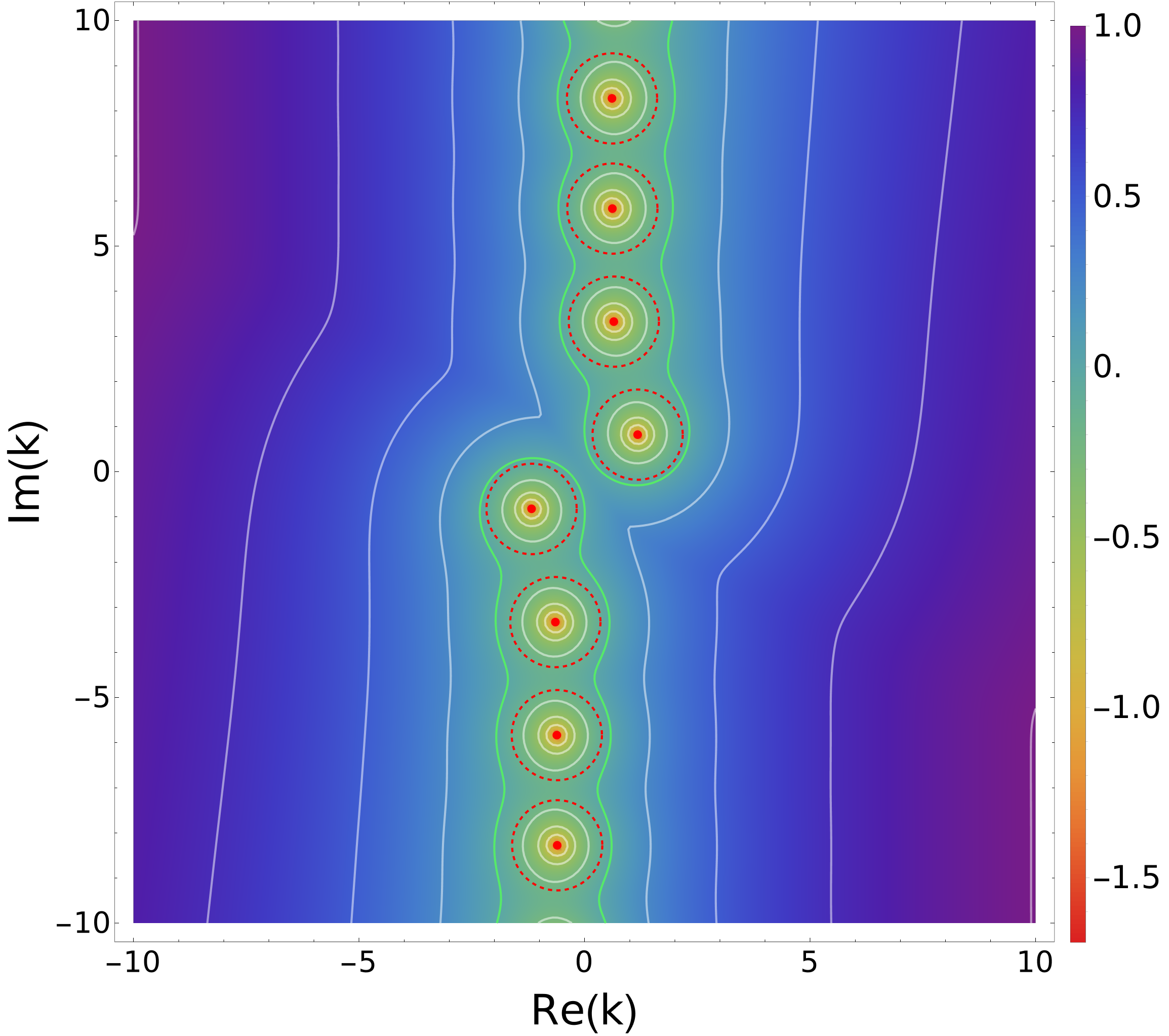}
    \label{fig:CLMPseudo_SAdS5Brane_GF_w1}
\end{subfigure}
\hfill
\begin{subfigure}{.49\textwidth}
    \includegraphics[height=.9\textwidth]{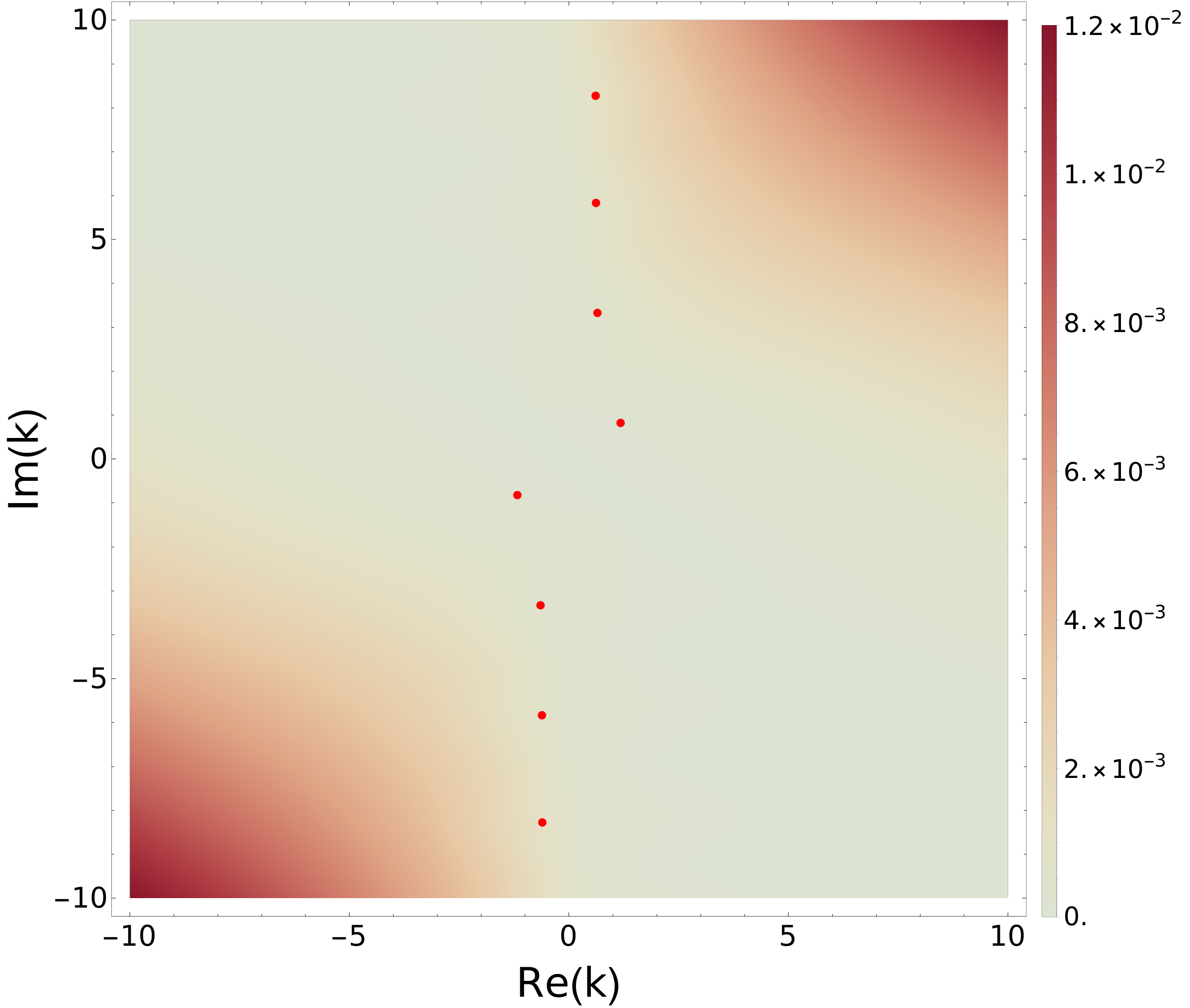}
    \label{fig:CLMPseudo_SAdS5Brane_Comparison_w1}
\end{subfigure}
\caption{(Left) \CLM pseudospectrum at $\omega=1$ for the SAdS$_{4+1}$ black brane. The white lines denote the boundaries of different $\varepsilon$-pseudospectra and the heat map corresponds to the logarithm in base 10 of the inverse of the norm of the resolvent. The green circles correspond to the boundaries of the $1$-pseudospectra and the dashed red circles are circles of radius $1$ centered on the \CLMs. Notably, the \CLMs showcase mild spectral instability. (Right) Heatmap of the percentage difference between MS and GF frameworks. The red dots correspond to the \CLMs.}
\label{fig:CLMPseudo_SAdS5Brane_w1}
\end{figure}

\begin{figure}[h!]
\centering
\begin{subfigure}{.49\textwidth}
    \includegraphics[height=.9\textwidth]{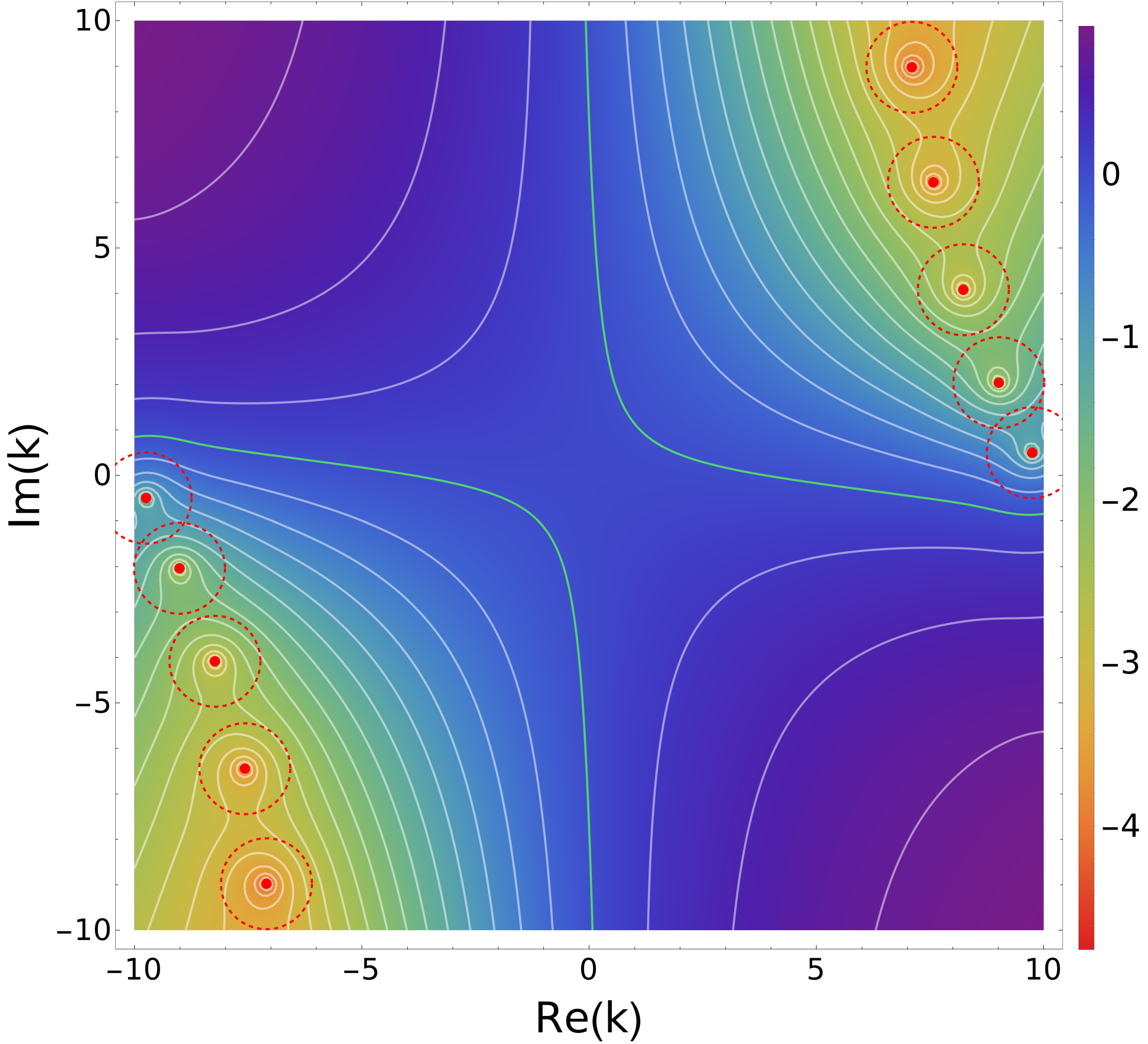}
    \label{fig:CLMPseudo_SAdS5Brane_GF_w10}
\end{subfigure}
\hfill
\begin{subfigure}{.49\textwidth}
    \includegraphics[height=.9\textwidth]{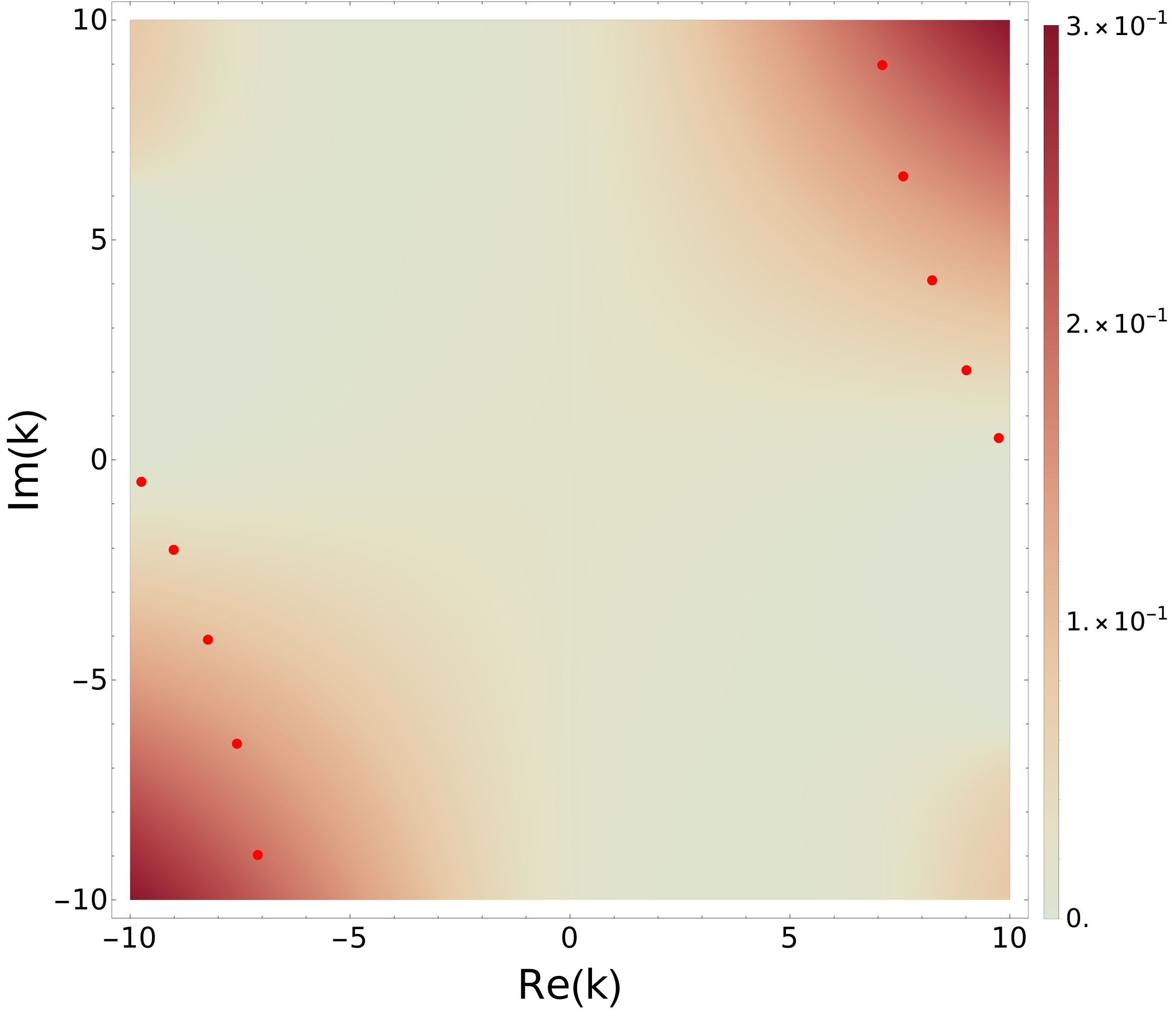}
    \label{fig:CLMPseudo_SAdS5Brane_Comparison_w10}
\end{subfigure}
\caption{(Left) \CLM pseudospectrum at $\omega=10$ for the SAdS$_{4+1}$ black brane. The white lines denote the boundaries of different $\varepsilon$-pseudospectra and the heat map corresponds to the logarithm in base 10 of the inverse of the norm of the resolvent. The green circles correspond to the boundaries of the $1$-pseudospectra and the dashed red circles are circles of radius $1$ centered on the \CLMs. Notably, the \CLMs showcase spectral instability. (Right) Heatmap of the percentage difference between MS and GF frameworks. The red dots correspond to the \CLMs.}
\label{fig:CLMPseudo_SAdS5Brane_w10}
\end{figure}

\begin{figure}[h!]
\centering
\begin{subfigure}{.49\textwidth}
    \includegraphics[height=.65\textwidth]{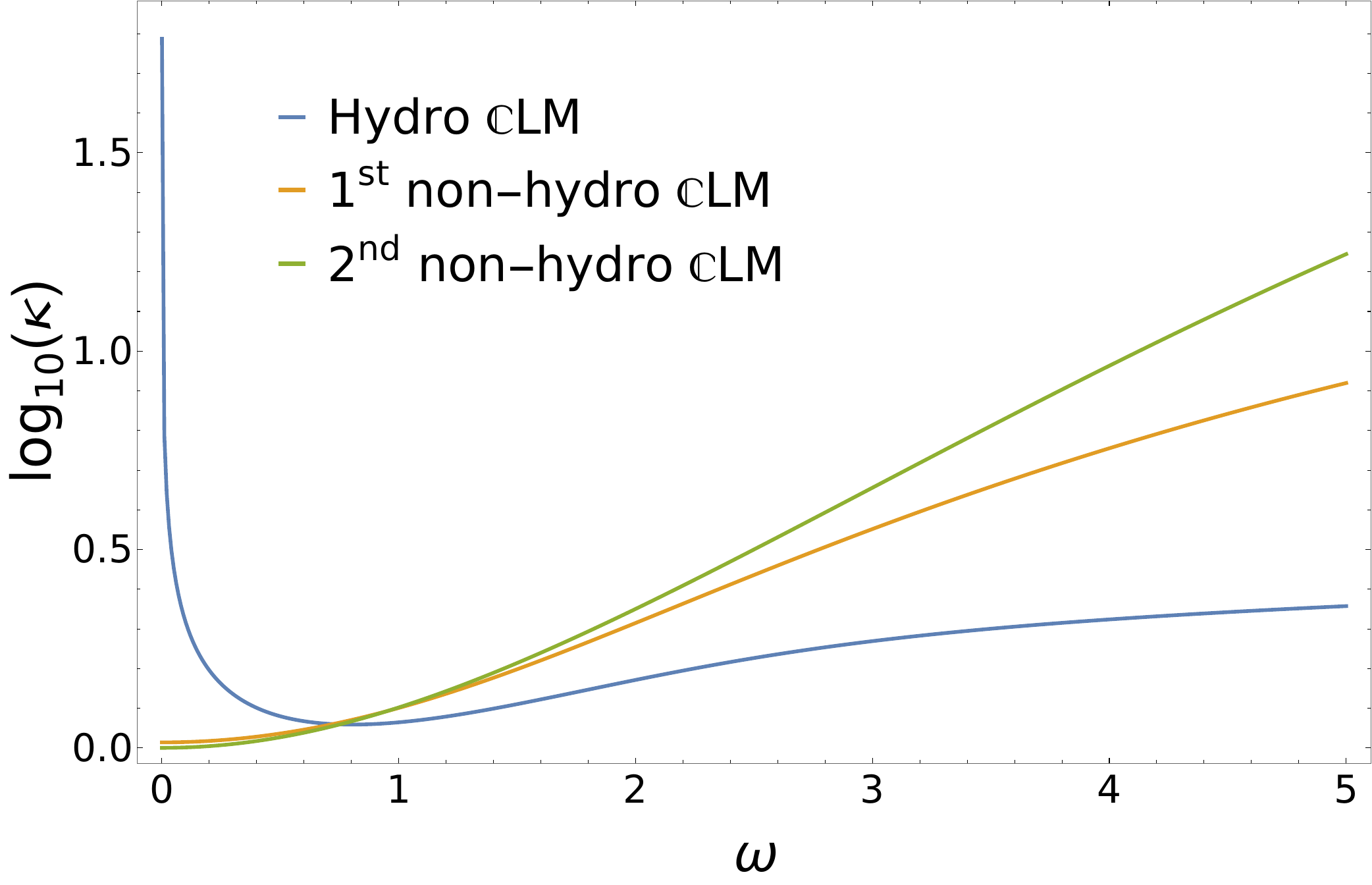}
\end{subfigure}
\hfill
\begin{subfigure}{.49\textwidth}
    \includegraphics[height=.65\textwidth]{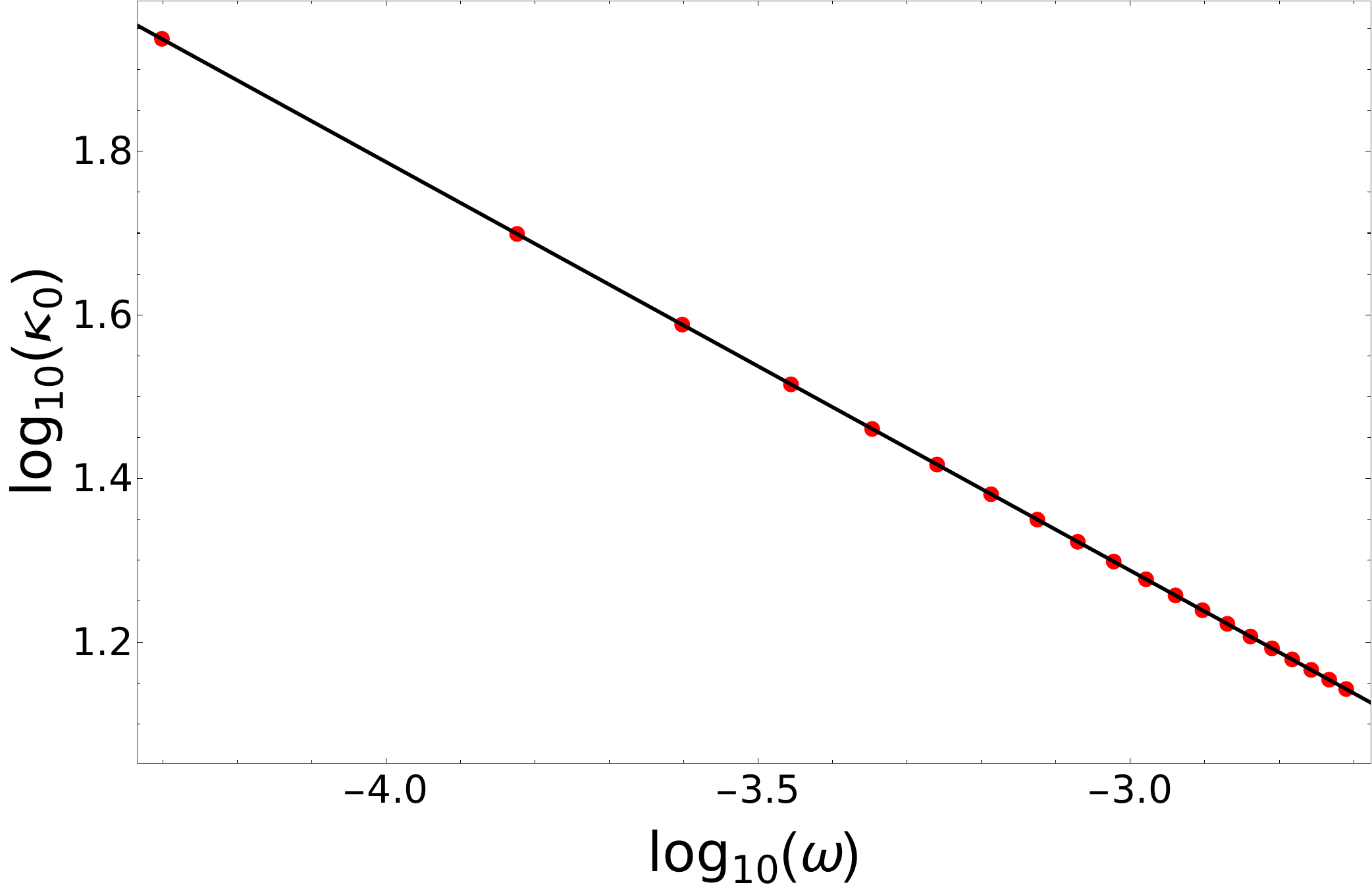}
\end{subfigure}
\caption{(Left) Condition numbers for the hydro \CLM and the first two non-hydro \CLMs as a function of the frequency $\omega$  for the SAdS$_{4+1}$ black brane. (Right) Plot of the hydrodynamic \CLM condition number at small frequency $\omega$ for the SAdS$_{4+1}$ black brane. The red dots denote the numerical values of the condition numbers and the black line is the fit $\log_{10}(\kappa_0)=-0.211-0.500\log_{10}(\omega)$.}
\label{fig:CLMKappaPlots_SAdS5Brane}
\end{figure}

We note that for the master scalar, in this case the
differential operator whose spectrum we are studying diverges as $(1-\rho)^{-1}$ towards the AdS boundary. This issue could be addressed by working with a generalized eigenvalue problem obtained from multiplying our standard one by $(1-\rho)$. However, to facilitate our analysis, we choose to maintain the standard eigenvalue problem and instead replace the rows of the discretized operator corresponding to the AdS boundary with an expression obtained by performing an asymptotic expansion around that boundary. Then, if we remove the terms that vanish when we fix our desired boundary conditions from the asymptotic expansion, we get a standard eigenvalue problem without any divergences. At any finite gridsize $N$, this procedure introduces numerical errors, which enhance the relative difference between MS and GF approaches with respect to the results found for SAdS$_{5+1}$. Nonetheless as $N$ increases the numerical error associated with changing the discretized operator becomes smaller and thus the relative difference between frameworks is also reduced (see appendix \ref{app:Convergence of the relative difference} for an explicit showcase of this phenomenon in the context of QNFs). This can be understood by noting that one can think that our procedure is altering the original operator in the region between the boundary and the second grid point. Hence when this region becomes smaller as $N$ increases, the change in the original operator decreases.

In figures \ref{fig:CLMPseudo_SAdS5Brane_w1em4}-\ref{fig:CLMPseudo_SAdS5Brane_w10} we plot the pseudospectra and the relative difference between GF and MS frameworks for $\omega=\{10^{-4},1,10\}$ and in figure \ref{fig:CLMKappaPlots_SAdS5Brane} we plot the condition numbers for the hydro and the first non-hydro \CLM. We observe great agreement between both approaches with relative differences of only $0.3\%$ in the worst case. As mentioned above, this increase in the relative difference is associated with the procedure employed to remove the divergence in the MS framework. 
The overall results are qualitatively identical to those found for AdS$_{5+1}$ in section \ref{subsubsec:CLMs of SAdS6 black brane - Numerical results}. Hence for the sake of brevity we do not repeat the discussion here.

\subsection{QNFs of  SAdS$_{4+1}$ black brane}\label{subsec:QNFs of SAdS5 black brane}
Now we analyze the spectral stability of the QNFs of the longitudinal sector of a gauge field with action \eqref{eq:Maxwell Action} in the (4+1)-dimensional black brane introduced in the previous subsection, whose metric is given by equation \eqref{eq:Generic Metric Branes} with $d=4$.

We note that this setup has already been considered in Eddington-Finkelstein coordinates in \cite{Cownden:2023dam}. In particular, in their notation our results correspond to the $Q=0$ results for the diffusive sector. We believe our presentation offers new insights into the problem. In particular, the stability analysis becomes more transparent as we have a standard eigenvalue problem rather than a generalized one. This allows us to properly quantify the stability of the hydro mode.

\subsubsection{Boundary conditions in the GF framework}\label{subsubsec:QNFs of SAdS5 black brane - Boundary conditions GF}

In the GF framework we study the pseudospectra of the eigenvalue problem \eqref{eq:GF_QNF_Eigenval_General} subject to the constraint \eqref{eq:GF_QNF_Constraint_General} with respect to the inner product \eqref{eq:GF_QNF_Energy_General}.
We impose regularity on the event horizon to select infalling modes and on the AdS boundary we demand that there is no source. Solving the equations of motion in the $\rho\rightarrow1$ region for modes with momentum $k$ and frequency $\omega$ we find the following asymptotic behavior of the fields
\begin{subequations}\label{eq: AdS5brane QNMs bcs GF} 
\begin{align}
    A_\rho&=\frac{i\omega s}{2}(1-\rho)+\frac{2i\omega v}{k^2-\omega^2}(1-\rho)+i\omega s(1-\rho)\log(1-\rho)+...\,,\\
    \alpha_\rho&=\frac{sk^2}{2}(1-\rho)+\frac{2vk^2}{k^2-\omega^2}(1-\rho)+sk^2(1-\rho)\log(1-\rho)+...\,,\\
    A_t&=s+v(1-\rho)^2+\frac{s}{2}(k^2-\omega^2)(1-\rho)^2\log(1-\rho)... \,,
\end{align}
\end{subequations}
where $s$ is the source which we should set to zero. We can impose this by defining 
\begin{subequations}
\begin{align}
    A_\rho&=\hat{A}_\rho\,,\\
    \alpha_\rho&=\hat{\alpha}_\rho\,,\\
    A_t&=(1-\rho)\hat{A}_t\,, 
\end{align}
\end{subequations}
and demanding Dirichlet boundary conditions for the rescaled fields $\{\hat{A}_\rho,\hat{\alpha}_\rho,\hat{A}_t\}$. We thus define the function space as the space of regular functions $\{\hat{A}_\rho,\hat{\alpha}_\rho,\hat{A}_t\}$ satisfying Dirichlet boundary conditions at $\rho=1$. This rescaling is consistent with demanding that any element in the function space has finite GF energy norm. We work directly in terms of the hatted variables and we rescale the original eigenvalue problem to maintain a standard eigenvalue problem for the hatted variables. 

\subsubsection{Boundary conditions in the MS framework}\label{subsubsec:QNFs of SAdS5 black brane - Boundary conditions MS}

We study the pseudospectra of the eigenvalue problem \eqref{eq:MSevproblem} with respect to the inner product \eqref{eq:MSinnerproduct}. Near the AdS boundary we have the following asymptotic behavior for the fields
\begin{subequations}\label{eq: AdS5brane QNMs bcs MS}
\begin{align}
    \psi&=\frac{\omega s}{2}\sqrt{1-\rho}+\frac{2\omega v}{k^2-\omega^2}\sqrt{1-\rho}+\omega s\sqrt{1-\rho}\log(1-\rho)+...\,,\\
    \phi&=-\frac{is\omega^2}{2}\sqrt{1-\rho}-\frac{2iv\omega^2}{k^2-\omega^2}\sqrt{1-\rho}-is\omega^2\sqrt{1-\rho}\log(1-\rho)+...\,,
\end{align}
\end{subequations}
where $s$ is the leading mode of $A_t$ in equation \eqref{eq: AdS5brane QNMs bcs GF}. Hence, we designate the function space as the space of regular functions $\{\hat\phi,\hat\psi\}$, where the hatted fields are defined with the non-analytic $\sqrt{1-\rho}$ factors removed 
\begin{subequations}
\begin{align}
    \psi=\sqrt{1-\rho}\,\hat{\psi}\,,\\
    \phi=\sqrt{1-\rho}\,\hat{\phi}\,.
\end{align}
\end{subequations}

\subsubsection{Numerical results}\label{subsubsec:QNFs of SAdS5 black brane - Numerical results}
Now we turn onto the discussion of our numerical results. In this section we use a grid of 150 points and work with $7\times$MachinePrecision in both frameworks. We note that as for the \CLMs in SAdS$_{4+1}$, we have a divergent operator in the MS framework and we deal with that divergence as explained previously in section \ref{subsubsec:CLMs of SAdS5 black brane - Numerical results}. 

\begin{figure}[h!]
\centering
\begin{subfigure}{.49\textwidth}
    \includegraphics[height=.87\textwidth]{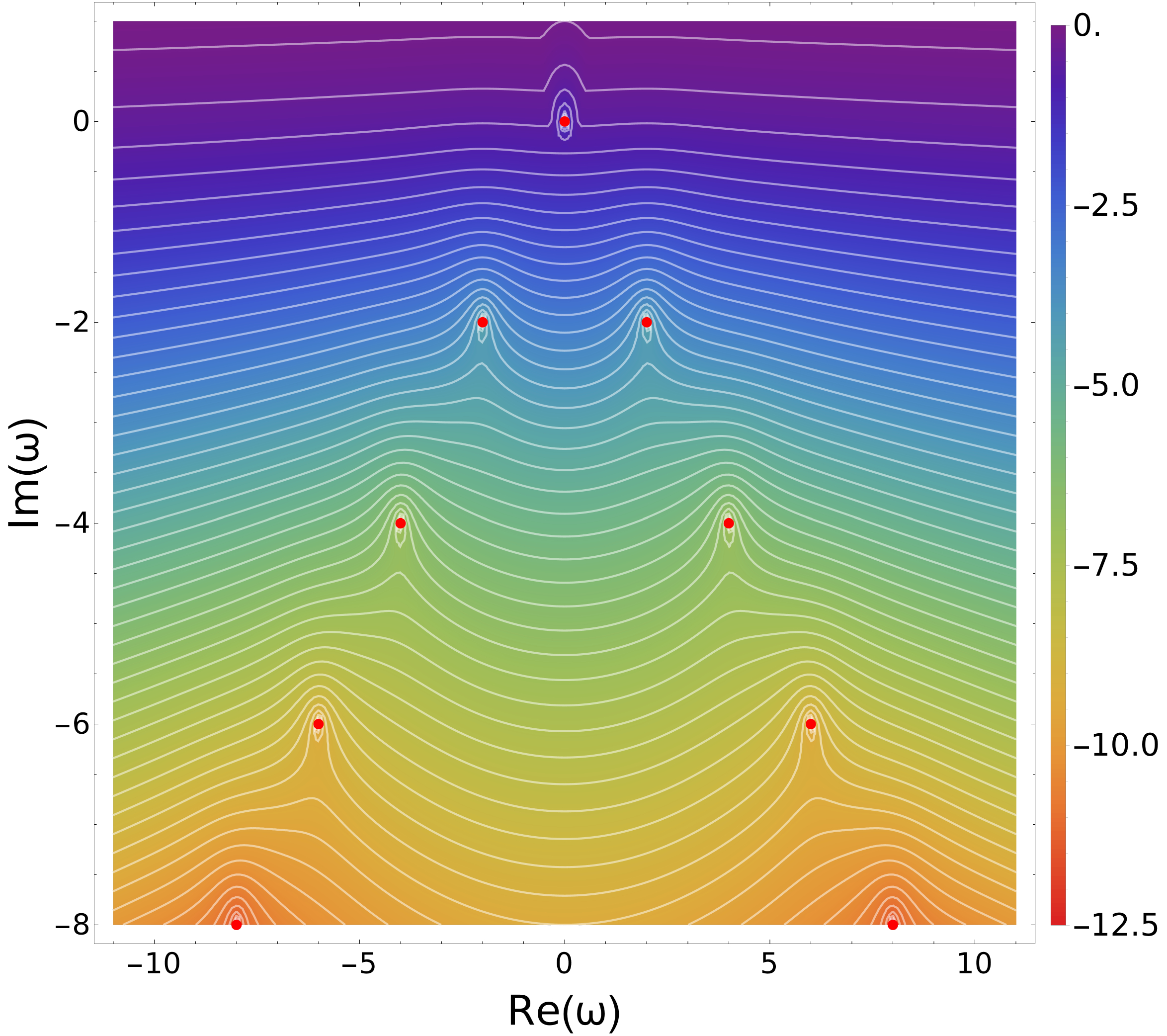}
    \label{fig:QNFPseudo_SAdS5Brane_GF_w1em4}
\end{subfigure}
\hfill
\begin{subfigure}{.49\textwidth}
    \includegraphics[height=.87\textwidth]{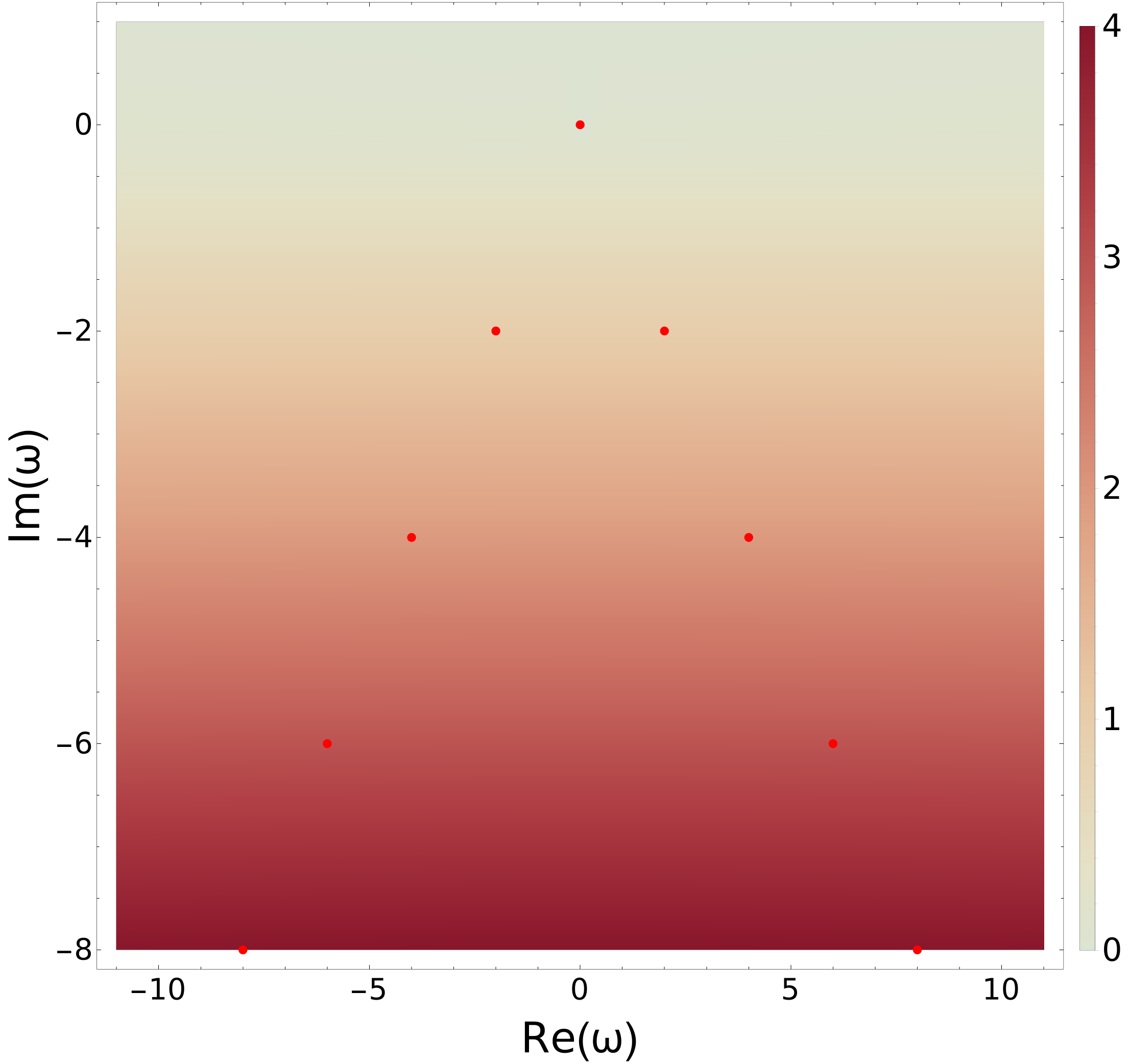}
    \label{fig:QNFPseudo_SAdS5Brane_Comparison_w1em4}
\end{subfigure}
\caption{(Left) QNF pseudospectrum at $k=10^{-4}$ for the SAdS$_{4+1}$ black brane. The white lines denote the boundaries of different $\varepsilon$-pseudospectra and the heat map corresponds to the logarithm in base 10 of the inverse of the norm of the resolvent. (Right) Heatmap of the percentage difference between MS and GF frameworks. The red dots correspond to the QNFs.}
\label{fig:QNFPseudo_SAdS5Brane_w1em4}
\end{figure}

\begin{figure}[h!]
\centering
\begin{subfigure}{.49\textwidth}
    \includegraphics[height=.87\textwidth]{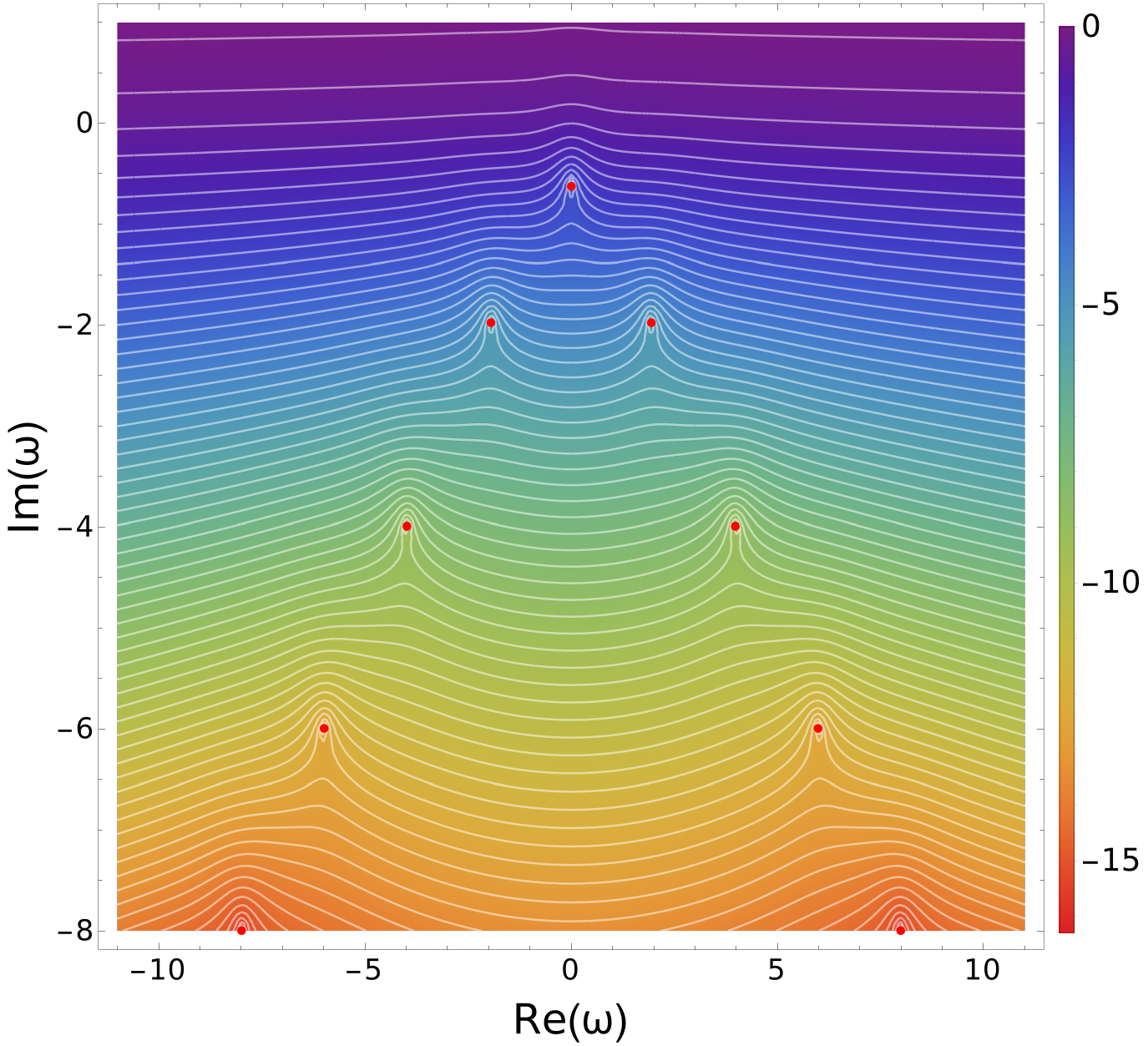}
    \label{fig:QNFPseudo_SAdS5Brane_GF_w1}
\end{subfigure}
\hfill
\begin{subfigure}{.49\textwidth}
    \includegraphics[height=.87\textwidth]{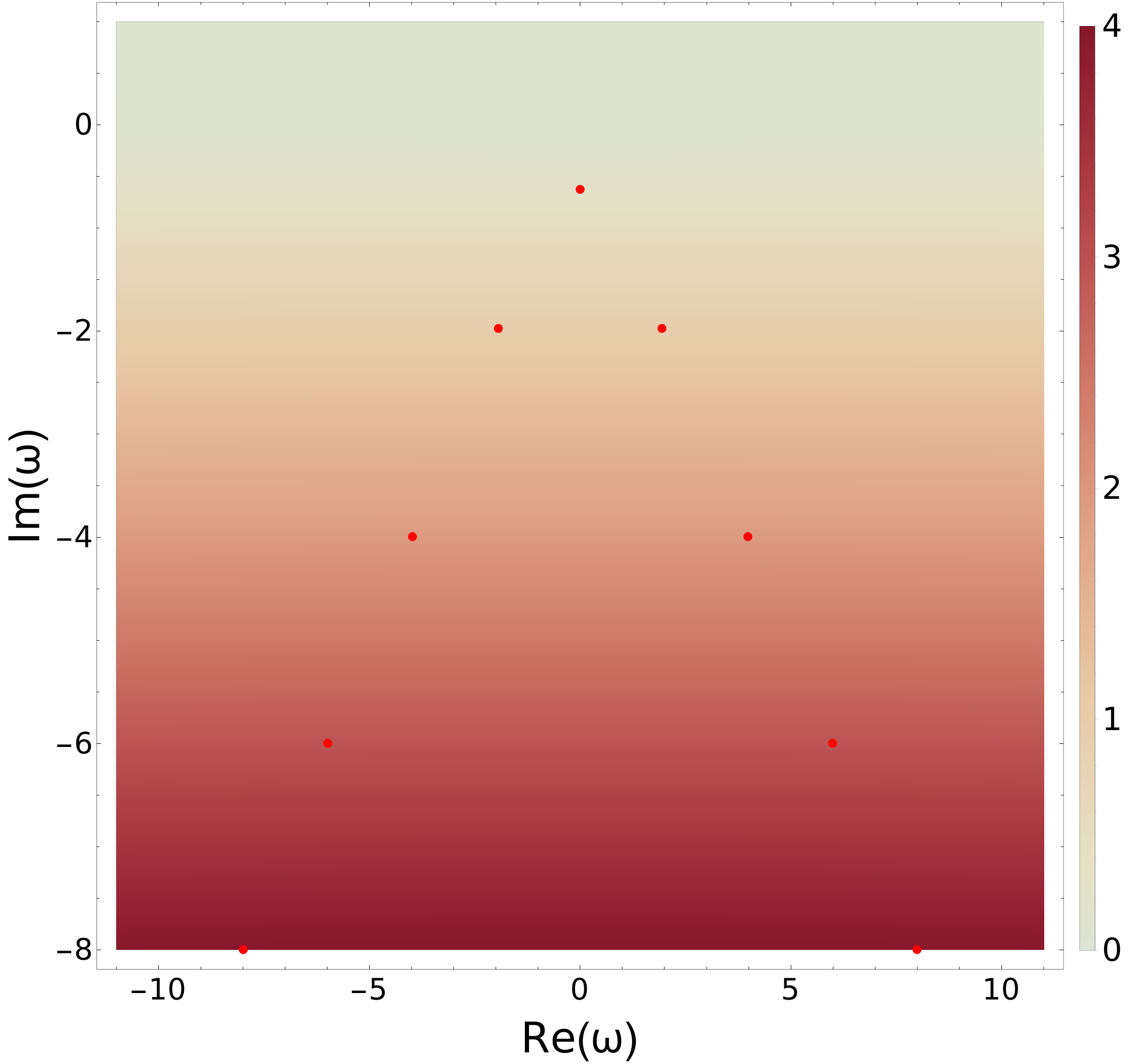}
    \label{fig:QNFPseudo_SAdS5Brane_Comparison_w1}
\end{subfigure}
\caption{(Left) QNF pseudospectrum at $k=1$ for the SAdS$_{4+1}$ black brane. The white lines denote the boundaries of different $\varepsilon$-pseudospectra and the heat map corresponds to the logarithm in base 10 of the inverse of the norm of the resolvent. (Right) Heatmap of the percentage difference between MS and GF frameworks. The red dots correspond to the QNFs.}
\label{fig:QNFPseudo_SAdS5Brane_w1}
\end{figure}

\begin{figure}[h!]
\centering
\begin{subfigure}{.49\textwidth}
    \includegraphics[height=.87\textwidth]{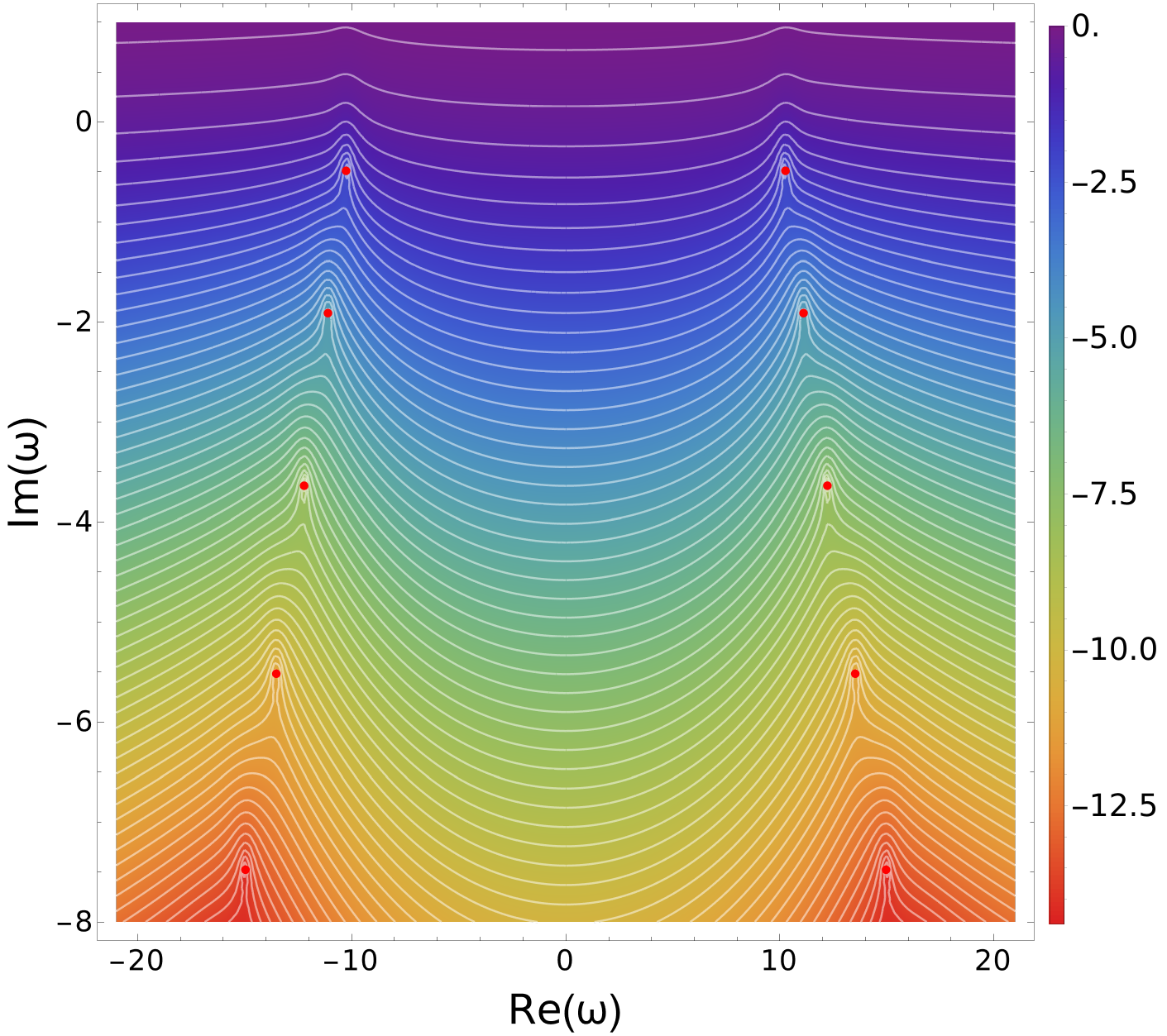}
    \label{fig:QNFPseudo_SAdS5Brane_GF_w10}
\end{subfigure}
\hfill
\begin{subfigure}{.49\textwidth}
    \includegraphics[height=.87\textwidth]{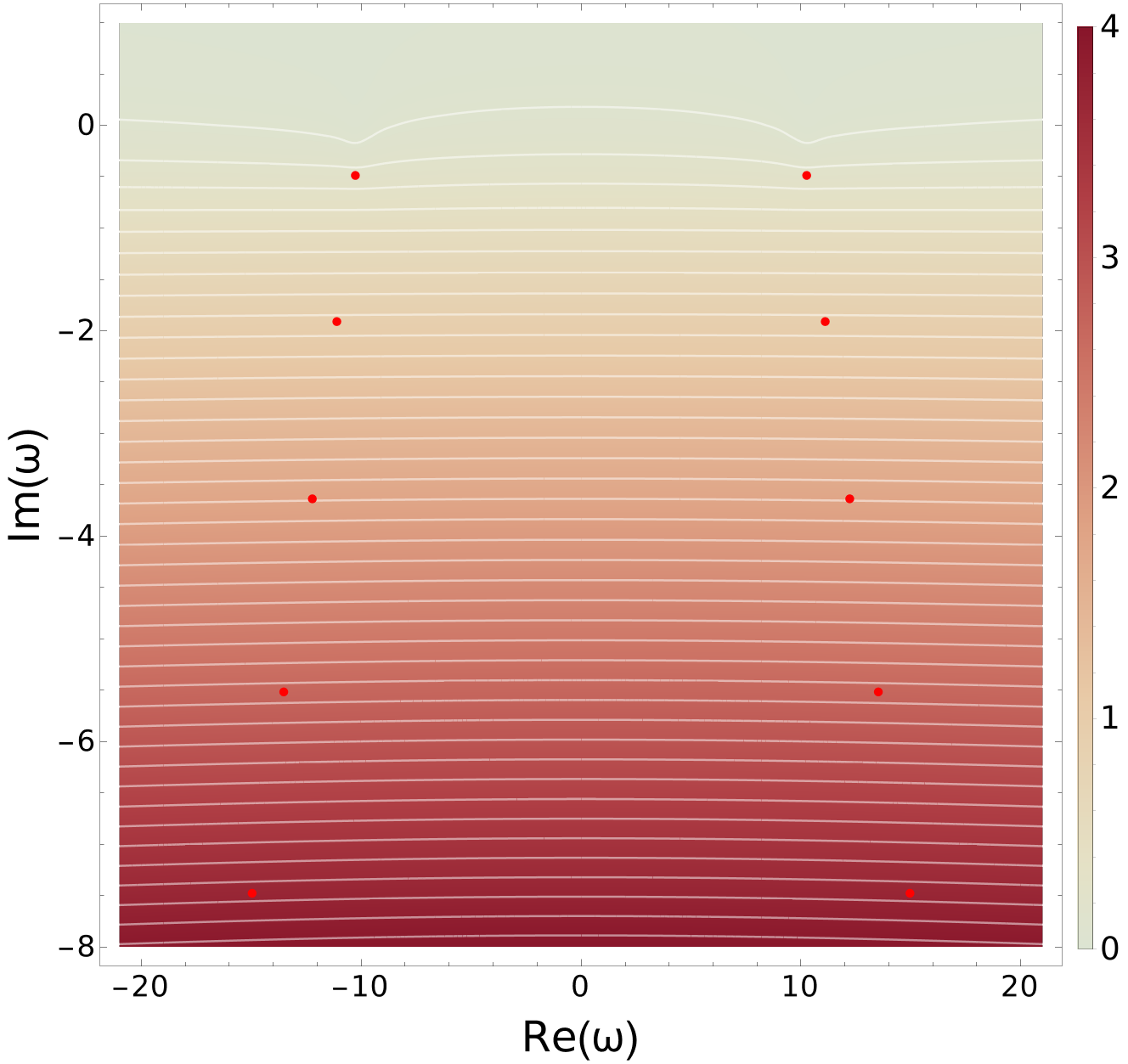}
    \label{fig:QNFPseudo_SAdS5Brane_Comparison_w10}
\end{subfigure}
\caption{(Left) QNF pseudospectrum at $k=10$ for the SAdS$_{4+1}$ black brane. The white lines denote the boundaries of different $\varepsilon$-pseudospectra and the heat map corresponds to the logarithm in base 10 of the inverse of the norm of the resolvent. (Right) Heatmap of the percentage difference between MS and GF frameworks. The red dots correspond to the QNFs.}
\label{fig:QNFPseudo_SAdS5Brane_w10}
\end{figure}

\begin{figure}[h!]
\centering
\begin{subfigure}{.49\textwidth}
    \includegraphics[width=\textwidth]{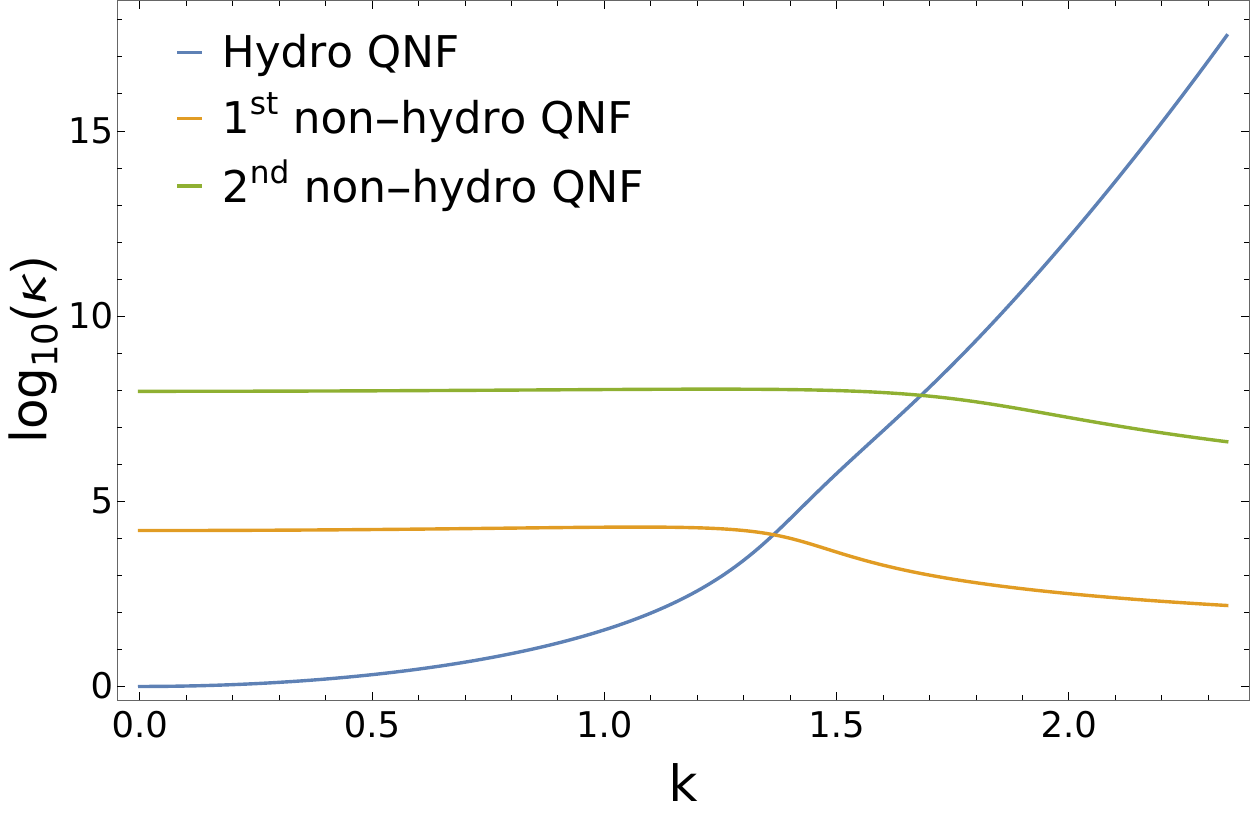}
    \caption{Condition numbers.}
    \label{fig:QNFKappa_SAdS5Brane}
\end{subfigure}
\hfill
\begin{subfigure}{.49\textwidth}
    \includegraphics[width=\textwidth]{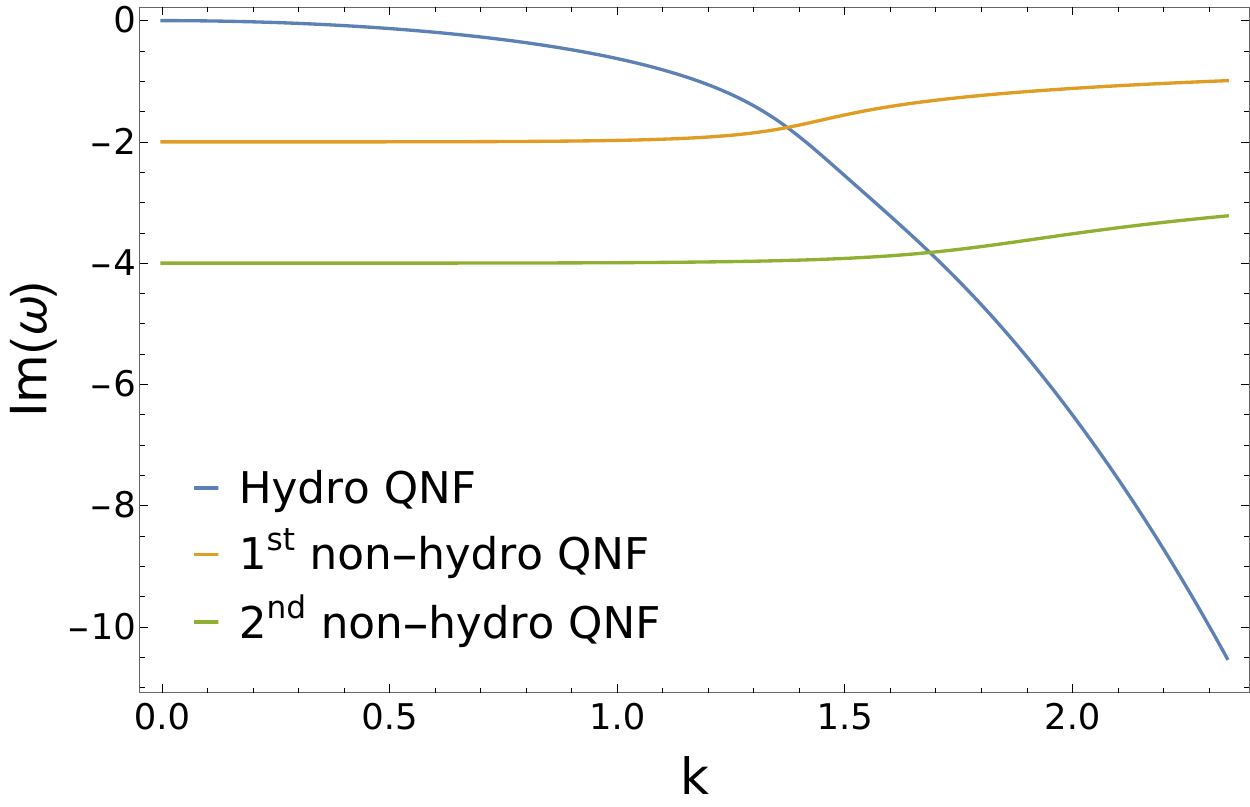}
    \caption{Imaginary parts.}
    \label{fig:ImQNF_SAdS5Brane}
\end{subfigure}
\caption{Condition numbers and imaginary part of the hydro QNF and the first two non-hydro QNFs as a function of the momentum $k$ for the SAdS$_{4+1}$ black brane. We see that the hydro QNF becomes more unstable than the non-hydro QNFs at values of the momentum similar to those for which the imaginary part of the former crosses that of the latter.}
\label{fig:QNF_SAdS5Brane}
\end{figure}
In figures \ref{fig:QNFPseudo_SAdS5Brane_w1em4}-\ref{fig:QNFPseudo_SAdS5Brane_w10} we plot the pseudospectra and the relative difference between GF and MS frameworks for $k=\{10^{-4},1,10\}$ and in figure \ref{fig:QNF_SAdS5Brane} we plot the condition numbers for the hydro and the first non-hydro QNF. We note that as for the \CLMs we have that our procedure of dealing with the divergence in the MS approach introduces numerical errors, which enhance the relative difference between MS and GF approaches. As before, the relative difference decreases as the grid becomes denser as we explicitly shown in appendix \ref{app:Convergence of the relative difference}.
Regarding the final results, we find qualitatively the same picture as for AdS$_{5+1}$ in section \ref{subsubsec:QNFs of SAdS6 black brane - Numerical results}, hence for the sake of brevity we do not repeat the discussion here.

\subsection{QNFs of  SAdS$_{4+1}$ black hole}\label{subsec:QNFs of SAdS5 black hole}
Finally we conclude by showcasing how our procedure can be applied to a spherical black hole. For concreteness we focus on the gauge field QNFs of spherical black hole in AdS$_{4+1}$. The corresponding metric in regular coordinates is given by:
\begin{align}
     ds^2&=\frac{1}{(1-\rho)^2}(-fdt^2+2r_h \beta dtd\rho+\frac{1-\beta^2}{f}r_h^2d\rho^2+r_h^2 d\Omega_2)\,,\nonumber\\
    f=\frac{r_h^2}{l^2}&+(1-\rho)^2-\left(1+\frac{r_h^2}{l^2}\right)(1-\rho)^4\,,\qquad \beta=\left[1+\frac{f}{(1-\rho)^{5}}\right]^{-1/2}\,,
\end{align}
where $l$ is the AdS length scale and $r_h$ is a constant with units of length. The Hawking temperature of these black holes is $T=(2\pi r_h)^{-1}+(r_h)/(\pi l^2)$. Unlike for the branes of the previous subsections, here $r_h$ and $l$ both enter non-trivially into the equations of motion and thus they cannot be simultaneously fixed without losing information. Hence we have a family of black holes labeled by the dimensionless ratio $r_h/l$. From the QFT point of view we see this as a consequence of the conformal symmetry of the UV. Now the dual QFT lives on a sphere and thus one can construct a dimensionless quantity to label state by combining the temperature and the radius of the sphere. Hence we have a continuous family of states labeled by a single parameter.

For the sake of brevity, here we focus on the $l=r_h=1$ case and postpone discussing the behavior of the QNFs and pseudospectra as a function of the ratio $r_h/l$ to future work. 

\subsubsection{Boundary conditions in the GF framework}\label{subsubsec:QNFs of SAdS5 black hole - Boundary conditions GF}

In the GF framework we study the pseudospectra of the eigenvalue problem \eqref{eq:GF_QNF_Eigenval_General} subject to the constraint \eqref{eq:GF_QNF_Constraint_General} with respect to the inner product \eqref{eq:GF_QNF_Energy_General}.
We define the boundary conditions following AdS/CFT as our guideline. We impose regularity on the event horizon to select infalling modes and on the AdS boundary we demand that there is no source. Solving the equations of motion in the $\rho\rightarrow1$ region for modes with total angular momentum $\lambda^2=L(2 + L)$ and frequency $\omega$ we find the following asymptotic behavior of the fields 
\begin{subequations}\label{eq: AdS5hole QNMs bcs GF}
\begin{align}
    A_\rho&=\frac{i\omega s}{2}(1-\rho)+\frac{2i\omega v}{\lambda^2-\omega^2}(1-\rho)+i\omega s(1-\rho)\log(1-\rho)+...\,,\\
    \alpha_\rho&=\frac{s\lambda^2}{2}(1-\rho)+\frac{2v\lambda^2}{\lambda^2-\omega^2}(1-\rho)+s\lambda^2(1-\rho)\log(1-\rho)+...\,,\\
    A_t&=s+v(1-\rho)^2+\frac{s}{2}(\lambda^2-\omega^2)(1-\rho)^2\log(1-\rho)+... \,,
\end{align}
\end{subequations}
where $s$ is the source which we should set to zero. We can impose this by defining 
\begin{subequations}
\begin{align}
    A_\rho&=\hat{A}_\rho\,,\\
    \alpha_\rho&=\hat{\alpha}_\rho\,,\\
    A_t&=(1-\rho)\hat{A}_t\,, 
\end{align}
\end{subequations}
and demanding Dirichlet boundary conditions for the rescaled fields $\{\hat{A}_\rho,\hat{\alpha}_\rho,\hat{A}_1\}$. We thus define the function space as the space of regular functions $\{\hat{A}_\rho,\hat{\alpha}_\rho,\hat{A}_1\}$ satisfying Dirichlet boundary conditions at $\rho=1$. This rescaling is consistent with demanding that any element in the function space has finite GF energy norm. We work directly in terms of the hatted variables and we rescale the original eigenvalue problem to maintain a standard eigenvalue problem for the hatted variables. 

\subsubsection{Boundary conditions in the MS framework}\label{subsubsec:QNFs of SAdS5 black hole - Boundary conditions MS}

We study the pseudospectra of the eigenvalue problem \eqref{eq:MSevproblem} with respect to the inner product \eqref{eq:MSinnerproduct}. Near the AdS boundary we have the following asymptotic behavior for the fields 
\begin{subequations}
\begin{align}
    \psi&=\frac{\omega s}{2}\sqrt{1-\rho}+\frac{2\omega v}{\lambda^2-\omega^2}\sqrt{1-\rho}+\omega s\sqrt{1-\rho}\log(1-\rho)+...\,,\\
    \phi&=-\frac{is\omega^2}{2}\sqrt{1-\rho}-\frac{2iv\omega^2}{\lambda^2-\omega^2}\sqrt{1-\rho}-is\omega^2\sqrt{1-\rho}\log(1-\rho)+...\,,
\end{align}
\end{subequations}
where $s$ is the leading mode of $A_t$ in equation \eqref{eq: AdS5hole QNMs bcs GF}. Hence, we designate the function space as the space of regular functions $\{\hat\psi,\hat\phi\}$, where the hatted fields are defined with the non-analytic $\sqrt{1-\rho}$ factors removed 
\begin{subequations}
\begin{align}
    \psi=\sqrt{1-\rho}\hat{\psi}\,,\\
    \phi=\sqrt{1-\rho}\hat{\phi}\,.
\end{align}
\end{subequations}

\subsubsection{Numerical results}\label{subsubsec:QNFs of SAdS5 black hole - Numerical results}
Now we turn onto the discussion of our numerical results. In this section we use a grid of 60 points and work with $6\times$MachinePrecision in both frameworks. As for the \CLMs and QNFs of SAdS$_{4+1}$ black brane, we have a divergent operator in the MS framework and we deal with that divergence as explained previously in section \ref{subsubsec:CLMs of SAdS5 black brane - Numerical results}. 

\begin{figure}[h!]
\centering
\begin{subfigure}{.49\textwidth}
    \includegraphics[height=.87\textwidth]{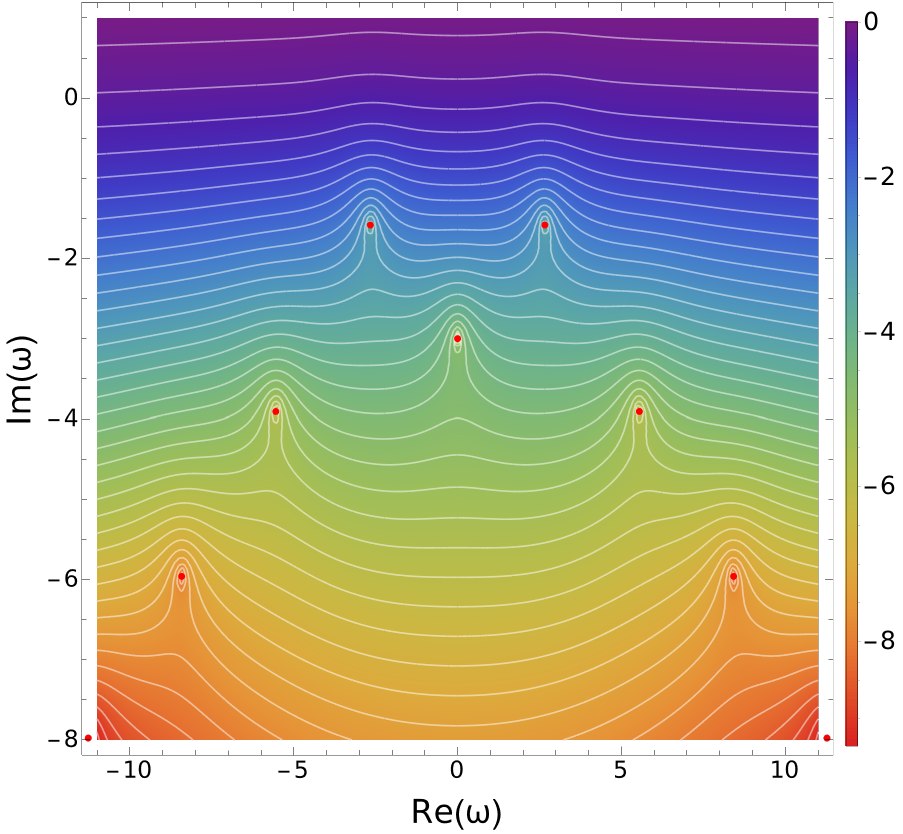}
    \label{fig:QNFPseudo_SAdS5BH_GF_L1_LoverRh1}
\end{subfigure}
\hfill
\begin{subfigure}{.49\textwidth}
    \includegraphics[height=.87\textwidth]{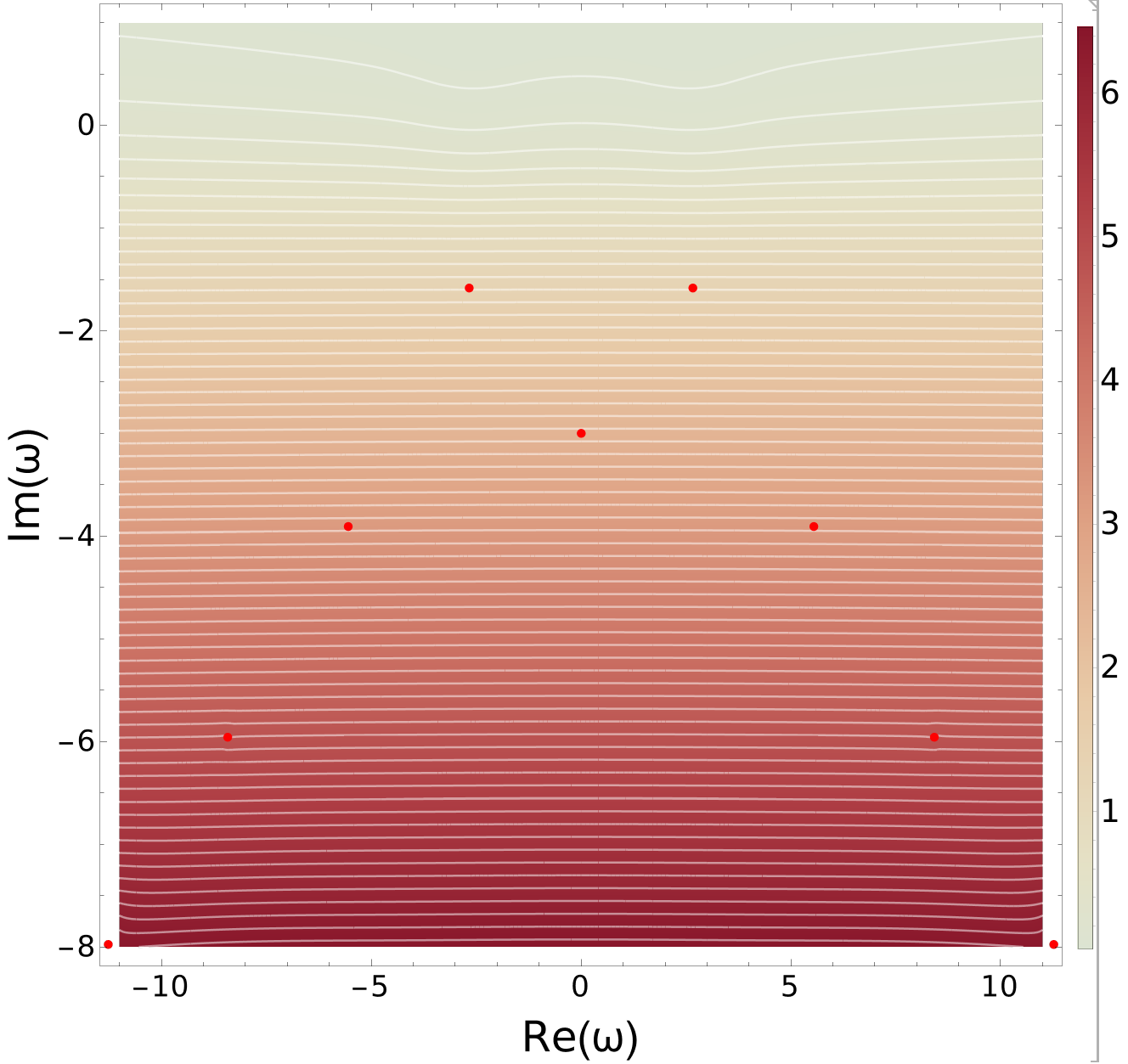}
    \label{fig:QNFPseudo_SAdS5BH_GF_L1_LoverRh1_RelativeError}
\end{subfigure}
\caption{(Left) QNF pseudospectrum at $L=1$ for the SAdS$_{4+1}$ black hole. The white lines denote the boundaries of different $\varepsilon$-pseudospectra and the heat map corresponds to the logarithm in base 10 of the inverse of the norm of the resolvent. (Right) Heatmap of the percentage difference between MS and GF frameworks. The red dots correspond to the QNFs.}
\label{fig:QNFPseudo_SAdS5Brane_L1_LoverRh1}
\end{figure}

\noindent For the sake of brevity we showcase results for the $L=1$ ($\lambda^2=L(2+L)=3$) sector only. This is the sector with the smallest possible value of the angular momenta for a Gauge field and thus it contains the mode closest to the real axis. In figure \ref{fig:QNFPseudo_SAdS5Brane_L1_LoverRh1} we plot the pseudospectra and the relative difference between GF and MS frameworks. We note that as for the black brane our procedure of dealing with the divergence in the MS approach introduces numerical errors, which enhance the relative difference between MS and GF approaches. Nonetheless, for the chosen grid the maximal difference is of only $7\%$, which could be lowered by making the grid denser as previously discussed. 

Unlike the case of the black brane, all QNFs are unstable, even the would-be hydrodynamic one (the one with no real part). This is a consequence of the large gap appearing for this mode, whose magnitude is related to the ratio $l/r_h$; 
 which we have chosen to be 1. This suggests that as the ratio $l/r_h$ decreases and the gap is reduced we should recover the spectral stability observed for the black brane at in the small momentum limit.

\

\section{Conclusions and outlook}\label{sec: Conclusions}

In this work we have studied the stability of complex linear momenta ($\mathbb{C}$LMs) and quasinormal frequencies (QNFs) of gauge fields using two different approaches: the gauge field (GF) approach and the master scalar (MS) approach.

The GF approach presents a physically well defined method to study numerically solutions of the gauge field in a determined background. In principle this method is totally inequivalent from the master scalar, presented as a not-so-well physically motivated approach to the same problem. However, we have shown that for black branes both approaches are related by Hodge duality in an effective 3-dimensional spacetime, giving physical meaning to the MS formulation.

Although recently it has been argued that ingoing Eddington Finkelstein (IEF) coordinates are better suited for numerically studying QNF pseudospectra \cite{Cownden:2023dam,Boyanov:2023qqf}, we work in regular coordinates \cite{Warnick:2013hba,Arean:2023ejh}. Unlike in \cite{Boyanov:2023qqf}, we find that the convergence properties match those expected from the results of \cite{Warnick:2013hba}. Moreover, when trying to give qualitative results in the convergent region of the pseudospectrum, IEF coordinates are ill-suited as they yield a generalized eigenvalue problem, for which the notion of stability is less transparent. In particular, this makes stable modes harder to identify.

The gauge field, and equivalently the Hodge dual scalar, present two clear advantages over the naively constructed master scalar. The first of them is that the energy norm built directly from the master scalar differs from the energy norm of the gauge field by a boundary term. In the 5-dimensional case, this term may allow for modes with negative energy, making the norm not positive definite. The other advantage comes when computing \CLMs. As the master scalar approach works with a 2-dimensional scalar field, it does not present a natural way to pose the problem of studying \CLMs. In this sense, Hodge duality gives a natural way of approaching this problem as one lands on a 3-dimensional space.

As was already seen in \cite{Garcia-Farina:2024pdd} pseudospectra of QNFs and \CLMs are qualitatively different. In particular the zero frequency complex momenta are normal, as they correspond to glueball masses of a dimensionally reduced theory. Here we have found, however a new phenomenon: the diffusion modes seen as a linear momentum modes show enhanced spectral instability (see also \cite{Cao:2025afs}). This is a direct consequence of an exceptional point sitting at zero frequency where the modes would collide. As is well known, the residue of the diffusion mode vanishes at zero frequency and strictly speaking this exceptional point is not present at $\omega=0$. Its influence is however detectable in the pseudospectrum in a neighborhood of $\omega=0$. 

There are of course several directions in which our work has to be extended. First one should study the pseudospectra of quasinormal modes and complex momentum modes of gauge fields in asymptotically flat space using the rectangular matrix approach. Another important question is how our approach can be generalized to gravitational modes. Finding an appropriate norm in terms of the metric fields itself seems difficult since there does not exist a gauge invariant local energy momentum tensor. Nevertheless it should be quite interesting and useful to use investigate this problem further in particular in view of the role of quasinormal modes in the ringdown phase of black hole mergers.
Lastly, one could also study pseudospectra in the vicinity of pole collisions. Our results for \CLMs suggests that there should be an enhancement of the spectral instability as one gets close to them. Hence, pseudospectrum analysis could serve as an useful tool to diagnose them.

\section*{Acknowledgments}
We thank D. Areán for valuable discussions. We thank B. Cownden, C. Pantelidou and M. Zilh\~ao for clarifications regarding their paper.
This work is supported through the grants CEX2020-001007-S and PID2021-123017NB-100, PID2021-127726NB-I00 funded by MCIN/AEI/10.13039/501100011033 and by ERDF ``A way of making Europe''. 
The work of D.G.F. is supported by FPI grant PRE2022-101810. 
The work of PSB is supported by Fundaci\'on S\'eneca, Agencia de Ciencia y Tecnolog\'ia de la Regi\'on de Murcia, grant 21609/FPI/21, Spanish MINECO grant PID2024-155685NB-C22 and Fundaci\'on S\'eneca de la Regi\'on de Murcia FSRM/10.13039/100007801 (22581/PI/24).

\appendix
\section{Stability of general eigenvalue problems}\label{app:Stability of general eigenvalue problems}
As discussed in \cite{Trefethen:2005}, one can generalize the notion of $\varepsilon$-pseudospectra to generalized eigenvalue problems of the form
\begin{equation}\label{eq:GeneralizedEigenvalueProblem}
    (A-\lambda B)u=0\,,
\end{equation}
with $A$ and $B$ differential operators. In this case one can find multiple possible definitions and should choose according to the particular problem under consideration (see e.g., Ch. 45 of \cite{Trefethen:2005} for a review). 

In the context of gravitational physics, it has been argued in \cite{Cownden:2023dam} that the adequate definition is given by
\begin{equation}
    \sigma_\varepsilon(A,B)=\left\{z\in\mathbb{C}: \left\Vert (A-z B)^{-1}\right\Vert<1/\varepsilon\right\}\,,
\end{equation}
which was proven in \cite{VanDorsselaer1997} to be equivalent to 
\begin{equation}
    \sigma_\varepsilon(A,B)=\left\{z\in\mathbb{C}\,, \exists V \text{ with } \Vert V\Vert <\varepsilon:z\in\sigma(A+V,B)\right\}\,.
\end{equation}
This definition indeed is the logical generalization of pseudospectra to generalized eigenvalue problems as it retains the notion of maximal region containing the eigenvalues after a perturbation. Moreover, with this definition one can define pseudospectra in infalling Eddington coordinates which were argued in \cite{Boyanov:2023qqf} to show better convergence properties.

\begin{figure}
\centering
\begin{subfigure}{.32\textwidth}
    \includegraphics[width=\textwidth]{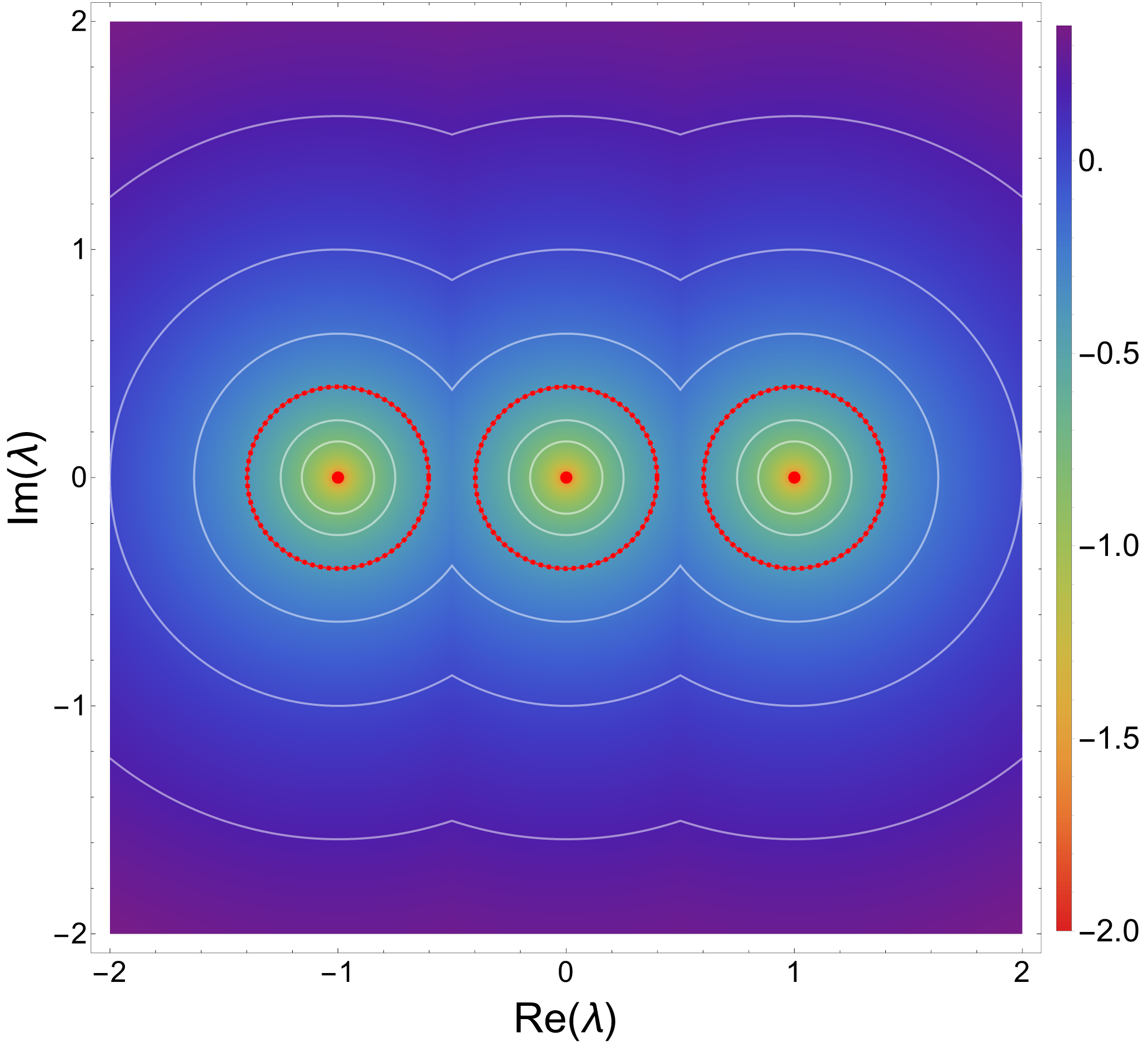}
    \caption{}
    \label{fig:GeneralizedEigenvaluePseudo_a}
\end{subfigure}
\hfill
\begin{subfigure}{.32\textwidth}
    \includegraphics[width=\textwidth]{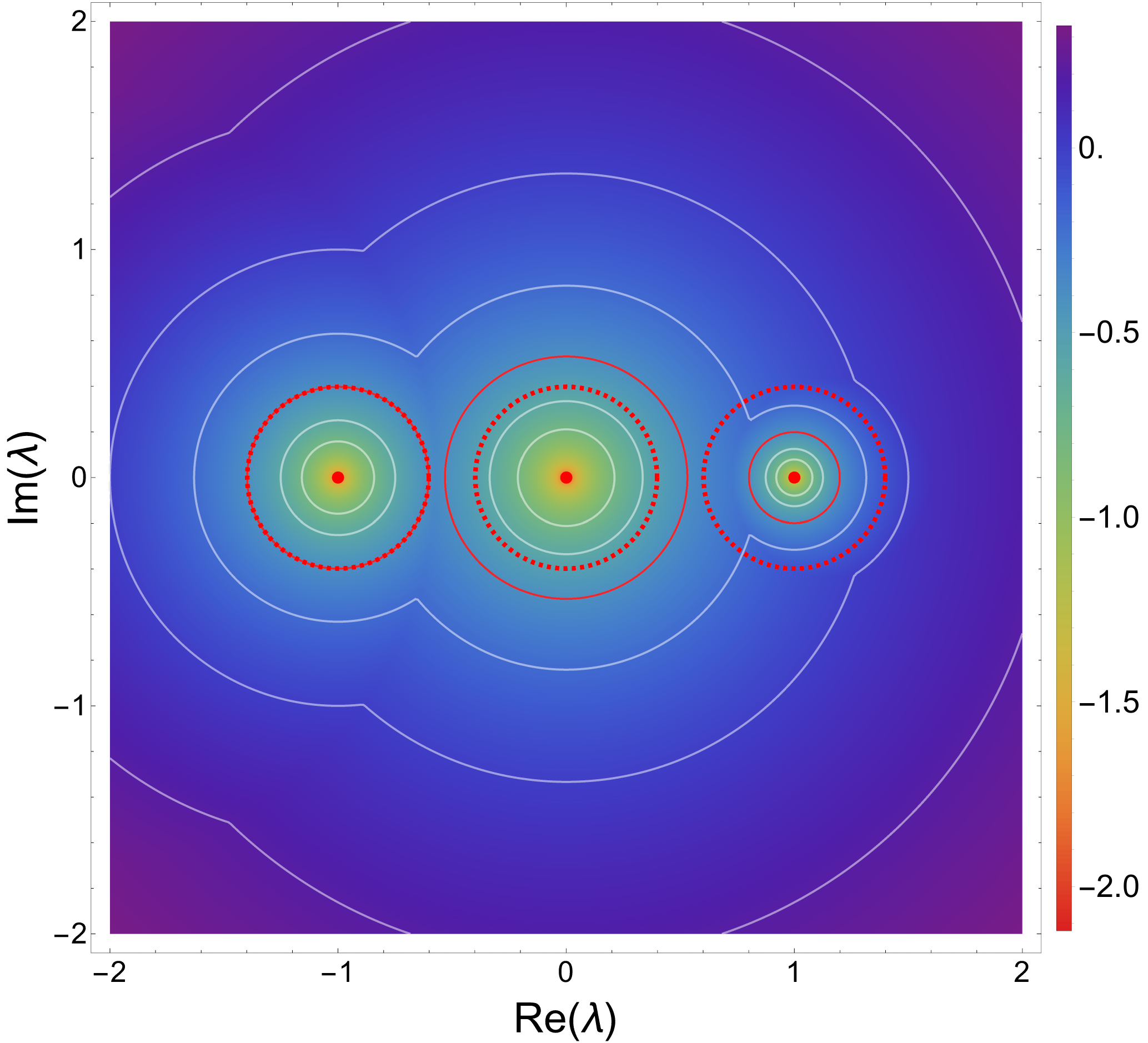}
    \caption{}
    \label{fig:GeneralizedEigenvaluePseudo_b}
\end{subfigure}
\hfill
\begin{subfigure}{.32\textwidth}
    \includegraphics[width=\textwidth]{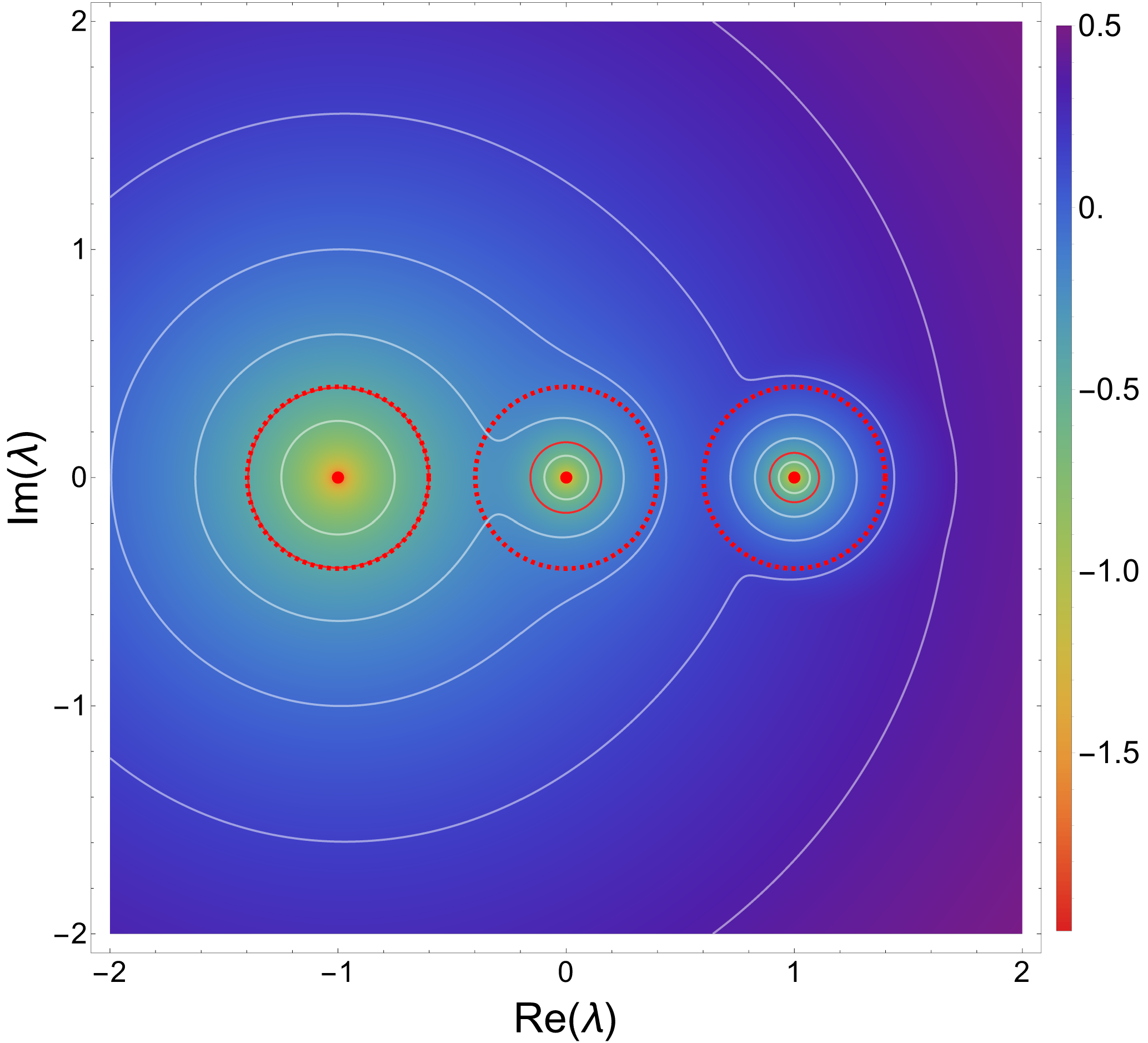}
    \caption{}
    \label{fig:GeneralizedEigenvaluePseudo_c}
\end{subfigure}
\caption{Pseudospectrum of $A-\lambda$ (a) and of $S(A-\lambda)$ with diagonal (b) and non-diagonal (c) $S$. The red dots correspond to the eigenvalues, the white lines represent the boundaries of various $\varepsilon$-pseudospectra, and the heat map represent to the logarithm in base 10 of the inverse of the resolvent. The solid red lines are the boundary of the $10^{-2/5}$-pseudospectrum and the dashed red circles are circles of radius $10^{-2/5}$ centered in the eigenvalues. For (a) we see that dashed and solid lines coincide while for (b) and (c) this is no longer the case.}
\label{fig:GeneralizedEigenvaluePseudo}
\end{figure}

We want to stress however one important subtlety in this formulation which is often overlooked. While for a standard eigenvalue problem, a spectrally stable operator has $\varepsilon$-pseudospectra whose boundaries are given by circles of radius $\varepsilon$; this is no longer the case for generalized eigenvalue problems. Let us illustrate this with a simple example. Take $A$ to be the following $3\times3$ matrix
\begin{equation}
    A=\text{diag}(-1,0,1)\,,
\end{equation}
and consider the standard euclidean inner product product
\begin{equation}\label{eq:l2_3-Norm}
    \langle u,v\rangle_2=\sum_{i=1}^{3}u_i^*v_i\,.
\end{equation}
Then, $A$ is a normal matrix and when studying the eigenvalue problem $(A-\lambda)u=0$, we observe a stable $\varepsilon$-pseudospectra comprised of circles of radius $\varepsilon$, as illustrated in figure \ref{fig:GeneralizedEigenvaluePseudo_a}. On the other hand, if we consider instead the eigenvalue problem $S(A-\lambda)u=0$ with $S=\text{diag}(1,3/4,2)$, we find a qualitatively different picture as shown in figure \ref{fig:GeneralizedEigenvaluePseudo_b}. We still observe that the $\varepsilon$-pseudospectrum is comprised of circles but now their radius is not only different from $\varepsilon$, but also not constant. Moreover, if we take a non-diagonal $S$
\begin{equation}
    S=\begin{pmatrix}1&0&1\\ 0&3/4&4\\ 1&4&2\end{pmatrix}
\end{equation}
as in figure \ref{fig:GeneralizedEigenvaluePseudo_c}, we see that even the circular structure is lost.
However, these generalized eigenvalue problems should be equivalent to the standard one discussed above. It is the non-trivial $B$ that deforms the notion of stability. This is can be explained by the lack of invariance of the pseudospectrum under multiplication by non-unitary operators
\begin{equation}
    \Vert(A-z)^{-1}\Vert\neq\Vert(SA-z S)^{-1}\Vert\,.
\end{equation}
We want to stress that this mismatch is not a flaw of pseudospectra, rather it is a physical feature. Clearly multiplying by a non-unitary operator changes the typical scales of the original operator, despite leaving the spectrum invariant. Thus, it makes sense that one should need larger/smaller perturbations in order to have an effect on the spectrum of the new shifted operator. This is particularly easy to understand if we take the $B$ operator in \eqref{eq:GeneralizedEigenvalueProblem} to be a constant. Then, clearly, a perturbation $V$ whose effect when $B=1$ is of order $O(1)$ will have a much smaller effect when $B=10^6$ as its relative size is lower. 

All in all, this does not mean that one should avoid generalized eigenvalue problems, or that they are ill-defined. It only suggests that one should be extra careful when interpreting pseudospectra of generalized eigenvalue problems. In particular, as seen in figure \ref{fig:GeneralizedEigenvaluePseudo_c}, it can lead to significant confusion as the notion of stability becomes deformed. Indeed, although naively figure \ref{fig:GeneralizedEigenvaluePseudo_c} suggest instability, in reality we know from the analysis of figure \ref{fig:GeneralizedEigenvaluePseudo_a} that it is an artifact of the non-trivial $B$.\footnote{An alternative viewpoint would be that multiplying by $S$ above corresponds to a linear transformation between Hilbert spaces and that the non-trivial pseudospectrum arises from taking the norm after the transformation to be \eqref{eq:l2_3-Norm}. This necessarily implies that, before the transformation the norm cannot be \eqref{eq:l2_3-Norm} and thus alters the original notion of normality making the operator unstable and rendering the comparison between figures \ref{fig:GeneralizedEigenvaluePseudo_a} and \ref{fig:GeneralizedEigenvaluePseudo_c} pointless.} For this reason we advocate for the use of regular coordinates where, in most cases, these issues can be avoided. Alternatively, one could also use a controlled case where the eigenvalues are known to be stable to calibrate (e.g. the $\omega\rightarrow0$ limit of \CLMs). 

To conclude, it is worth mentioning that for rectangular eigenvalue problems, a non-trivial $B$ matrix need not spoil the naive intuition of stability. If $B$ is chosen such that $(A-\lambda B)$ is equivalent to a standard eigenvalue problem $(L-\lambda)$ restricted to a region of the whole function space, then the standard notion of stability is trivially inherited from that of $(L-\lambda)$. We explicitly see this in our numerical computations performed in the GF framework in section \ref{sec:Holographic setup}.

\section{Numerical methods}\label{app:Numerical methods}
In this appendix we briefly discuss the numerical procedure allowing us to compute the pseudospectra of section \ref{sec:Holographic setup}. The core idea is to discretize the system and use theorem compute the pseudospectra of the resulting matrices. As the computation of the pseudospectra of matrices has already been explained in section \ref{sec:Pseudospectra and stability} here we briefly discuss the discretization procedure. For a more detailed discussion we refer the reader to \cite{Jaramillo:2020tuu,Arean:2023ejh}

We start by discretizing the radial direction $\rho$ in a Chebyshev-Gauss-Lobatto grid with points
\begin{equation}\label{eq:Chebyshev grid}
    \rho_j=\frac{1}{2}\left[1-\cos\left(\frac{j\pi}{N}\right)\right]\,,\qquad j=0,1,...,N\,,
\end{equation}
Then, in this grid we construct the discretized versions of our numerical operators by replacing the derivatives with the corresponding Chebyshev derivatives \cite{Trefethen:2000}. This corresponds to approximating the modes by sums of Chebyshev polynomials of degree $N$. Remarkably, as discussed in \cite{Jaramillo:2020tuu}, consistency with the discretization, requires us to construct the energy norm in a grid with twice as many points and then interpolate it back to the original grid. This ensures that the norm is exact for polynomials of degree $N$. The explicit details on how to perform the interpolation can be found in \cite{Jaramillo:2020tuu}.

Regarding the function space, as indicated in the main text we need to work in terms of the hatted variables satisfying regularity on the horizon and Dirchlet/regularity on the AdS boundary. Remarkably, as a byproduct of the discretization, regularity is immediately guaranteed as any finite sum of Chebyshev polynomials is always a regular function.  On the other hand, Dirichlet boundary conditions are imposed by removing from the operators the rows and columns corresponding to the AdS boundary.

\section{Convergence of the relative difference between MS and GF approaches for AdS$_{4+1}$}\label{app:Convergence of the relative difference}
As argued in the main text, in the setups with AdS$_{4+1}$ asymptotics (sections \ref{subsec:CLMs of SAdS5 black brane}-\ref{subsec:QNFs of SAdS5 black hole}), we have to deal with the divergences in the operator appearing in the MS approach in a subtle manner which introduces numerical errors which are reduced as the grid becomes denser. Hence, even if the pseudospectra do not converge (as for the QNFs) we observe that the relative difference between MS and GF approaches is lowered as the aforementioned numerical errors in the MS approach become less prominent. 

\begin{figure}[h!]
\centering
\begin{subfigure}{.32\textwidth}
    \includegraphics[width=\textwidth]{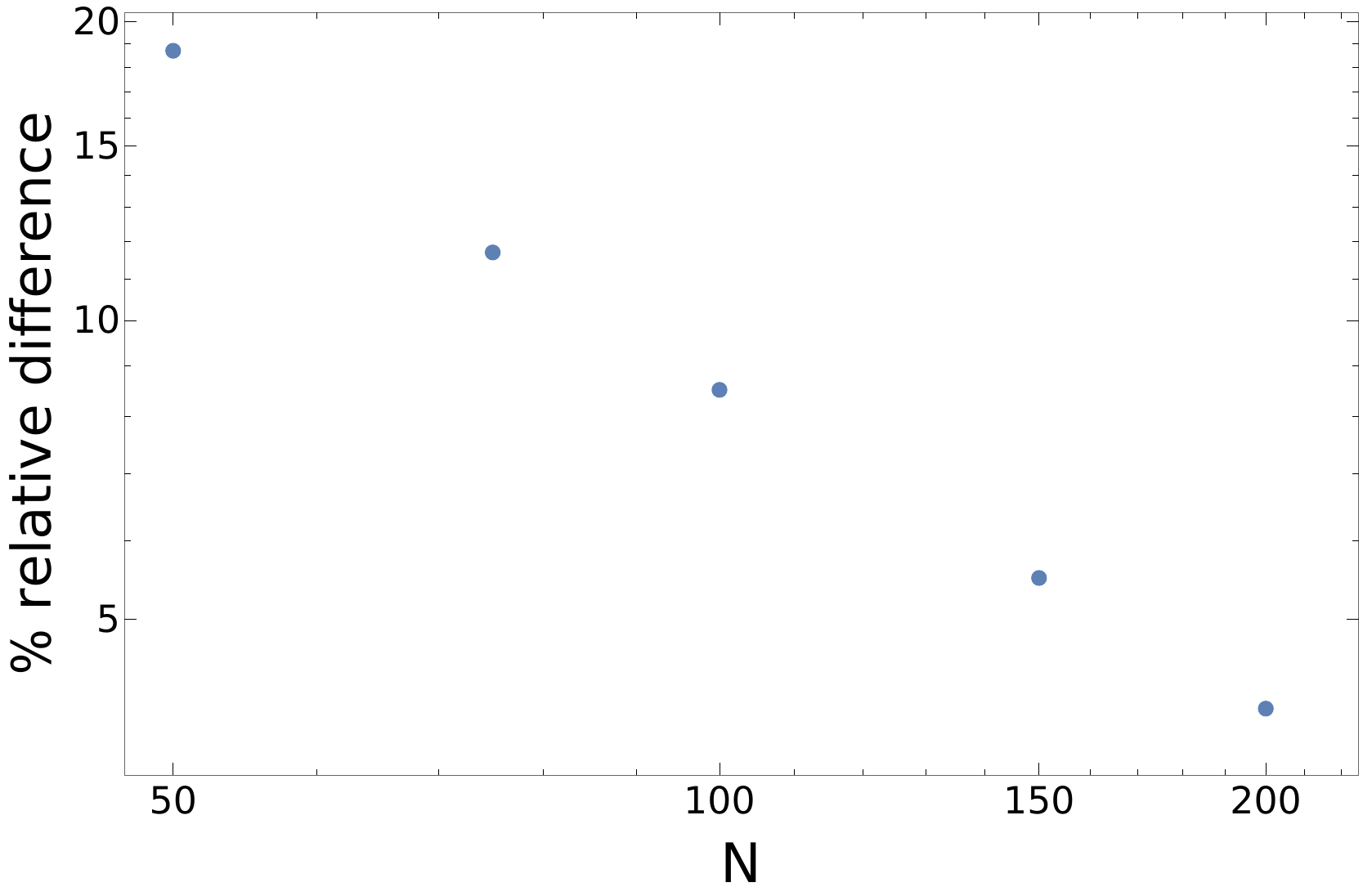}
    \caption{$k=10^{-4}$}
    \label{fig:QNFPseudo_SAdS5Brane_w1em4_RelativeError}
\end{subfigure}
\hfill
\begin{subfigure}{.32\textwidth}
    \includegraphics[width=\textwidth]{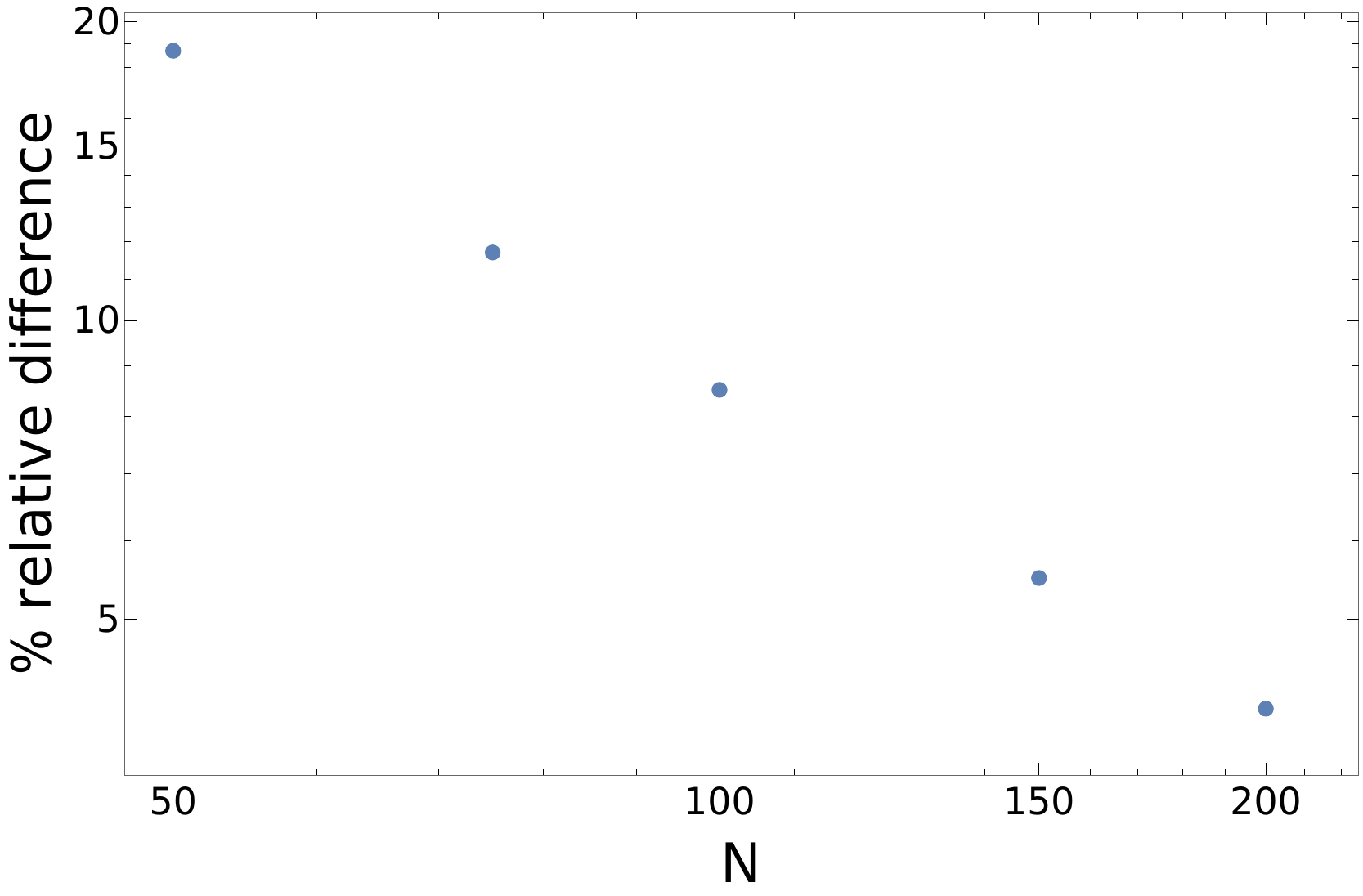}
    \caption{$k=1$}
    \label{fig:QNFPseudo_SAdS5Brane_w1_RelativeError}
\end{subfigure}
\hfill
\begin{subfigure}{.32\textwidth}
    \includegraphics[width=\textwidth]{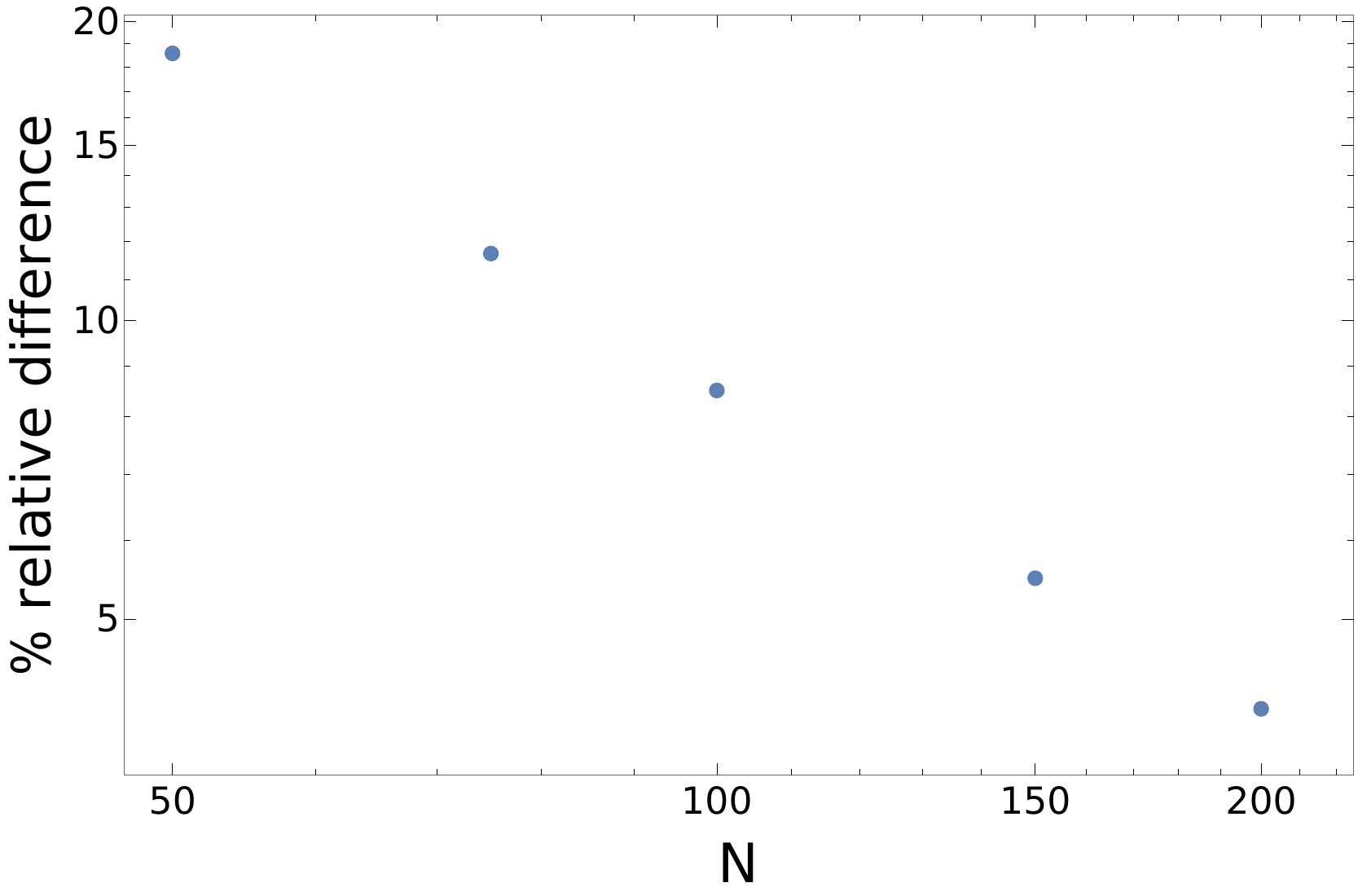}
    \caption{$k=10$}
    \label{fig:QNFPseudo_SAdS5Brane_w10_RelativeError}
\end{subfigure}
\caption{Relative difference as a function of the gridsize $N$ for the QNFs of the SAdS$_{4+1}$ black brane discussed in section \ref{subsec:QNFs of SAdS5 black brane} for different values of the momentum $k$. We find power-law behavior as indicated by the fact that the data points seem to follow straight lines in log-log scale. }
\label{fig:NFPseudo_SAdS5Brane_RelativeError}
\end{figure}

In this appendix we illustrate this fact for the QNFs of the SAdS$_{4+1}$ black brane of section \ref{subsec:QNFs of SAdS5 black brane}. In figure \ref{fig:NFPseudo_SAdS5Brane_RelativeError} we plot the relative difference against the grid size $N$ in a log-log plot for different values of the momentum $k$. Remarkably, we do observe that the relative difference is reduced and as $N$ increases, with power-law behavior.

\section{Convergence tests for QNFs and \CLMs}\label{app:Convergence tests pseudo}
\begin{figure}
\centering
\begin{subfigure}{.49\textwidth}
    \includegraphics[width=\textwidth]{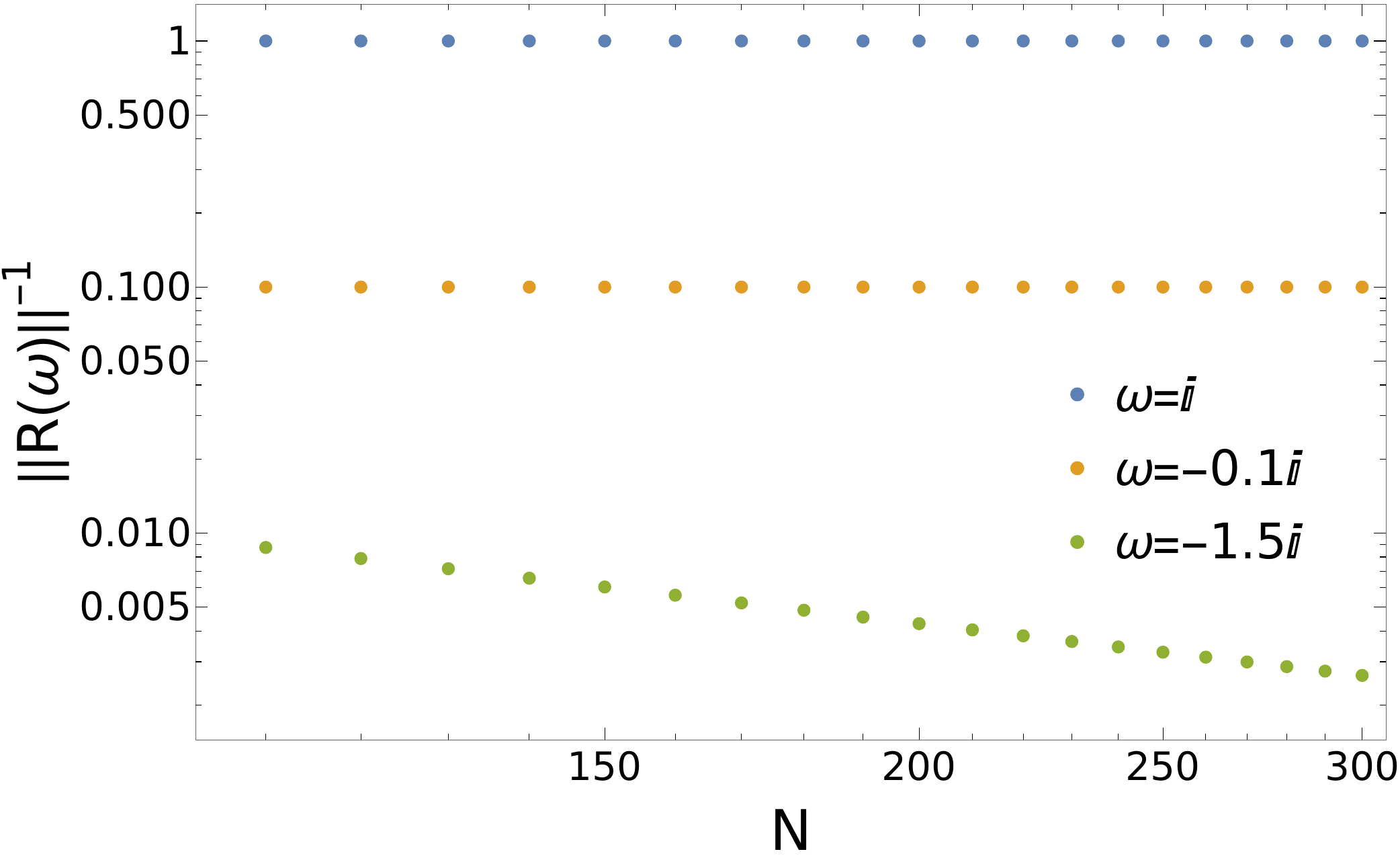}
    \caption{$k=10^{-4}$}
    \label{fig:QNF_convergenceAdS6k1em4}
\end{subfigure}
\hfill
\begin{subfigure}{.49\textwidth}
    \includegraphics[width=\textwidth]{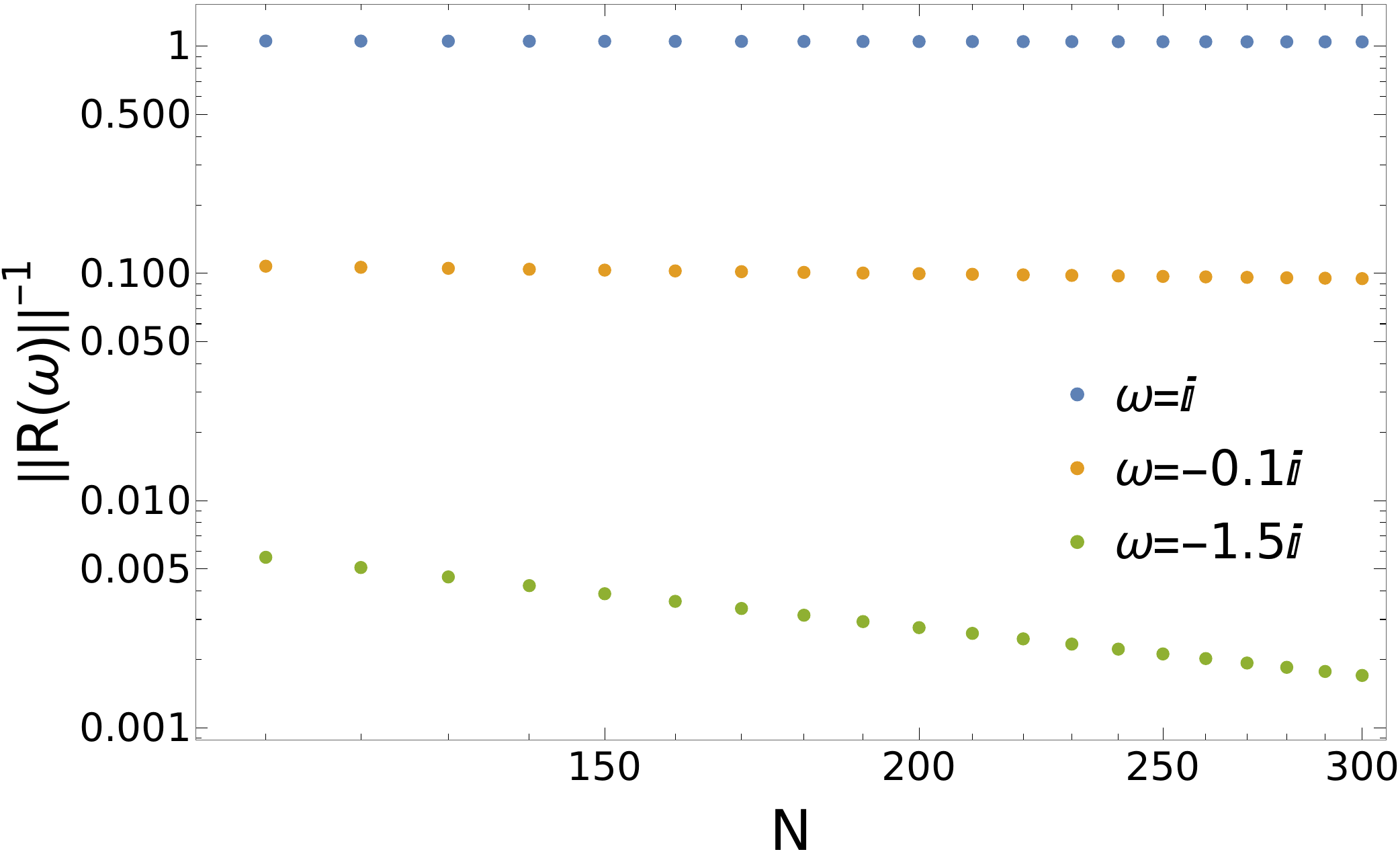}
    \caption{$k=1$}
    \label{fig:QNF_convergenceAdS6k1}
\end{subfigure}
\caption{Inverse of the norm of the resolvent of the QNF pseudospectrum for the SAdS$_{5+1}$ black brane as a function of the grid size $N$ evaluated at the representative points $\omega=\{i,-0.1i,-1.5i\}$. The results agree with the prediction that the pseudospectrum should converge in the region $\Im(\omega)>-5/4$.}
\label{fig:QNF_convergenceAdS6}
\end{figure}

In sections \ref{sec:Longitudinal gauge field} and \ref{subsec:Enorm MS} we found that in the continuum limit the region $\Im(\omega)<-\pi T$ belonged to the spectrum of the operator. Hence we argued that the pseudospectra will only converge above this band. In this section we check this prediction for the QNFs of the SAdS$_{5+1}$ brane discussed in section \ref{subsec:QNFs of SAdS6 black brane}, for which, in our units, $T=5/(4\pi)$ so that the pseudospectra will converge in the region $\Im(\omega)>-5/4$. In figure \ref{fig:QNF_convergenceAdS6} we plot the inverse of the norm of the resolvent as a function of the grid size $N$ evaluated at $\omega={1 i,-0.1 i}$ (convergent region) and at $\omega=-1.5i$ (non-convergent region). We observe that indeed the pseudospectrum converges at $\omega=-0.1 i$, thus ensuring that for sufficiently small momenta, the results concerning the stability of the hydrodynamic mode are independent of $N$. On the other hand, for $\omega=-1.5i$ we see that the resolvent is not converging. This signals that its value tends to zero as $N$ increases, in agreement with the fact that in the continuum limit this point belongs to the spectrum of the operator and thus the inverse of the norm of the resolvent should be zero.

It is worth mentioning that even in the region $\Im(\omega)>-\pi T$, the numerical results converge more slowly the closer one gets to the line $\Im(\omega)=-\pi T$. For this reason, one generically expects to need to go to very large grids to numerically identify the boundary of the convergent region. 

\begin{figure}
\centering
\begin{subfigure}{.49\textwidth}
    \includegraphics[height=.61\textwidth]{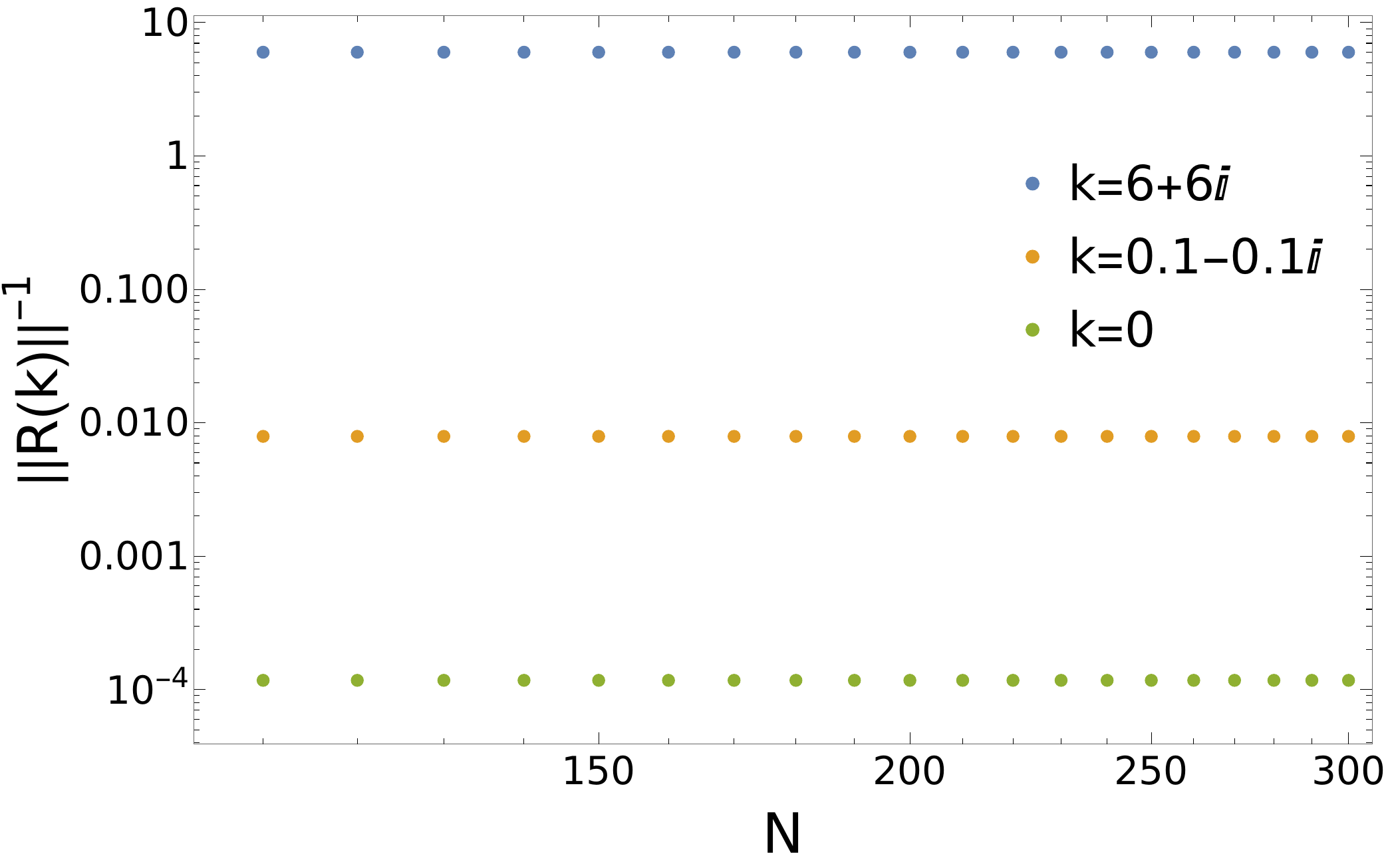}
    \caption{$\omega=10^{-4}$}
    \label{fig:CLM_convergenceAdS6k1em4}
\end{subfigure}
\hfill
\begin{subfigure}{.49\textwidth}
    \includegraphics[height=0.6\textwidth]{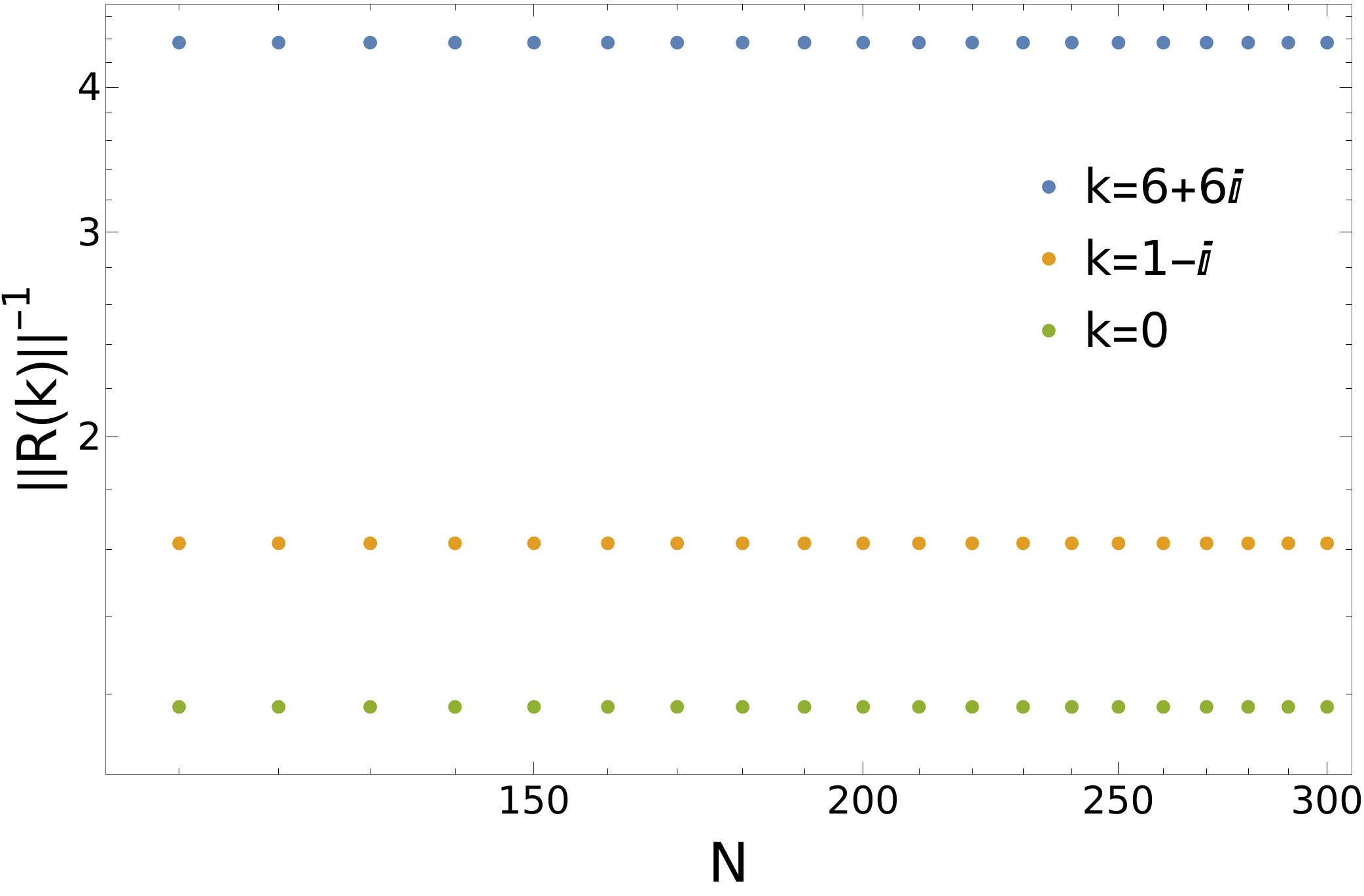}
    \caption{$\omega=1$}
    \label{fig:CLM_convergenceAdS6k1}
\end{subfigure}
\caption{Inverse of the norm of the resolvent of the \CLM pseudospectrum for the SAdS$_{5+1}$ black brane as a function of the grid size $N$ evaluated at the representative points $k=\{6+6i,1-1i,0\}$. The results are convergent, in agreement with the prediction that the \CLM pseudospectrum should be convergent everywhere in the complex $k$ plane.}
\label{fig:CLM_convergenceAdS6}
\end{figure}

In sections \ref{subsec: Modifications for CLMs} and \ref{sec:hodgeclm} we argued that the \CLM pseudospectra converges everywhere as for fixed real frequency $\omega$ outgoing modes are always removed from the spectrum. In this section we check this prediction for the \CLMs of the SAdS$_{5+1}$ brane discussed in section \ref{subsec:CLMs of SAdS6 black brane}. In figure \ref{fig:CLM_convergenceAdS6} we plot the inverse of the norm of the resolvent as a function of the grid size $N$ for different points in the complex momentum plane. We observe that indeed the pseudospectrum converges for all point concluding that, as argued in \cite{Garcia-Farina:2024pdd}, \CLMs do not suffer the convergence issues present for QNFs.

\bibliographystyle{JHEP}

\bibliography{biblio}
\end{document}